\documentclass[lettersize,journal]{IEEEtran}
\IEEEoverridecommandlockouts
\usepackage{cite}
\usepackage{amsmath, amssymb, amsthm}
\usepackage{amsfonts}
\usepackage{graphicx}
\usepackage{textcomp}
\usepackage{xcolor}
\usepackage[colorlinks=true, linkcolor=black, citecolor=black]{hyperref}
\usepackage{diagbox}

\usepackage{caption} 
\usepackage{algorithmicx,algorithm}

\usepackage[noend]{algpseudocode}
\usepackage{subfigure}

\usepackage{enumitem}

\usepackage{lineno} 
\begin{document}	
\switchlinenumbers

\title{FairSort: Learning to Fair Rank for Personalized Recommendations 
in Two-Sided Platforms}

\markboth{Journal of \LaTeX\ Class Files,~Vol.~14, No.~8, August~2023}%
{Shell \MakeLowercase{\textit{et al.}}: A Sample Article Using IEEEtran.cls for IEEE Journals}


\author{\IEEEauthorblockN{Guoli Wu\textsuperscript{1}, Zhiyong Feng\textsuperscript{1}, Shizhan Chen\textsuperscript{1}, Hongyue Wu\textsuperscript{1,*\thanks{*Corresponding author: Hongyue Wu.}}, Xiao Xue\textsuperscript{1}\\Jianmao Xiao\textsuperscript{2}, Guodong Fan\textsuperscript{1},  Hongqi Chen\textsuperscript{1}, Jingyu Li\textsuperscript{1}}\\ 
\IEEEauthorblockA{\textsuperscript{1}\textit{College of Intelligence and Computing, Tianjin University, Tianjin, China} \\
\textit{\textsuperscript{2}School of Software, Jiangxi Normal University, Nanchang, China}, jm\textunderscore xiao@jxnu.edu.cn\\
\{wuguoli\textunderscore it999, zyfeng, shizhan, hongyue.wu, jzxuexiao, guodongfan, hongqi, lijingyu\textunderscore working\}@tju.edu.cn}\vspace{-1.8\baselineskip} 
}
 
\maketitle

\begin{abstract}

Traditional recommendation systems focus on maximizing user satisfaction by suggesting their favorite items.  This user-centric approach may lead to unfair exposure distribution among the providers.  On the contrary, a provider-centric design might become unfair to the users. Therefore, this paper proposes a re-ranking model FairSort\footnote{\textbf{Reproducibility:}The code and datasets are available at  \url{https://github.com/WuGL-CS/FairSort}} to find a trade-off solution among user-side fairness, provider-side fairness, and personalized recommendations utility. Previous works habitually treat this issue as a knapsack problem, incorporating both-side fairness as constraints.

In this paper, we adopt a novel perspective, treating each recommendation list as a runway rather than a knapsack. In this perspective, each item on the runway gains a velocity and runs within a specific time, achieving re-ranking for both-side fairness. Meanwhile, we ensure the Minimum Utility Guarantee for personalized recommendations by designing a  Binary Search approach. This can provide more reliable recommendations compared to the conventional greedy strategy based on the knapsack problem.   We further broaden the applicability of FairSort, designing two versions for online and offline recommendation scenarios. Theoretical analysis and extensive experiments on real-world datasets indicate that FairSort can ensure more reliable personalized recommendations while considering fairness for both the provider and user.

\end{abstract}


\section{Introduction}

   Although bringing great convenience for users by mining their interests to provide personalized recommendations, the recommender system also introduces certain fairness problems. This is because recommender systems typically have inherent biases, such as data and algorithmic bias \cite{1,4,5,6,7}, resulting in unfairness in both the recommendation process and its outcomes. During the process, models may embed biased information (e.g. sensitive attributes like race, and gender), resulting in discriminatory practices like offering more technical job opportunities to men over women \cite{zhao2023fair,zeng2021fair}. Regarding outcomes, models may generate biased recommendations for different user groups like the unfair allocation of exposure opportunities among various providers \cite{ge2021towards,li2021user}.
   
       \begin{figure}[ht]
	    \centering
	    \centerline{\resizebox{0.25\width}{!}{\includegraphics{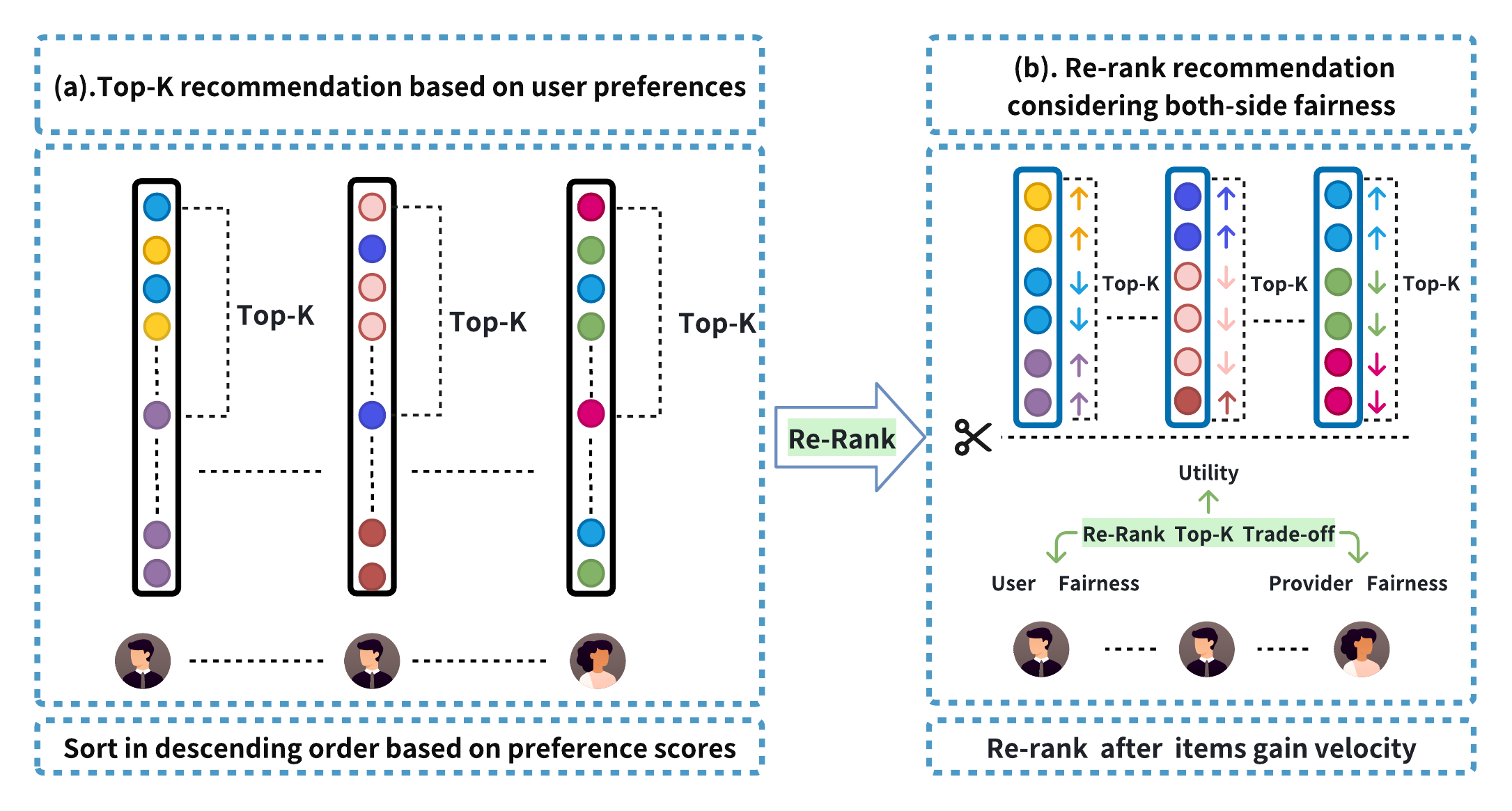}}}
        \centering
	    \caption{\textbf{Re-rank problem to ensure both-side fairness.  Items of identical colours signify they originate from the same provider.}}
	    \label{re-rank problem}
     \vspace{-1.\baselineskip} 
	\end{figure}
                     While many methods aim to ensure fairness in recommendation systems, most address user or provider fairness separately \cite{fu2020fairness,zhu2021fairness,beutel2019fairness,xu2023p}, with few considering both sides simultaneously. An undeniable fact is that enhancing fairness on one side often harms the other side. In this paper, we study the two-sided fairness problem in recommender systems, with the specific scenarios illustrated as follows. If we adopt the traditional Top-K recommendation method as shown in (Fig\ref{re-rank problem} a),  it is often seen as maximizing user satisfaction. whereas, Patro et al.\cite{8} reveal that this user-centric approach can lead to unfair exposure distribution among providers. This means that a few providers monopolize most of the platform's exposure, leaving others with very little. Providers naturally seek greater exposure as it often results in more orders and revenue. Conversely, inadequate exposure may drive them to leave the platform, which could undermine the recommender system's sustainability. However, efforts to fairly distribute exposure among providers can negatively impact recommendation quality for users, with the degree of impact varying among different users. \cite{8}. This is because re-ranking the Top-K list to improve fairness inevitably sacrifices some user satisfaction, and this impact is often difficult to control. This inconsistency in recommendation quality among users further exacerbates user-side unfairness. Therefore, an urgent need exists for a re-ranking algorithm that ensures fairness for both providers and users while maintaining high-quality personalized recommendations, as shown in (Fig\ref{re-rank problem} b).

          It is challenging to find a trade-off solution among three objects. The essence of this problem lies in the combinatorially exploded solution space (\textit{solution space: all possible combinations of picking K items to reform the user's recommendation list}), to re-select a non-direct Top-K recommendation list for each user that can satisfy the both-side fairness while remaining a high recommendation utility. The main challenges are twofold. Firstly, the solution space is enormous (combinatorially exploded), making it difficult to efficiently find a resolution that satisfies both-side fairness. Second, there are inherent conflicts between user-side fairness, provider-side fairness, and personalized recommendation utility, meaning that pursuing one aspect may compromise the others. 

         Currently, only a few studies consider both-side fairness, with TFROM \cite{9}, FairRec \cite{8}, and CPFair \cite{naghiaei2022cpfair} being representative works. Previous research habitually analogizes this problem to a knapsack problem, which is known to be NP-complete \cite{Karp1972}. In this analogy, the length of the recommendation list is seen as the knapsack's capacity, the items to be recommended as the knapsack's contents, and both-side fairness as the constraints. Building on this perspective, these studies employ a greedy strategy to select high-score items for the knapsack to maximize recommendation quality. However, the reliability of relying on a greedy algorithm to ensure recommendation quality is questionable. We empirically observe that the greedy strategy results in an uneven distribution of recommendation quality among users, with some recommendation lists experiencing significant quality degradation due to the introduction of both-side fairness. This inconsistency further exacerbates user-side unfairness.
   

      In this paper, we offer a novel perspective, as illustrated in (Fig\ref{re-rank problem} b), where each recommendation list is viewed as a runway rather than adhering to the traditional knapsack. Specifically, each item on the runway gains a velocity (either upward $ \uparrow $ or downward $ \downarrow $ ) and runs within a specific time, achieving a re-ranking Top-K list for both user and provider fairness. In this perspective, we ensure the \textbf{Minimum Utility Guarantee} for each re-ranked list in terms of personalized recommendations utility by designing a Binary Search approach based on a discovered theorem 1.  The \textbf{Minimum Utility Guarantee} strategy can provide more reliable personalized recommendations compared to the conventional greedy strategy under the knapsack problem perspective, as it effectively tackles the problem of unpredictable loss in personalized recommendation utility when considering both-sided fairness.  In summary, our re-ranking model {FairSort} is designed to guarantee the Minimum Utility of recommendation quality while simultaneously ensuring fairness for both sides. The major contributions of this paper are as follows: 


   
        \begin{itemize}
           \item \textbf{Motivating multi-stakeholder fairness in RS}. We propose a novel perspective to address multi-stakeholder fairness in recommendation systems. We view each recommendation list as a runway rather than adhering to the traditional knapsack problem.
            \item \textbf{Minimum Utility Guarantee.} We propose the Minimum Utility Guarantee for personalized recommendations. It effectively tackles the problem of uncontrollable potential loss in personalized recommendation utility when the algorithm considers both-sides fairness. We prove and guarantee this theoretically and experimentally (§\ref{Proof}).
            \item \textbf{Experiment.} In addition to the theoretical guarantees (§\ref{Proof}), extensive experimentation and evaluation over three real-world datasets deliver strong evidence of the effectiveness of our proposed FairSort. It can ensure more reliable personalized recommendations while considering both-side fairness (§\ref{Experiment}), which is implemented in two versions for online and offline scenarios.
        \end{itemize}

    The paper is structured as follows: Section \ref{section II} is related work, and Section \ref{section III} formalizes the two-sided fairness problem. Section \ref{section IV} presents the FairSort model. The experiment results are detailed in Section \ref{section V}. We conclude the paper in Section \ref{section VI}. Finally, theoretical support is given in Section \ref{section VII}.
  
\section{RELATED WORK}\label{section II}
    We survey related works in three directions: (i) Single-side Fairness vs Multi-side Fairness, (ii) A Taxonomy of Fairness Notions in Recommendation (iii) Minimum Utility Guarantee. 
    
    \textbf{Single-side Fairness vs Multi-side Fairness.\ }Currently, most research focuses on guaranteeing single-side fairness such as item-side fairness \cite{li2022fairgan,wu2021learning,wu2022selective}. Some studies indicate that traditional methods of estimating item utility and exposure exhibit biases, lead to unfair distribution of item-side exposure.\cite{wang2022make,heuss2022fairness}. However, recommender systems need address the challenge of balancing fairness requirements across multiple stakeholders. Only a few studies have addressed the multi-side fairness\cite{8,9,naghiaei2022cpfair,liu2023toward,wang2024intersectional,wang2023two}. FairRec \cite{8} guarantees minimum exposure for each item from the provider side and achieves the EF-1 \cite{14} fairness from the user side.  TFROM \cite{9} on the user side guarantees the consistent user satisfaction of the recommendation list, and on the provider side, ensures the fair allocation of exposure\cite{15}. CPFair\cite{naghiaei2022cpfair} uses a mixed-integer linear programming approach that seamlessly integrates fairness constraints from both sides in a joint objective framework. It divides users and items into two groups (active and inactive), respectively, aiming to achieve group fairness. These approaches are tailored to deterministic re-ranking problems and rely on greedy algorithms. In contrast, TSFD\cite{wang2021user} is also dedicated to addressing both-sided fairness, but specifically targeting scenarios involving stochastic ranking, employing a combination of greedy strategies and convex optimization. In this paper, we neither adopt a greedy strategy nor utilize optimization techniques. We offer another perspective, viewing each recommendation list as a runway, instead of the traditional approach that regards it as a knapsack. 

    \textbf{A Taxonomy of Fairness Notions in Recommendation.} Researchers classify the fairness work of recommendation algorithms into pre-processing\cite{zafar2019fairness}, mid-processing\cite{goh2016satisfying}, and post-processing\cite{zehlike2017fa,biega2018equity} according to the timing of solving the fairness problem. Pre-processing refers to correcting the data bias that leads to the unfairness problem. Mid-processing refers to preventing the model from learning biased information. Post-processing is agnostic to the model, intervenes fairly in the output results of the model, and adjusts the results to make them fair. The approach taken in this paper is a post-processing mechanism. It is independent of classical recommendation algorithm models and can be applied to various recommendation algorithms. Related research classifies recommendation algorithms' fairness into group fairness and individual fairness \cite{wan2021modeling,caton2020fairness}. Group fairness requires that dominant and inferior groups be treated fairly in some aspect, while individual fairness requires similar individuals to have similar treatment. This paper considers both:  provider-side fairness focuses on the fairness of the groups of items provided by each provider, while from the user side, we focus on fairness between individual users.

    \textbf{Minimum Utility Guarantee.} It is crucial in personalized recommendation systems to ensure users receive a certain level of utility. Some researchers have proposed minimum wage guarantee as a fairness standard \cite{pollin2008measure,green2010minimum,falk2006fairness}, research \cite{engbom2022earnings,lin2016effects} showed evidence of how minimum wage guarantee decreases income inequality. Inspired by these works, we propose the notion of a minimum utility guarantee for user-side fairness.

\section{PRELIMINARIES}\label{section III}
   
We address the following questions to gradually analyze the two-sided fairness problem: (i) Quantifying the exposure. (ii) Quantifying the recommendation quality. (iii) Defining user-side fairness. (iv) Defining provider-side fairness. (v) Exploring the relationship among user-side fairness, provider-side fairness, and personalized recommendation utility. Finally, we present a formal problem description.

    \subsection{\textbf{Notations}}
        $\begin{array}{l}
            \mathbf{U = \left\{u_{1}, u_{2}, \ldots, u_{m}\right\}}  \text{ is a set of users.}\\
            \mathbf{I = \left\{i_{1}, i_{2}, \ldots, i_{n}\right\}}  \text{ is a set of recommended items.}\\
             \mathbf{P = \left\{p_{1}, p_{2}, \ldots, p_{l}\right\}}  \text{ is a set of providers supplying items.}\\
             \mathbf{I_{p}  }\text{ is the set of items provided by provider  p .}\\
             \mathbf{V = \left[v_{u_{1}, i_{1}}, v_{u_{1}, i_{2}}, \ldots, v_{u_{m}, i_{n}}\right]} \text{ is a relevant rating matrix} \\\text{produced by the  recommendation algorithm.}\\
             \mathbf{L^{\text {ori }} = \left\{\mathbf{l}_{u_{1}}^{\text {ori }}, \mathbf{l}_{u_{2}}^{\text {ori }}, \ldots, \mathbf{l}_{u_{m}}^{\text {ori }}\right\}} \text{ is a set of original Sorted List} \\
             \text{including all items in $\mathbf{l}^{\text {ori }}$,ranking based on V.} \\ 
             \mathbf{L = \left\{\mathbf{l}_{u_{1}}, \mathbf{l}_{u_{2}}, \ldots, \mathbf{l}_{u_{m}}\right\} }\text{ is a set of recommendation lists} \\ \text{ finally outputted to users, including K items in $\mathbf{l}$.}
              
        \end{array}$

    \subsection{\textbf{Quantify Exposure Resources}}
      The position of each item in the recommendation list is distributed from high to low, and related research shows that the higher the position of the item, the more attention it receives from the user and the more exposure it obtains. Especially an item in position 5 is largely ignored \cite{10}. Therefore, we will consider the position bias to quantify the exposure. We use a decay function, and the exposure of an item $i$ can be defined as:
        \begin{equation}
            \label{E1}
            e_{i}=\sum_{u \in U} \frac{\mathbf{1}_{\mathbf{l}_{u}(i)}}{\log _{2}\left(r_{u, i}+1\right)}
        \end{equation}
        Where $\mathbf{1}_{\mathbf{l}_{u}(i)}$ is the indicator function when the item is successfully recommended, that is, the item is in $\mathbf{l}_{u}$, then the indicator function is 1, otherwise it is 0. The symbol $r_{u,i}$ represents the position of item i in $\mathbf{l}_{u}$. The total exposure gained by a provider is the sum of the exposures of items it offers as\ \ $e_{p}=\sum_{i \in I_{p}} e_{i}$.

    \subsection{\textbf{Quantify The Recommendation Quality}}
        From the rating matrix V, we can pick out  K items with the highest preference scores for each user. i.e., the traditional Top-K recommendation. Although this Top-K list may still not be absolutely in line with users' preferences in practice, it can be assumed that it has already reflected their preferences as much as possible. Thus, this Top-K list is the optimal recommendation list. The idea of measuring the recommendation quality of $\mathbf{l}_{u}$ is to compare it with the optimal recommendation list. We take two classical metrics in information retrieval, namely discounted cumulative gain (DCG) and normalized discounted cumulative gain (NDCG), to measure the quality of recommendation\cite{jarvelin2002cumulated}. DCG quantifies recommendation quality by summing the value of each item in the recommendation list. This value is gained by multiplying the preference score of the item by the function value that is logarithmically decayed based on the position of the item. It is also consistent with the decay function used to quantify exposure. 
       \begin{equation}
            \label{E2}
            D C G_{u,\mathbf{l_{u}}} = \sum_{i = 1}^{\mathbf{K}} \frac{v_{u, \mathbf{l_{u}}[i]}}{\log _{2}(i+1)}
       \end{equation}

       Where $\mathbf{l_{u}}[i]$ represents the i-th item selected from the recommendation list $\mathbf{l_{u}}$. Then, the DCG  of the optimal recommendation list is the maximum value, which can not be exceeded by the DCG  of the list formed by other permutations, thus, the quality measure of any recommendation list is normalized to [0,1].
    
      \begin{equation}
        N D C G_{u}=\frac{D C G_{u, \mathbf{l}_{u}}}{D C G_{u, \mathbf{l}_{u}^{\text {ori }}}}
      \end{equation}
      the larger the NDCG  is, the closer the recommendation list is to the optimal recommendation list. In particular, when the NDCG  is equal to 1, the recommendation list is the optimal recommendation list. On the contrary, the smaller the NDCG  is, the larger the recommendation quality loss is.

\subsection{\textbf{Defining User-Side Fairness}}

     Due to the introduction of provider-side fairness, it is difficult for users to get the optimal recommendation list, which means that recommendation quality will be sacrificed in exchange for fairness on the provider side. We then define the user-side fairness guarantee as follows: The recommendation quality loss should be consistent across users and ensure recommendation quality is not below a certain threshold: Minimum Utility Guarantee.
    
    
    \textbf{Definition 1 \textit{(Fair recommendation for users)}}: The recommendation is fair for users if each user receives recommendation results with the same NDCG  and which will not fall below the threshold.
    \begin{equation}
        \label{E4}
        \resizebox{0.82\width}{!}{$
        (NDCG_{u_{1}}=NDCG_{u_{2}}, \forall u_{1}, u_{2} \in U )\wedge (NDCG_{u_{i}} \geq threshold ,\forall u_{i} \in \mathbf{U})$}
    \end{equation} \vspace{-2\baselineskip} 

\subsection{\textbf{Defining Provider-Side Fairness}}\label{SCM}
    To ensure provider-side fairness, the definition can be based on two ideas: First, the amount of exposure allocated to the provider should be proportional to their quality \cite{biega2018equity}. Second, a provider's exposure should be proportional to the number of items it provides.

    \textbf{Definition 2 \textit{ (Uniform Fairness)}}: The provider exposure distribution satisfies Uniform Fairness if the conversion rate of exposure resources based on the number of items they provide is equal between any two providers, abbreviated as UF.
    \begin{equation}
        \label{E5}
        \frac{e_{p_{1}}}{\mathbf{\left|I_{p_{1}}\right|}}=\frac{e_{p_{2}}}{\mathbf{\left|I_{p_{2}}\right|}}, \forall p_{1}, p_{2} \in P
    \end{equation}

    \textbf{Definition 3 \textit{ (Quality Weighted Fairness)}}: The provider exposure distribution meets Quality Weighted Fairness if the conversion rate of exposure based on their own quality is equal between any two providers, abbreviated as QF.
    \begin{equation}
         \label{E6}
        \frac{e_{p_{1}}}{\sum_{i \in \mathbf{I}_{p_{1}}} \sum_{u \in U} v_{u, i}}=\frac{e_{p_{2}}}{\sum_{i \in \mathbf{I}_{p_{2}}} \sum_{u \in U} v_{u, i}}, \forall p_{1}, p_{2} \in P .
    \end{equation}

\subsection{\textbf{The Game  Between Users and Providers}}
       We first deeply grasp the game relationship between both sides of fairness. In the paper \cite{8}, these two goals are often difficult to achieve simultaneously. Suppose the Top-K recommendation is implemented directly for each user. In that case, each user gets the recommendation list with an NDCG value of 1, and it can also meet the \textbf{Minimum Utility Guarantee} of recommendation quality. Thus, the fairness of the user side is completely guaranteed, but the fairness metrics of the provider side may be inferior. However, to improve the provider-side fairness metrics, we must reorder each Top-K result and regenerate a new recommendation list for each user. But it means that each user's recommendation quality will certainly decrease. Especially the distribution of the loss of the recommendation quality may be diverse for different users. After grasping their game relationship, we summarize three optimization objectives as our algorithm's guiding idea.

        \begin{enumerate}
            \item \textbf{Provider-Side Fairness}: sacrifice the recommendation quality in exchange for fairness on the provider's side.
           \item \textbf{Personalized Recommendations Utility}: maintain a high recommendation quality.
           
            \item \textbf{User-Side Fairness}: the loss of recommendation quality is evenly distributed to all users.
        \end{enumerate}

         We propose a re-ranking strategy denoted as $\mathbf{\pi}$, such that $\mathbf{\pi(\mathbf{L^{ori}})= }\mathbf{L}$ and $\mathbf{L}$ satisfy the  three objectives above.

\section{A TWO-SIDED FAIRNESS-AWARE Learning To Rank MODEL (FairSort)}\label{section IV}

\begin{figure}[htbp]
            
        \centering
               
        \centerline{ 
    
     \resizebox{\hsize}{!}{
       \includegraphics{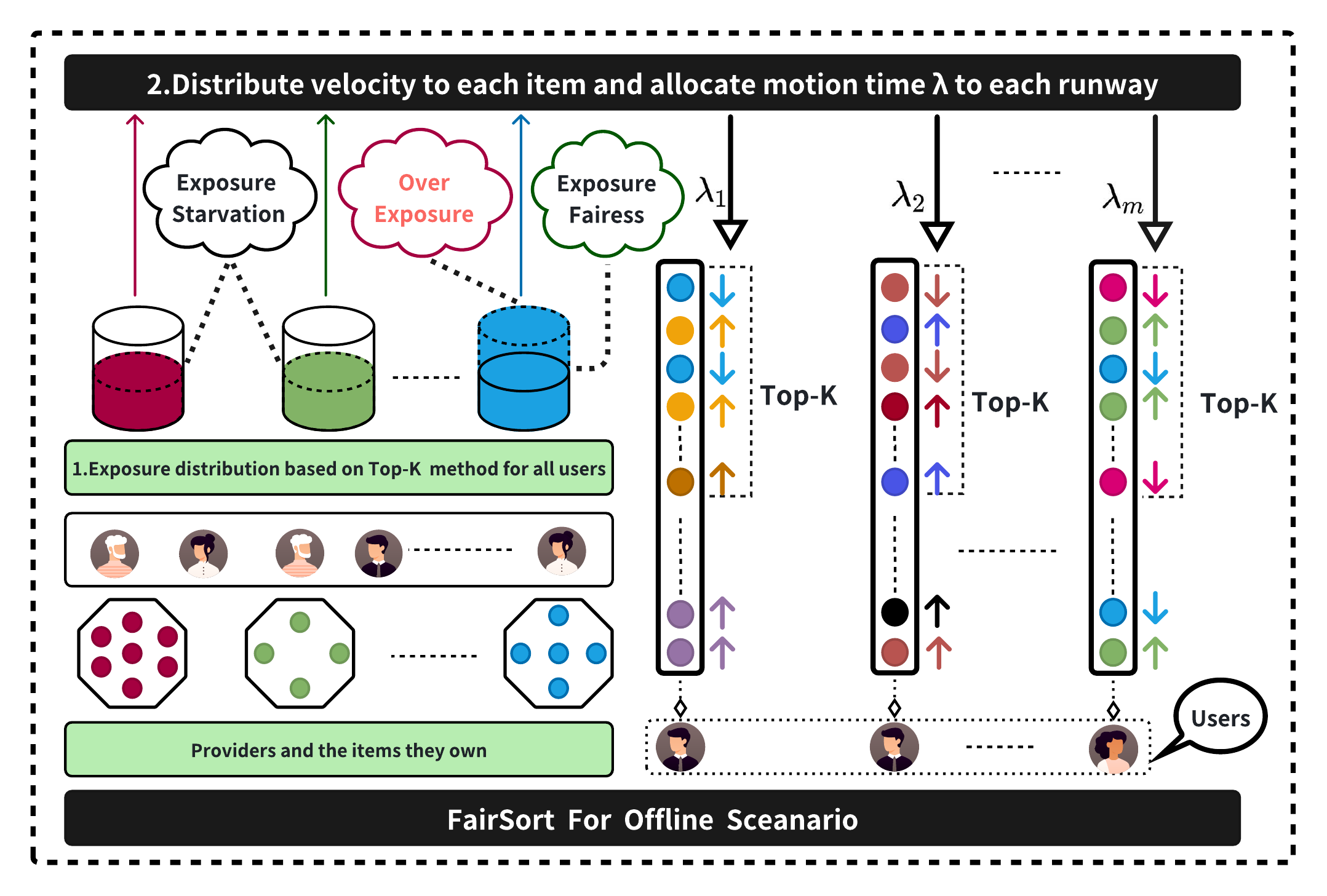}}  
        }
       
        \caption{FairSort Offline Diagram.}
    	 \label{fairSortOff}
       \vspace{-1.5\baselineskip} 
    \end{figure}

In Section \ref{section III}, we model the fairness problem on both the user and provider sides, summarized in figure \ref{fairSortOff}. Each list in the figure represents $\mathbf{l_{u} ^{ori}}$, where items are sorted in descending order based on user preference scores from V. The Top-K items are highlighted, and the left section shows the distribution of exposure across different providers when using the traditional Top-K method, where some providers receive more exposure while others receive less. Items of the same color represent those from the same provider: $\mathbf{I}_{p}$. We consider $\mathbf{l_{u} ^{ori}}$ as a runway, where each item is assigned an initial velocity designed to account for provider-side fairness, with directions indicated by the arrows, either upwards or downwards. Once a velocity is assigned, all items run according to a time parameter $\lambda$, achieving a fair re-ranking that ensures fairness on both sides. This leads to several key questions: how to assign velocity to each item? And how to determine the running time $\lambda$?

 In this section, we propose that FairSort solve the two-sided fairness problem. FairSort consists of two algorithms designed for two scenarios. One is an offline scenario in which the system makes recommendations to all users once at the same time, such as via an advertising push. The other is an online scenario where users’ requests arrive randomly.The system needs to respond to each request quickly and provide the recommendation results, for example, online purchases.
  \vspace{-0.1\baselineskip} 

\subsection{\textbf{FairSort For Offline Scenario }}

    \textbf{Total Exposure Resources:} In the offline scenario, it is a one-time recommendation service for all users. Then, when the number of users $m$ and the length of the recommendation list K are determined, we can calculate the total exposure as follows.
    \begin{equation}
          \label{E7}
          E_{\text {total }}=m \times \sum_{\text {rank }=1}^{\mathbf{K}} \frac{1}{\log _{2}(\operatorname{rank}+1)}
    \end{equation}
    \textbf{Fair Exposure Resources:} Further, since we define exposure fairness on the provider side. Thus, when the total exposure is determined, we can calculate how much exposure each provider should obtain to reach a fair state. 
    \begin{equation}
        \label{E8}
        e_{p_{j}}^{\text {Fair }}=\frac{E_{\text {total }} \times\left|\mathbf{I}_{p_{j}}\right|}{\sum_{p \in P}\left|\mathbf{I}_{p}\right|} \text { (Uniform Fairness) }
    \end{equation}
    \begin{equation}\label{E9}
		e_{p_j}^{Fair} = \frac{E_{total} \times \sum_{i\in \mathbf{I}_{p_j}}\sum_{u \in U}v_{u,i}}{\sum_{p\in P}\sum_{i\in \mathbf{I}_p}\sum_{u \in U}v_{u,i}} \text{(Quality Weight Fairness)}
    \end{equation}

    If all providers receive exposure equal to $e_{p_{j}}^{Fair}$, then the provider-side distribution of exposure is absolutely fair. However, the exposure that each recommendation list position can bring is fixed. So often absolute fairness can not be achieved, but $e_{p_{j}}^{Fair}$ can be used as our benchmark, and as long as each provider gets exposure infinitely close to $e_{p_{j}}^{Fair}$, then relative fairness can be achieved.

    \textbf{Allocate Fairness Lift Velocity: }From figure \ref{fairSortOff}, we observe that some providers receive more exposure through the Top-K recommendation, while others receive less. To address this, we should assign a downward velocity to the items of providers with higher exposure, thereby naturally reducing their overall exposure. Conversely, items from providers with lower exposure should be assigned an upward velocity, allowing their exposure to increase and approach $e_{p_j}^{Fair}$. Thus, this mechanism can ensure fair exposure distribution on the provider side. 
    We ensure that items from the same provider receive identical velocities, lifting the entire group and preserving the \textit{\textbf{Partial order relations} } within each group.


       We have determined the velocity direction, but how should the velocity value be determined? Our core goal is to maintain a fair exposure distribution on the provider side. Further abstracting the essence of the problem is to converge the consistency of conversion rate on the provider side, according to equation (\ref{E5}), (\ref{E6}). Thus, we calculate the error  \textit{ErrRate} for each provider relative to the fair conversion rate, taking the sign of \textit{ErrRate} as the direction of the velocity, and the value of \textit{ErrRate}  also introduces the information of provider's quality or the total number of items it provides.
    \begin{equation}
       \operatorname{ErrRate}_{p_j}=\frac{e_{p_j}^{\mathrm{Fair}}-e_{p_j}}{\left|\mathbf{I}_{p_j}\right|} \quad \forall p_j \in P \quad \text { (Uniform Fairness) }
       \label{E10}
    \end{equation}
     \begin{equation}\label{E11}
     \resizebox{0.9\width}{!}{
		$\operatorname{ErrRate}_{p_j} = \frac{e_{p_j}^{\mathrm{Fair}}-e_{p_j}}{\sum_{i \in \mathbf{I}_{p_j}} \sum_{u \in U} v_{u, i}}\ \forall p_j \in P$  (Quality Weighted Fairness) }
    \end{equation}

    ErrRate will be normalized as follows (\ref{E12}), then its lift velocity will be determined. \textit{J} is a certain filter function. The output is the original value if the parameter and the numerator have the same sign. Otherwise, the output is 0, and then take the absolute value to sum up. The calculation formula is as follows.

    \begin{equation}
        \label{E12}
        \begin{array}{c}
        \forall p_{j} \in P  \text { Lift }_{p_{j}} = \left\{\begin{array}{ll}
        \frac{\operatorname{ErrRate}_{p_{j}}}{\sum_{i = 1}^{|P|}\left|J\left(\operatorname{ErrRate}_{p_{i}}\right)\right|} & \text { ErrRate }_{p_{j}} \neq 0 \\
        0 & \text { otherwise }
        \end{array}\right.
        \end{array}
    \end{equation}

    \begin{equation}
        getFair(i_{k})=\label{E13}\operatorname{Lift}_{i_{k}}=\operatorname{Lift}_{p_{j}} \quad\left(\forall i_{k} \in \mathbf{I} \wedge i_{k} \in \mathbf{I}_{p_{j}}\right)
    \end{equation}

   We calculate the error value between each provider's current conversion rate and the fair conversion rate. Further, normalize the process and assign the lift velocity for each item. The essence of this process is that the conversion rate of each provider is converging to equal the fair conversion rate.
    
    \textbf{Learning To Rank:} To guarantee the fair allocation of exposure on the provider side, we propose the above Re-Ranking mechanism,  a formulation described as follows.
    \begin{equation}
        \label{E14}
        \forall u_i \in U \ \ \ \ \mathbf{l_{u_i}}=\operatorname{argsort}_{i_{k} \in R_{u}}\left(V_{\mathrm{ui_{k}}}+\lambda \operatorname{getFair}(i_{k})\right)
    \end{equation}

    As can be seen from the formula, the result of any user's ranking considers two dimensions of information. The first is the user's preference score for the item, based on matrix \textbf{V}. The second dimension is the velocity of the item obtained by equation (\ref{E13}), considering the provider-side fairness. Therefore, with time $\lambda$ determined, items move up or down in the sorted list in exchange for fairness on both sides. The set \textit{R}, formed by the hyperparameter $ratio$, which is used to take out the top-ranked items from $\mathbf{l^{ori}_u}$ according to a certain ratio, then items in $R$ participate in the Re-Ranking process.

    \textbf{Binary Search Fairness Weighting Factors:} In a physical sense, the larger the fairness weighting factor is, the more importance is attached to fairness on the provider side. But at the same time, the recommendation quality on the user side will be sacrificed more. Thus, $\lambda$ will undertake the trade-off between the interests of both sides. We need a mechanism to determine a more appropriate $\lambda$ value, but this task is challenging. We first introduce a theorem that we discovered, based on which we can cleverly calculate a reasonable $\lambda$ and achieve fairness for both sides by using a Binary Search strategy.

    \textbf{Theorem 1:\label{theorem}} $\lambda$  $\in$ [0, +$\infty$],  $\forall u\in$ U, the initial $NDCG_{u}$ of $\mathbf{l_{u}^{ori}}$ is 1 and then decreases monotonically as the value of $\lambda$ increases. The proof is presented in section \ref{Proof}.
    
    	    

    We use this theorem property to design a Binary Search strategy to accelerate the computation of $\lambda$ so that the NDCG value is highly approximated to the threshold. $\lambda_{target}$  is the value that can make the NDCG value absolutely equal to the threshold. We summarize the binary search process as follows.
    \begin{equation}
    \resizebox{0.96\width}{!}{
        $\begin{array}{c}
          \lambda_{\text {target }}\in\left[0, \lambda_{\text {max }}\right], \lambda_{\text {max }}>\text{gap} \mathop{\longrightarrow} \limits_{\small Time}^{\small BinarySearch}\lambda_{\text {target }} \in\left[m_{1}, m_{2}\right]
        \end{array}$
        }
    \end{equation}

    \textbf{ When Algorithm termination: }
    \begin{equation}
         \lambda=\left\{\begin{array}{ll}
        \lambda_{\text {target }} & \left(m_{2}-m_{1}\right)>g a p \\
        m_1 & \left(m_{2}-m_{1}\right) \leq g a p
        \end{array}\right.
    \end{equation}

  In the process of Binary Search, we need two hyper-parameters. One is $\lambda_{max}$, which is the maximum that $\lambda$ can take, and it can ensure the NDCG  $\le$ threshold. The other is $gap$, which is the accuracy between the $\lambda$ and $\lambda_{target}$, and also describes the interval length containing $\lambda_{target}$. As shown in the formula above, the interval $[0,\lambda_{max}]$ contains  $\lambda_{target}$ at the beginning, and due to binary search can make the interval continuously folded in half without missing  $\lambda_{target}$, then if the search process can directly find  $\lambda_{target}$, it can directly stop the algorithm. Otherwise, the algorithm terminates when the interval's length $\le gap$ and  $\lambda$ take the left endpoint of the interval, that is, $\lambda=m_1$. Thus, if the $gap$ is small enough, it can ensure that the NDCG  is close to and not below the threshold. $Time$ is the round of binary search, and $\frac{\lambda_{max}}{2^{Time}}\le gap$ can trigger the termination conditions of the algorithm, so $Time\leq(\left \lceil \log_{2}(\lambda_{max}/gap)  \right \rceil )$.

   The specific pseudocode is shown in Algorithm \ref{alg1}. In the offline scenario, FairSort first allocates exposure to all users through Top-K recommendations (line 1). Then, for each user, it sequentially performs re-ranking. The specific approach is as follows: Utilizing the current allocation of exposure, we calculate the fair lift velocity for each item using a heuristic function (line 4). Subsequently, a binary search is employed to determine the fairness weighting factor, thus obtaining the re-ranked recommendation list (lines 5-6). Finally, based on the re-ranked recommendation list, the exposure allocation on the provider side is readjusted (line 7).

\begin{algorithm}
		
		\caption{FairSort  Model for Offline Scenario}
            \label{alg1}
		\begin{algorithmic}[1]
			\Require \ \ \newline
                $\textbf{K}$: Number of items recommended for users\newline
			 \textbf{gap}: Hyper-parameters gap\newline 
            $\mathbf{\lambda_{max}}$:Hyper-parameters $\lambda_{max}$ \newline 
            \textbf{V}: Rating matrix;\newline
           $\textbf{threshold}:$ NDCG Minimum Guarantee Value\newline
            $\textbf{l}_{u_1}^{ori}$, $\textbf{l}_{u_2}^{ori}$,..., $\textbf{l}_{u_m}^{ori}$: Original  list for $m$ users;\newline
            $\textbf{ratio}$: Percentage of top-ranked items from $\textbf{l}_{u}^{ori}$ participating in the re-ranking process.
			\Ensure \ \ \newline
			$\textbf{l}_{u_1}$, $\textbf{l}_{u_2}$,..., $\textbf{l}_{u_m}$: Recommendation results for users
         \newline
                \State{$\mathbf{E=\left[e_{p_1}, e_{p_2},..., e_{p_l}\right]\gets\phantom{t}} \forall u \in \mathbf{U}$ according to $\mathbf{l_{u}^{ori}}$ to Top-K allocate exposure to providers.}
                
                \State{$\mathbf{E^{fair}=\left[e_{p_1}^{Fair}, e_{p_2}^{Fair},..., e_{p_l}^{Fair}\right]\gets}$ 
                Compute the $\mathbf{E^{fair}}$ \phantom{textss}according to equation (\ref{E8}),(\ref{E9})}.
                
			\For{$u\_temp$ in U}
                        \State{$\mathbf{Lifts} \gets$ Get a fair lift velocity per provider based on \phantom{text}equation (\ref{E10},\ref{E11},\ref{E12},\ref{E13})}
                       
                         \State{$ \mathbf{R_{u\_temp}}\gets \mathbf{l_{u\_temp}^{ori}}\left[0\text{\textbf{ : }}len\left(\mathbf{l_{u\_temp}^{ori}}\right)*ratio\right]$ }
                        
                         \State{$\mathbf{l_{u\_temp} }\gets\operatorname{argsort}_{i_{k} \in R_{u\_temp}}\left(V_{\mathrm{ui_{k}}}+\lambda \operatorname{getFair}(i_{k})\right)$ \phantom{text}Through  BinarySearch determines $\lambda$}
                       
                        \State{$\mathbf{E=\left[e_{p_1}, e_{p_2},..., e_{p_l}\right]} \gets$ Update exposure allocation \phantom{text}based on Re-ranking results $\textbf{l}_{u\_temp}$}
                \EndFor
                
        \State{\textbf{end for}}
    
	\State\Return $\textbf{l}_{u_1}$, $\textbf{l}_{u_2}$,..., $\textbf{l}_{u_m}$; 

\end{algorithmic}	
\end{algorithm}
\begin{algorithm}[!h]
		
		\caption{FairSort  Model for Online Scenario}
            \label{alg2}
		\begin{algorithmic}[1]
			\Require \ \ \newline
                $\textbf{K}$: Number of items recommended for users\newline
			 \textbf{gap}: Hyper-parameters gap\newline 
             $\mathbf{\lambda_{max}}$:Hyper-parameters $\lambda_{max}$ \newline
             $\textbf{threshold}:$ NDCG Minimum Guarantee Value\newline
			$\textbf{u}$: The coming user;\newline
             \textbf{V}: Rating matrix;\newline
              $\mathbf{l_{u}^{ori}}$: Original recommendation list of the coming user
            \newline
		   $\textbf{ratio}$: Percentage of top-ranked items from $\textbf{l}_{u}^{ori}$ participating in the re-ranking process.
			\Ensure \ \ \newline
			$\textbf{l}_u$: Recommendation results for the coming user $u$;
\newline
              \State{$\mathbf{E^{fair}=\left[e_{p_1}^{Fair}, e_{p_2}^{Fair},..., e_{p_l}^{Fair}\right]\gets}$ 
                Compute the $\mathbf{E^{fair}}$  \phantom{textss}according to equation (\ref{E17}, \ref{E8}, \ref{E9})}.
                
			 \State{$\mathbf{E=\left[e_{p_1}, e_{p_2},..., e_{p_l}\right]\gets\phantom{t}}$ according to $\mathbf{l_{u}^{ori}}$ to Top-K \phantom{textss}Allocate Exposure to providers.}
   
                \State{$\mathbf{Lifts} \gets$ Get a fair lift velocity per provider based \phantom{textaa}on equation (\ref{E10},\ref{E11},\ref{E12},\ref{E13})}
                       
                         \State{$ \mathbf{R_{u}}\gets \mathbf{l_{u}^{ori}}\left[0\text{\textbf{ : }}len\left(\mathbf{l_{u}^{ori}}\right)*ratio\right]$ }
                        
                         \State{$\mathbf{l_{u}} \gets\operatorname{argsort}_{i_{k} \in R_{u}}\left(V_{\mathrm{ui_{k}}}+\lambda \operatorname{getFair}(i_{k})\right)$ Through \par BinarySearch determines $\lambda$}
                       
                        \State{$\mathbf{E=\left[e_{p_1}, e_{p_2},..., e_{p_l}\right]} \gets$ Update exposure allocation \phantom{textss}based on Re-ranking results $\textbf{l}_{u}$}
                  
			\State\Return $\textbf{l}_u$; 
		\end{algorithmic}	
\end{algorithm}        

    \begin{figure}[htbp]
            
        \centering
               
        \centerline{ 
    
     \resizebox{\hsize}{!}{
       \includegraphics{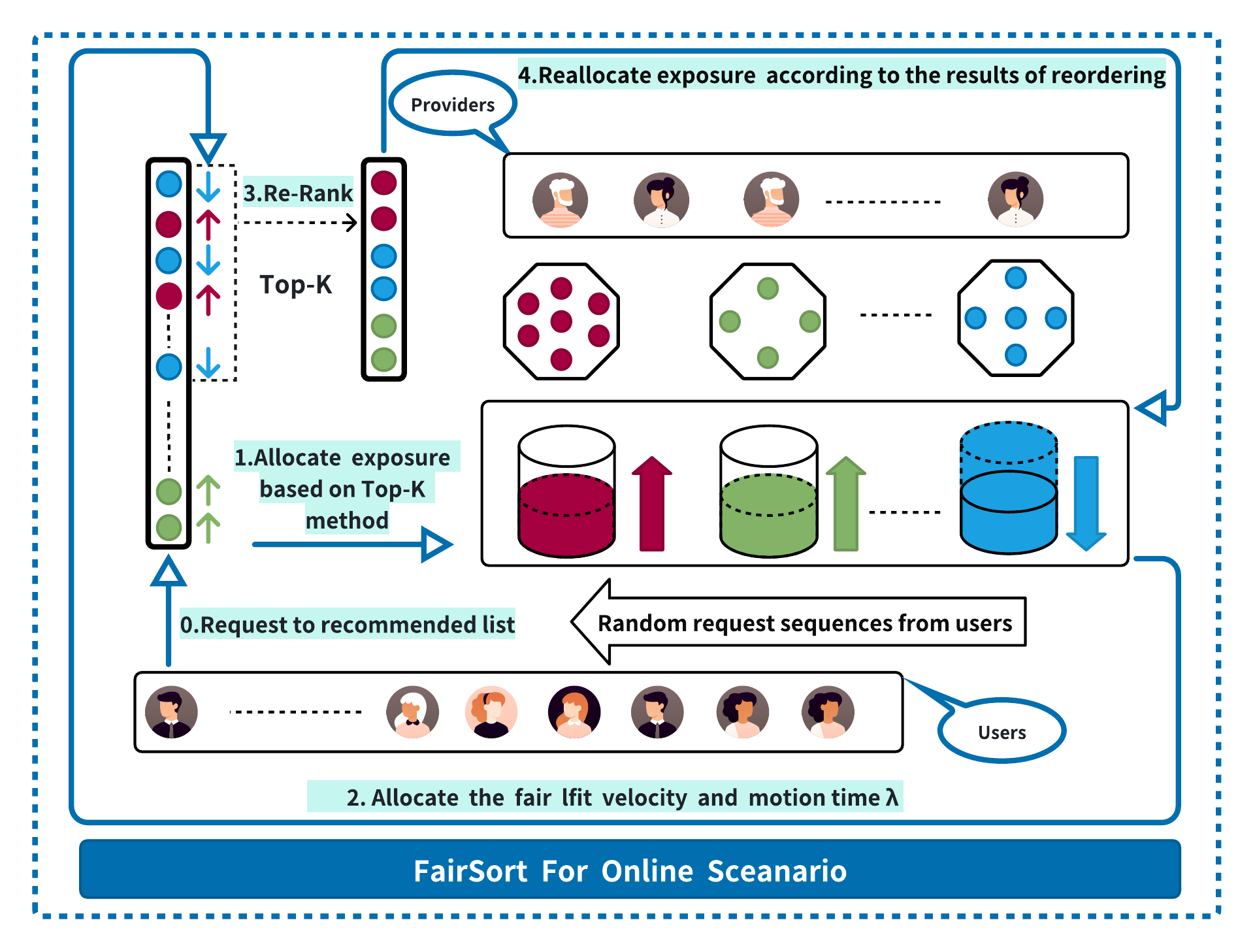}}  
        }
       
        \caption{FairSort Online  Diagram.}
    	 \label{fairSortOn}
       \vspace{-1.5\baselineskip} 
    \end{figure}
    \subsection{\textbf{FairSort For Online Scenario }}
        In the online scenario, user requests arrive randomly. It is necessary to provide a recommendation service for each request within a certain time and ensure both sides' fairness. 
         
         In this process, the user's recommendation quality and the provider's total exposure must be considered process quantity. Thus, the user side calculates the average recommendation quality. 
         While the total exposure  can be calculated as follows: 
        \begin{equation}
            \label{E17}
             \resizebox{0.9\width}{!}{
		$\begin{aligned}
			&E_{total} = c\_num\times \sum_{rank=1}^{\mathbf{K}}\frac{1}{log_2(rank+1)}\\
			&\textit{where } c\_num \textit{ is the number of user requests.}
		\end{aligned}$	}
	\end{equation} 
   
    Pseudocode is shown in Algorithm \ref{alg2}. We can calculate the total exposure when each user request arrives, according to the above formula. This is followed by the calculation of each provider's $e^{fair}$, which allows us to establish a benchmark for fair allocation of exposure (line 1). At this time, according to the Top-K recommendation method, allocate the exposure for each provider (line
     2), next through the above formula (\ref{E13}), to inspire the fair lift velocity for each item (line 3). And then through the Binary Search Strategy, achieve a fair Re-Ranking (lines 4-5). Finally, the allocation of exposure will be adjusted (line 6). Figure \ref{fairSortOn} shows this whole process.

    \vspace{-1\baselineskip} 
    \subsection{\textbf{Time Complexity Of FairSort}}
        The time complexity of \textit{FairSort} is analyzed as follows. First, we analyze the offline version, which provides a Re-Ranking service for each user in order. Thus, the outer loop has $O(m)$, and each time we perform a re-ranking service for a user. It needs to call the heuristic function that assigns fair lifting velocity for each provider. The time complexity is $O(l)$, and $l$ is the number of providers. Then, the number of items involved in Re-Ranking is $g$, determined jointly with $\mathbf{l_{u}^{ori}}$'s length and $ratio$. By binary search, the quick sort algorithm is performed, and the time complexity is $O(g\log(g))$\cite{hoare1962quicksort}. Thus, the time complexity of the offline version is $O(m(l+g\log(g)))$, and the online version has no outer loop, so the time complexity of each request is $O(l+g\log(g))$.

\section{Experiment}\label{section V}
    \label{Experiment}
    \subsection{\textbf{Datasets and Metrics}}

        \textbf{Ctrip Flight Dataset.} The entire dataset contains data from 3,814 users, 6,006 kinds of air tickets, and 25,190 orders. It also provides basic information on users, air ticket class, air ticket price, flight time, airline company of the ticket, and other  information. we adopt the state-of-the-art collaborative filtering air ticket recommendation algorithm\cite{gu2019addressing} to process the data and obtain a  preference matrix.

        \textbf{Amazon Review Dataset.} We used data from \cite{he2016ups}, which has the largest data due to its large number of reviews. We pre-filtered items and users with less than 10 reviews or being reviewed and only consider reviews of items in the “Clothing Shoes and Jewelry” category, which has the largest number of reviews. And using the well-known matrix decomposition model \cite{13} to estimate users' preference scores for items, the dataset does not provide information between items and providers. We then model the providers by clustering methods, with 1-100 items clustered into one category. The processed dataset contains 1,851 users, 7,538 items, 161 providers.

        \textbf{Google Local DataSet}. This dataset is unique, where each item represents an individual provider. This dataset contains reviews about local businesses from Google Maps. It also filtered items and users who participated in reviews less than 10 times, then obtained a dataset containing 3,335 users, 4,927 items (providers), and 97,658 reviews. The data is processed by using an implicit decomposition algorithm \cite{he2014location} based on location information. 

        \textbf{Metrics and Hyperparameter Setting:} On the provider side, \textit{Quality Weighted Fairness} (QF) is measured by calculating the variance of the ratio between their exposure $e_{p}$ and their own quality (named \textit{Variance of the ratio of exposure and relevance}).  For \textit{Uniform Fairness} (UF), we calculate the variance of the ratio between their exposure  $e_{p}$ and the number of items they provide (named \textit{Variance of exposure}). On the user side, we measure the (\textit{Variance of NDCG}) of the user's recommendation list and compute the distribution of the NDCG. We count the user's total recommendation quality to evaluate the loss situation. The smaller the variance, the fairer the recommendation results. The greater the sum of $NDCG_{u}$, the smaller the loss of the recommendation quality\cite{9}. Due to space limitations, only a formal description of the metric \textit{QF} is given here, and the others are similar. We use DCF (Deviation from User (Customer) Fairness) and DPF (Deviation from Provider Fairness) as the abbreviations for fairness metrics on the user and provider sides respectively. The $\lambda_{max}$ is selected from \{$2^2$, $2^3$, $2^4$\}. The $gap$ is selected from\{$2^{-5}$, $2^{-6}$, $2^{-7}$\}. The \textit{threshold} is selected from\{0.85, 0.90, 0.95\}. The \textit{ratio} is selected from\{0.1, 0.2, 1\}.

\begin{equation}
\resizebox{.89\linewidth}{!}{
    $DPF = \mathbb{E}_{p\sim P}\left[\frac{e_{p}}{\sum_{i \in I_{p}} \sum_{u \in U} v_{u, i}}-\mathbb{E}_{p\sim P}(\frac{e_{p}}{\sum_{i \in I_{p}} \sum_{u \in U} v_{u, i}})\right]^{2}$
}
\end{equation}

        \begin{equation}
             UIR = \frac{w_{1}\frac{DCF}{\mu_{1}}+w_{2}\frac{DPF}{\mu_{2}}}{Utility} 
             \label{UIR}
        \end{equation}

 \begin{figure*}[!h]
		\centering
		\subfigure[Total recommendation quality ($ \uparrow $)]{
			\begin{minipage}[t]{0.25\linewidth}
				\centering
				\includegraphics[width=\textwidth,height=2.8cm]{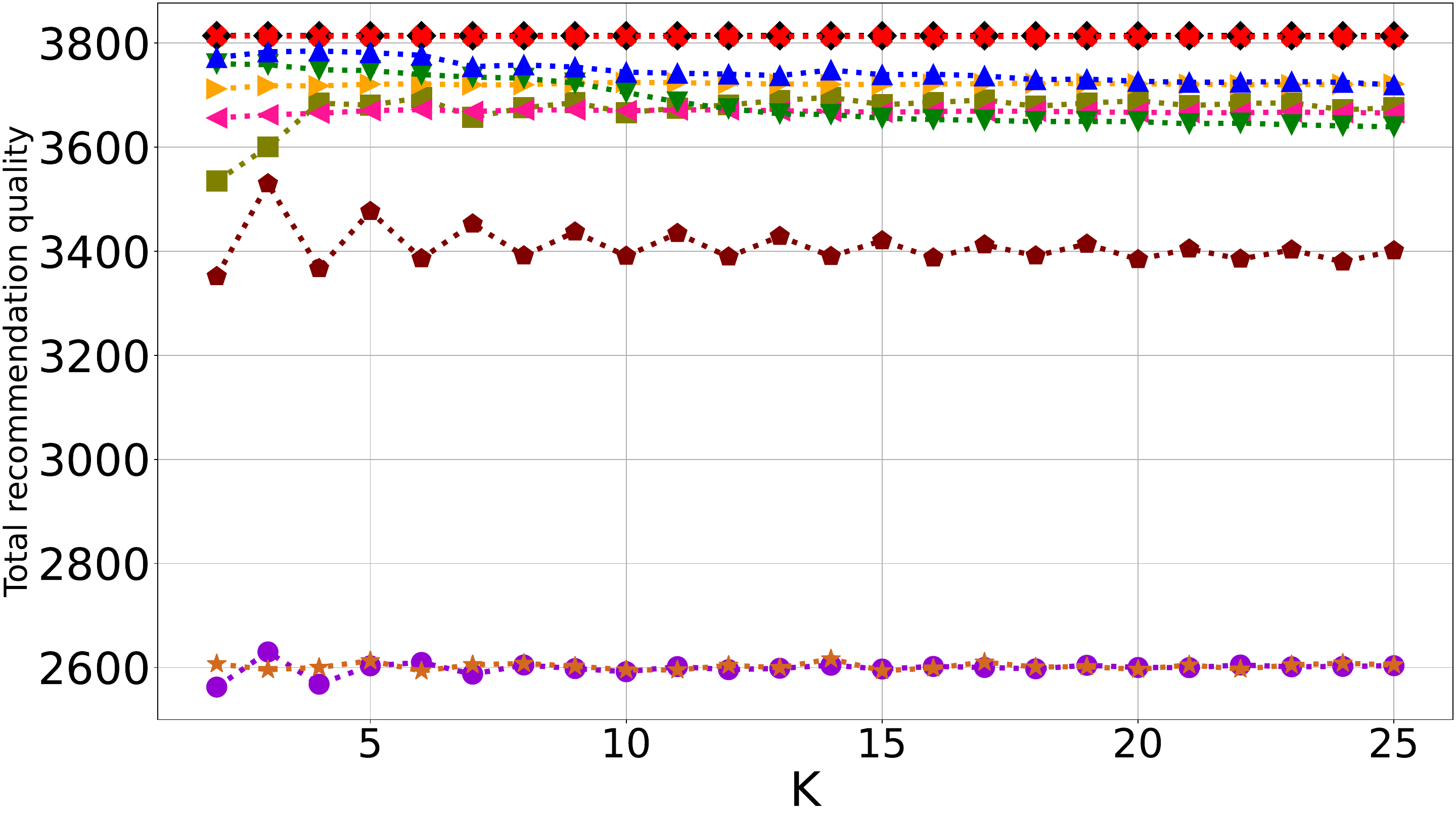}
			\end{minipage}%
		}%
		\subfigure[DCF--Variance of NDCG ($ \downarrow $)]{
			\begin{minipage}[t]{0.25\linewidth}
				\centering
				\includegraphics[width=\textwidth,height=2.8cm]{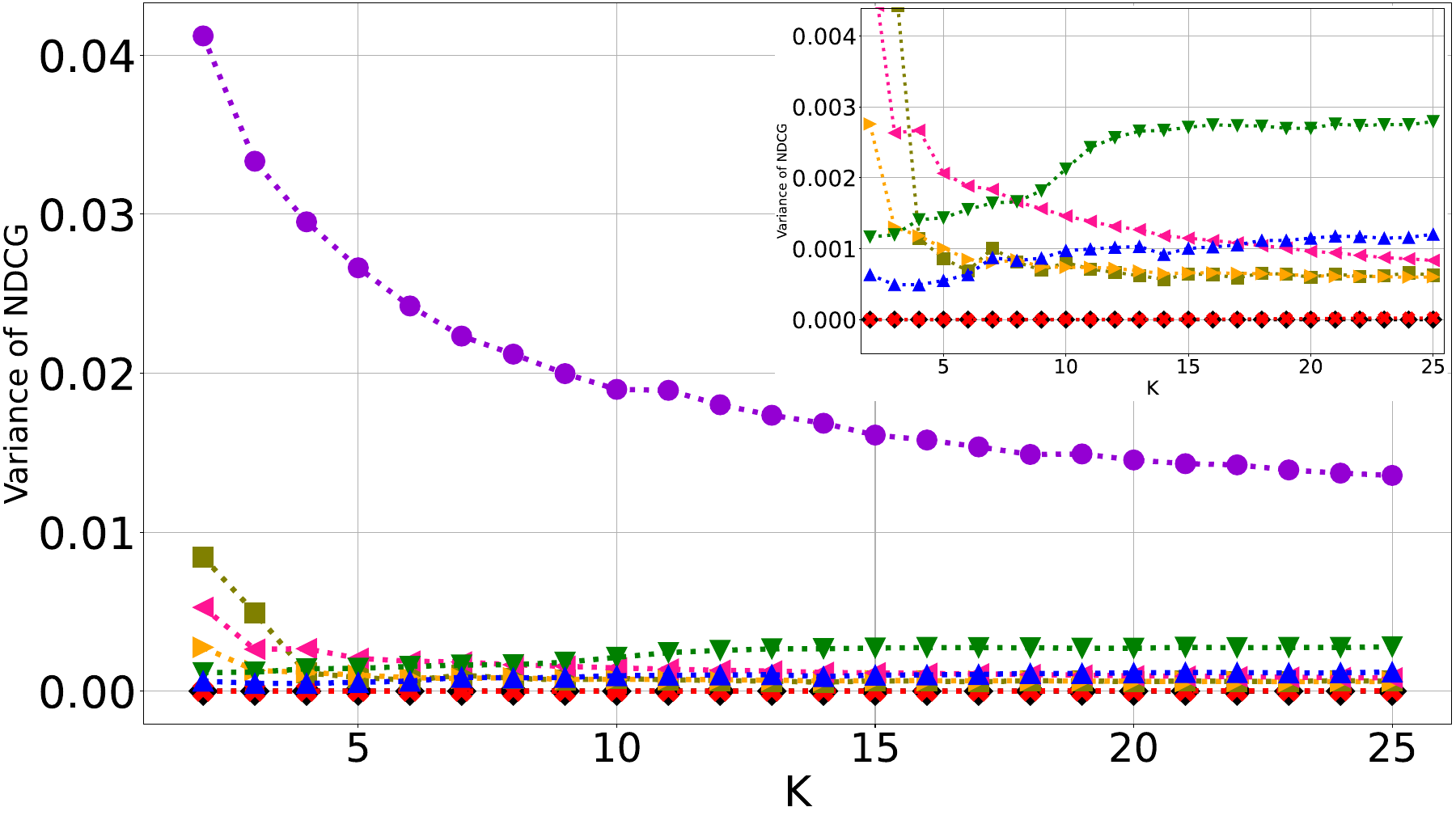}
			\end{minipage}
		}%
		\subfigure[DPF--Uniform Weight Fairness ($ \downarrow $)]{
			\begin{minipage}[t]{0.25\linewidth}
				\centering
				\includegraphics[width=\textwidth,height=2.8cm]{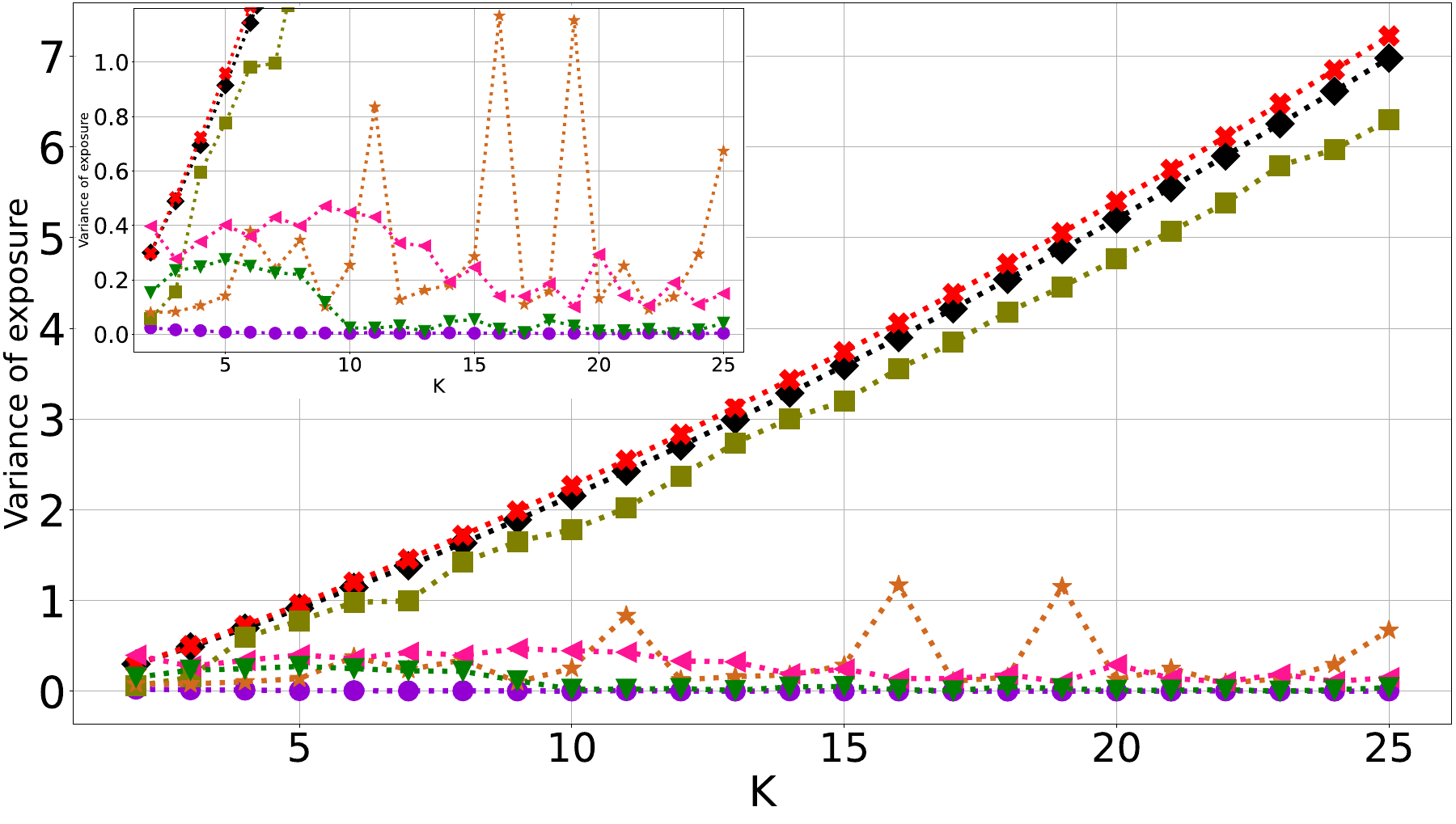}
			\end{minipage}
		}%
		\subfigure[DPF--Quality Weight Fairness ($ \downarrow $)]{
			\begin{minipage}[t]{0.25\linewidth}
				\centering
				\includegraphics[width=\textwidth,height=2.8cm]{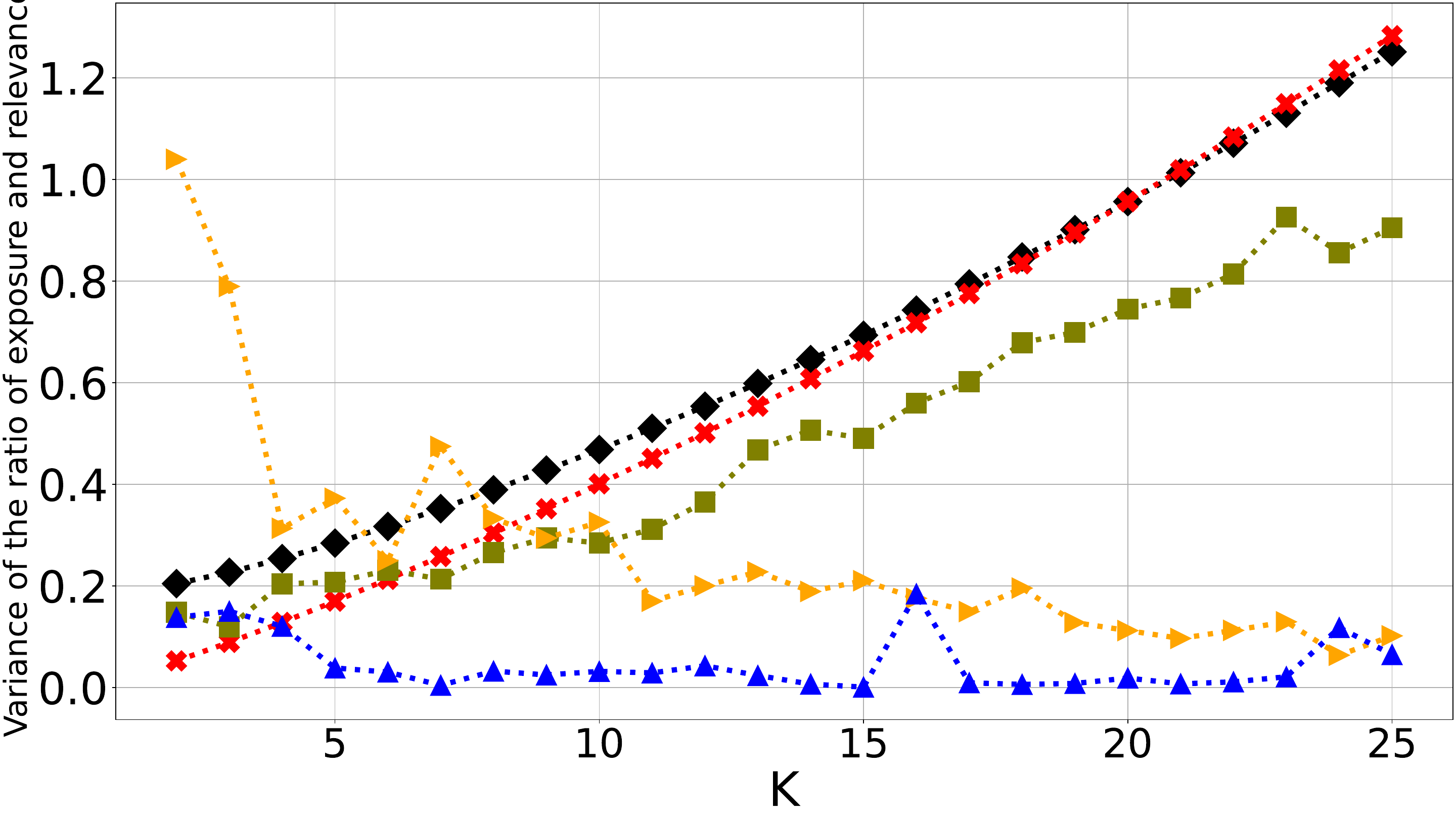}
			\end{minipage}
		}%
		\centering
		\caption{Experiment Results on Ctrip Dataset in the Offline Scenario.}
		\label{offEfig1}
	\end{figure*}

	\begin{figure*}[!h]
		\centering
		\subfigure[Total recommendation quality ($ \uparrow $)]{
			\begin{minipage}[t]{0.25\linewidth}
				\centering
				\includegraphics[width=\textwidth,height=2.8cm]{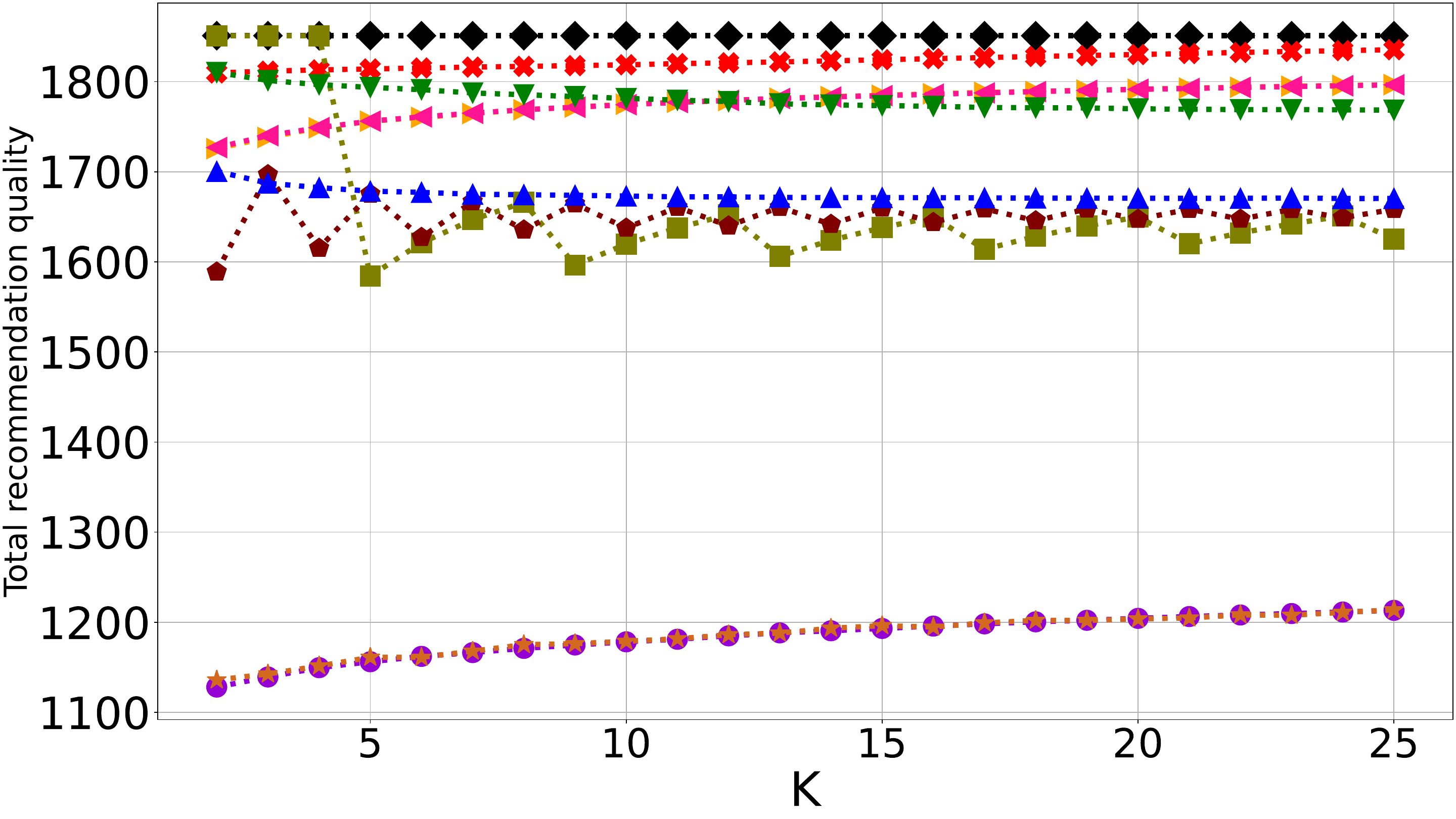}
			\end{minipage}%
		}%
		\subfigure[DCF--Variance of NDCG ($ \downarrow $)]{
			\begin{minipage}[t]{0.25\linewidth}
				\centering
				\includegraphics[width=\textwidth,height=2.8cm]{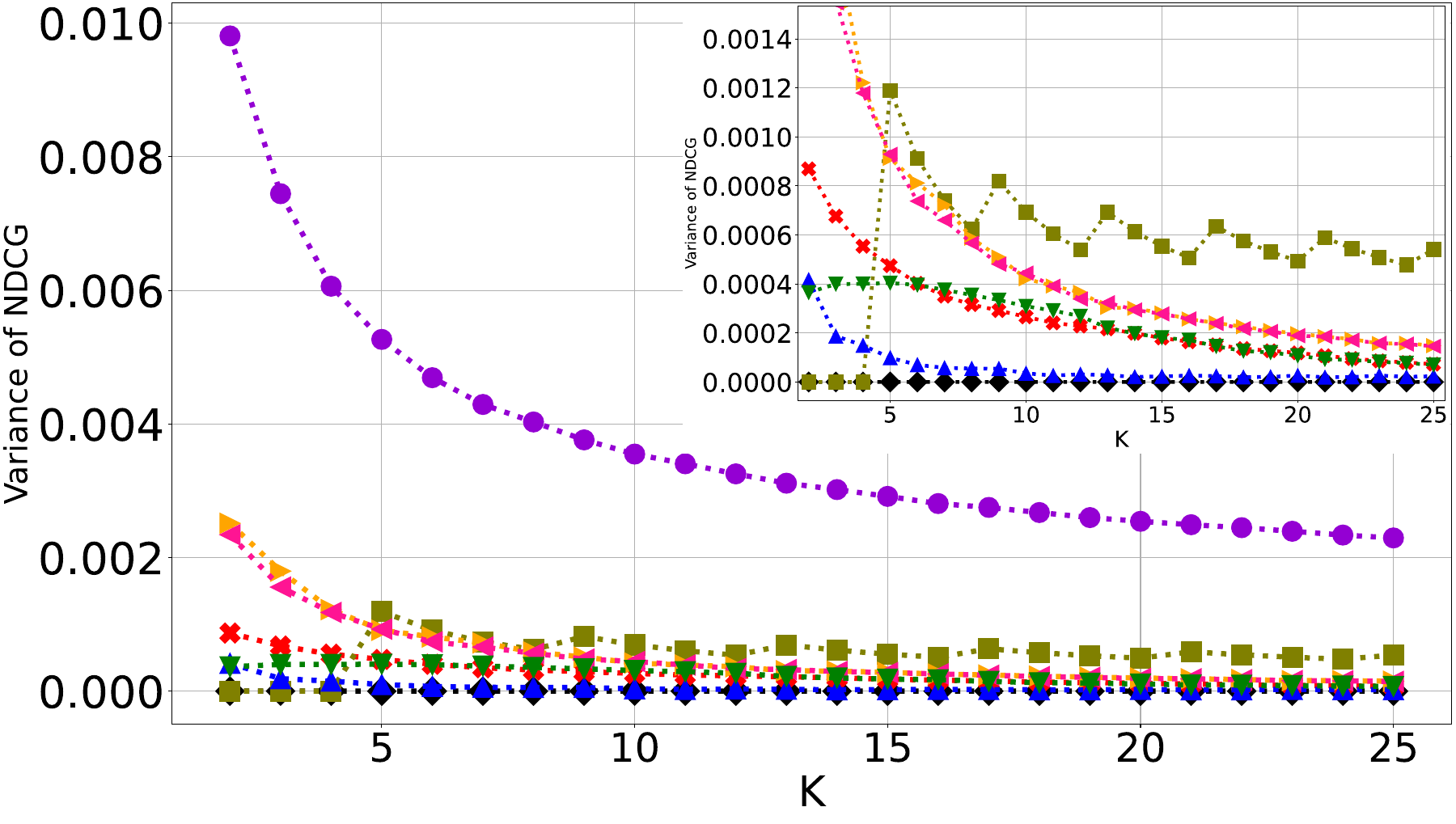}
			\end{minipage}
		}%
		\subfigure[DPF--Uniform Weight Fairness ($ \downarrow $)]{
			\begin{minipage}[t]{0.25\linewidth}
				\centering
				\includegraphics[width=\textwidth,height=2.8cm]{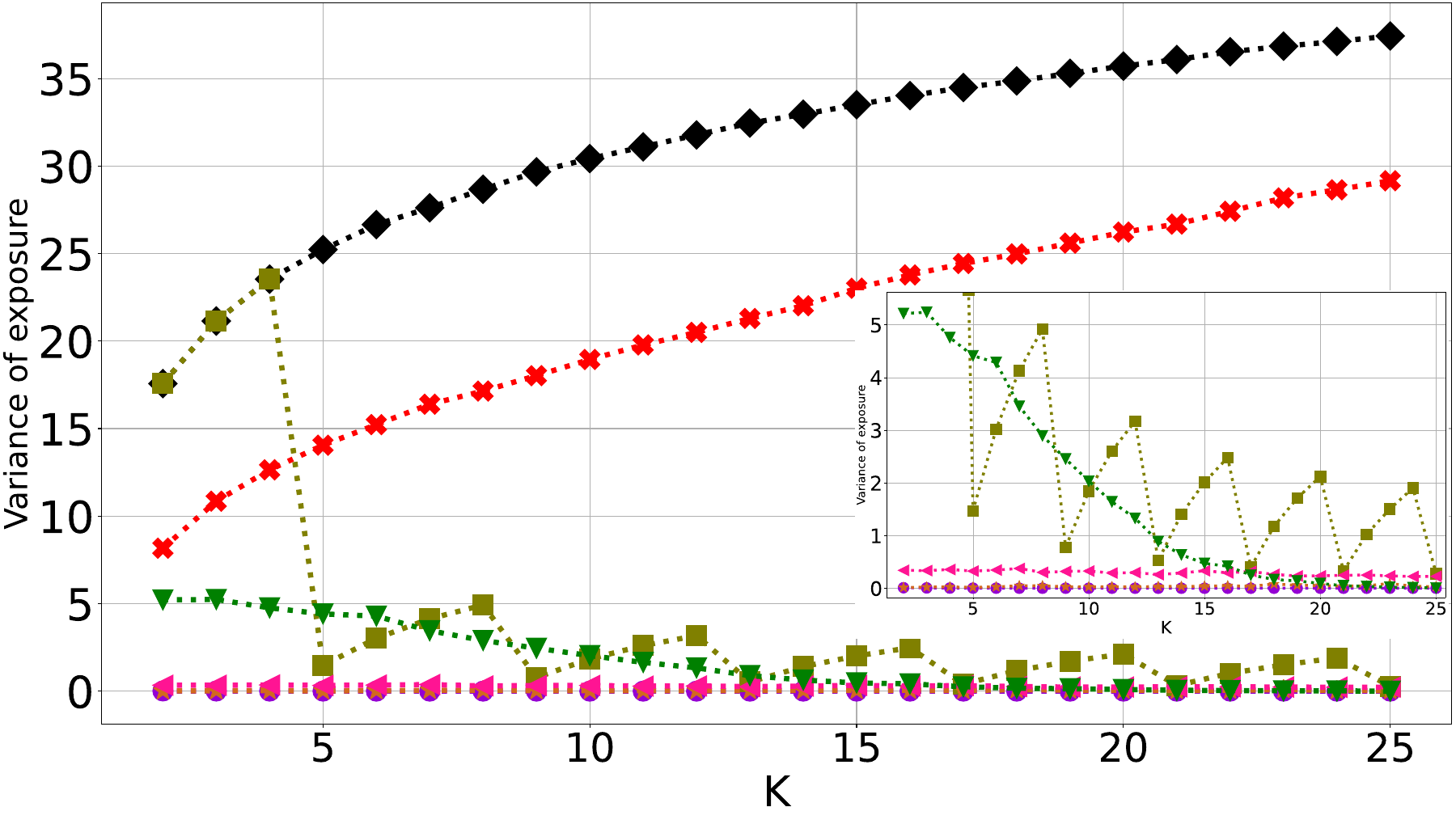}
			\end{minipage}
		}%
		\subfigure[DPF--Quality Weight Fairness ($ \downarrow $)]{
			\begin{minipage}[t]{0.25\linewidth}
				\centering
				\includegraphics[width=\textwidth,height=2.8cm]{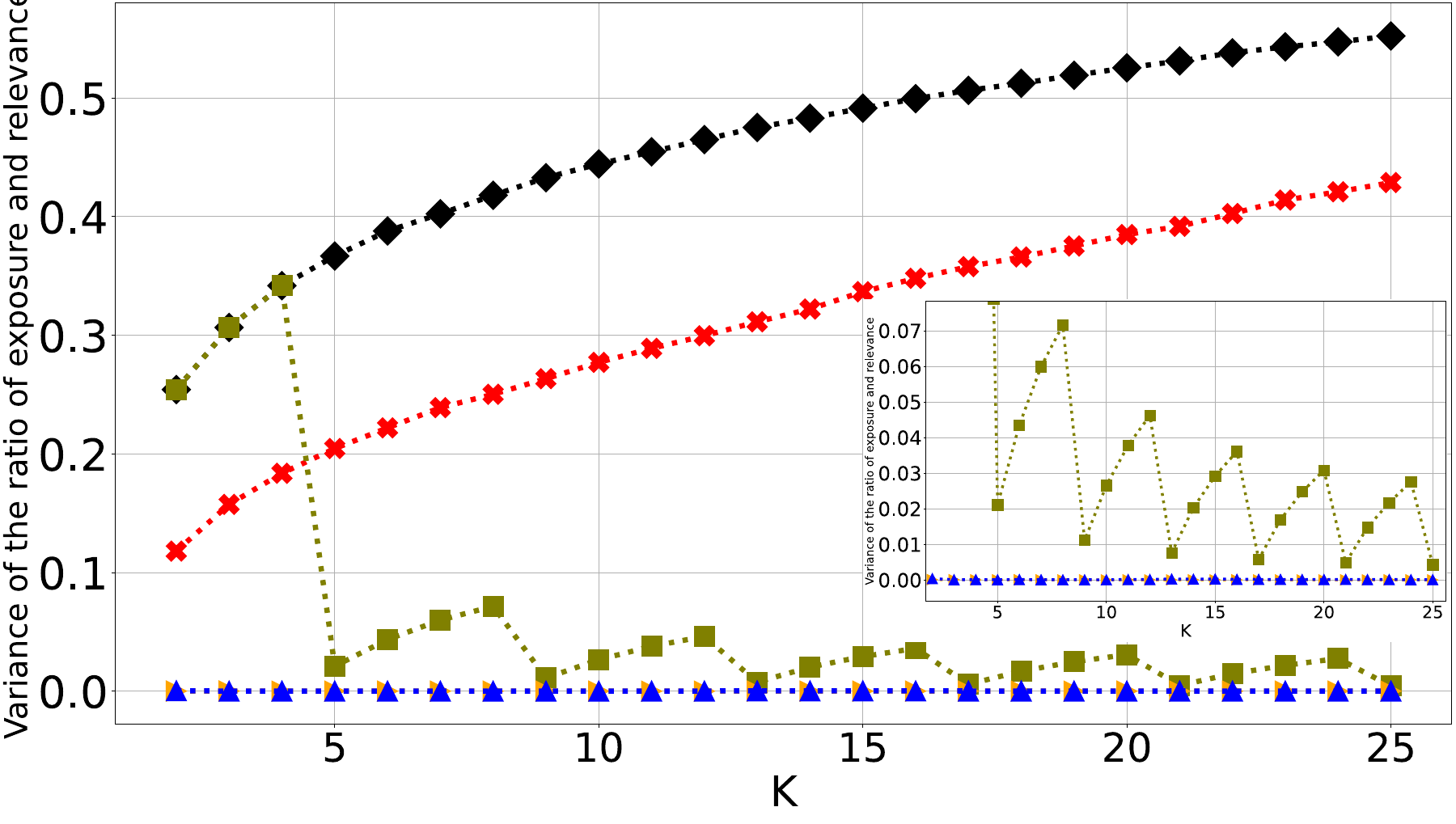}
			\end{minipage}
		}%
		\centering
		\caption{Experiment Results on Amazon Dataset in the Offline Scenario.}
		\label{offEfig2}
	\end{figure*}
	
	\begin{figure*}[!h]
		\centering
		\subfigure[Total recommendation quality ($ \uparrow $)]{
			\begin{minipage}[t]{0.25\linewidth}
				\centering
				\includegraphics[width=\textwidth,height=2.8cm]{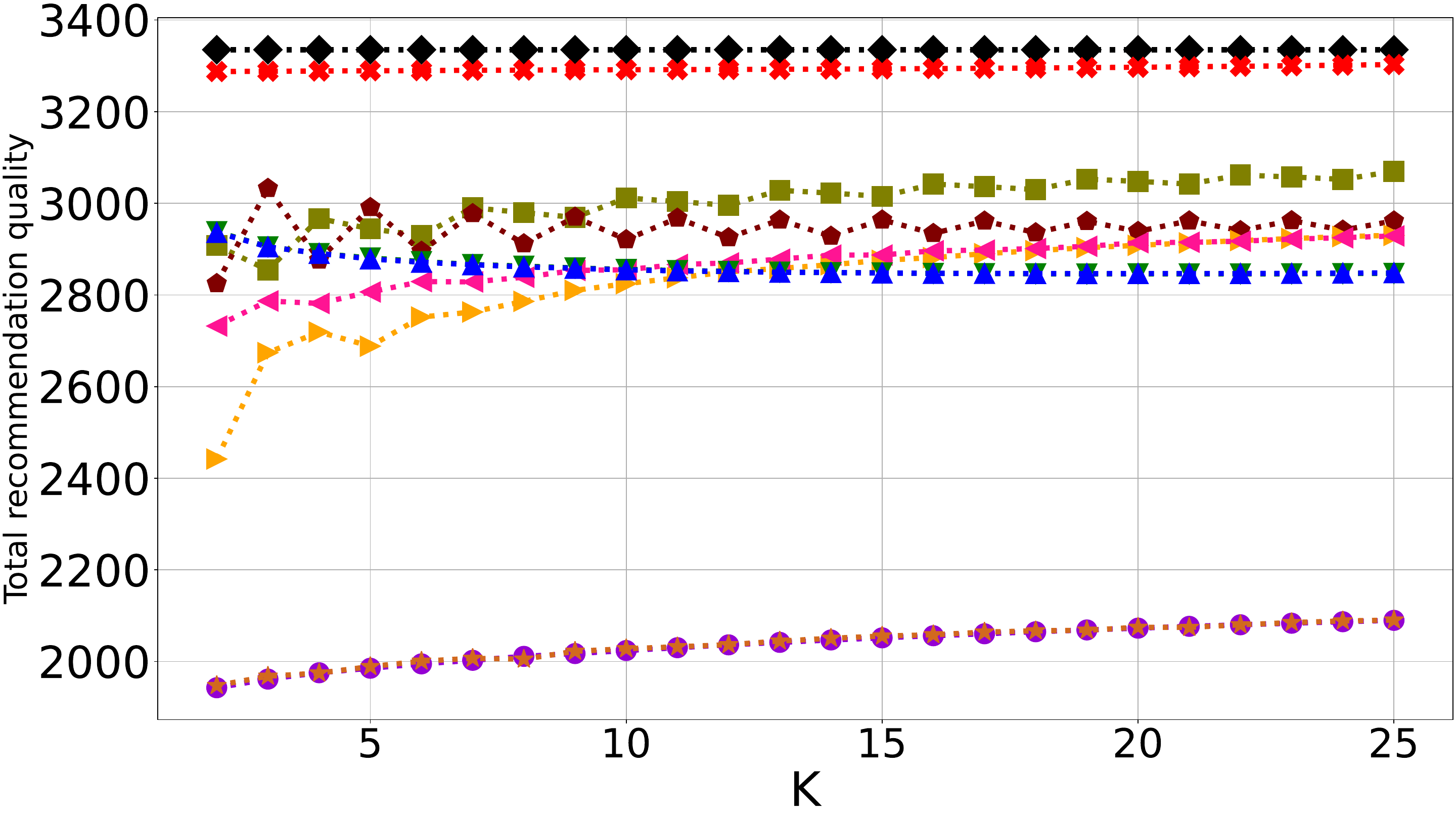}
			\end{minipage}%
		}%
		\subfigure[DCF--Variance of NDCG ($ \downarrow $)]{
			\begin{minipage}[t]{0.25\linewidth}
				\centering
				\includegraphics[width=\textwidth,height=2.8cm]{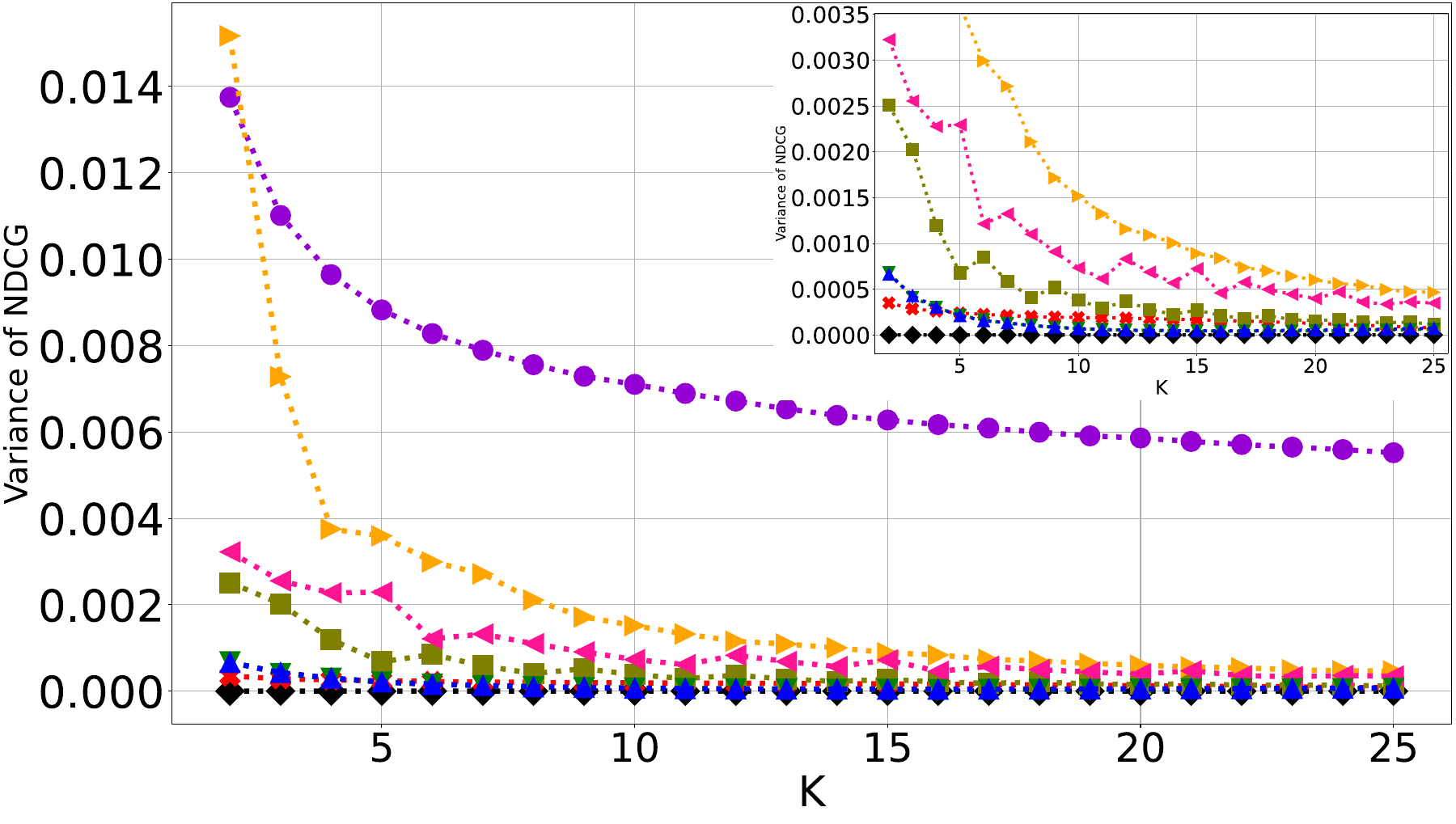}
			\end{minipage}
		}%
		\subfigure[DPF--Uniform Weight Fairness ($ \downarrow $)]{
			\begin{minipage}[t]{0.25\linewidth}
				\centering
				\includegraphics[width=\textwidth,height=2.8cm]{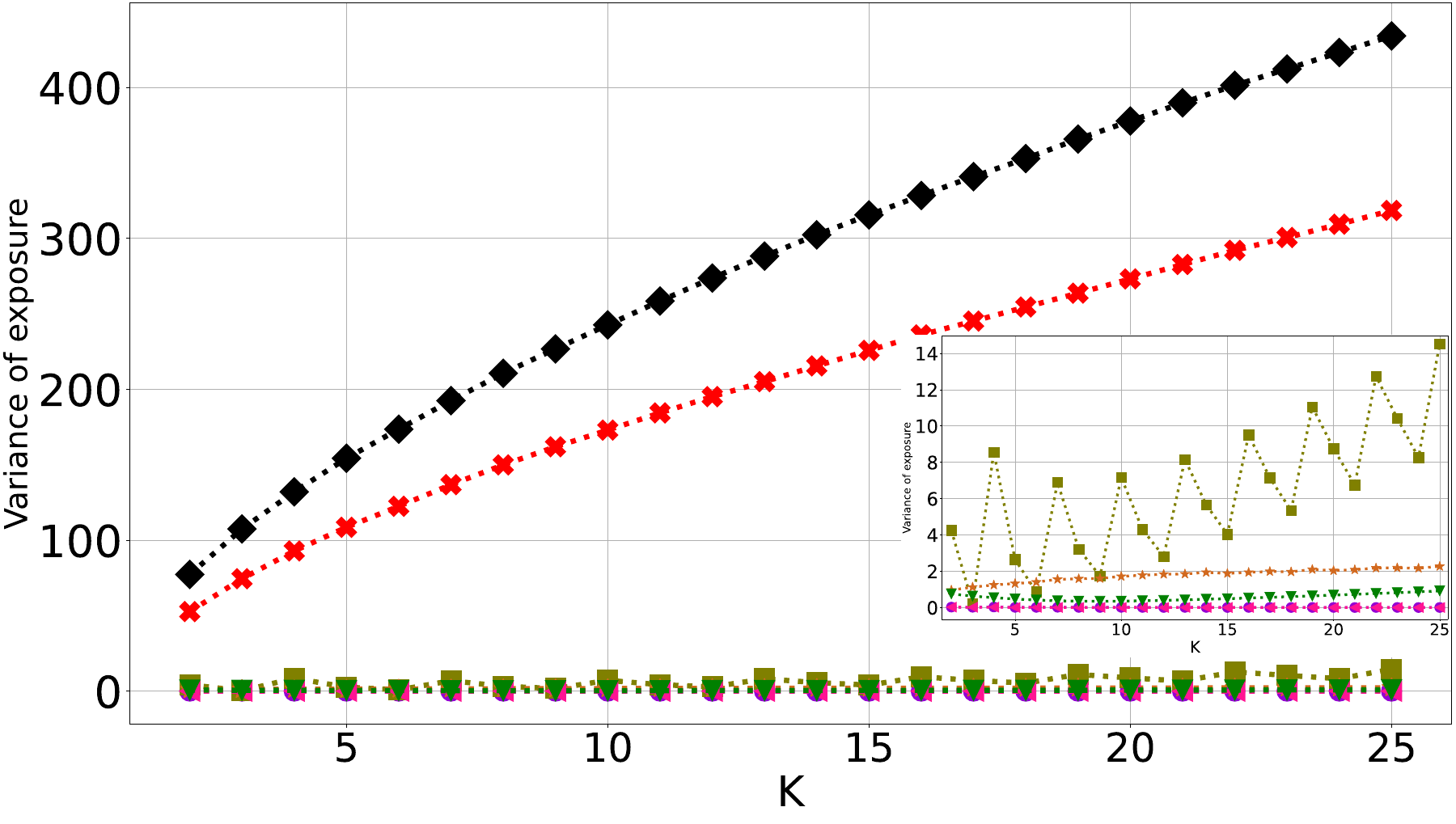}
			\end{minipage}
		}%
		\subfigure[DPF--Quality Weight Fairness ($ \downarrow $)]{
			\begin{minipage}[t]{0.25\linewidth}
				\centering
				\includegraphics[width=\textwidth,height=2.8cm]{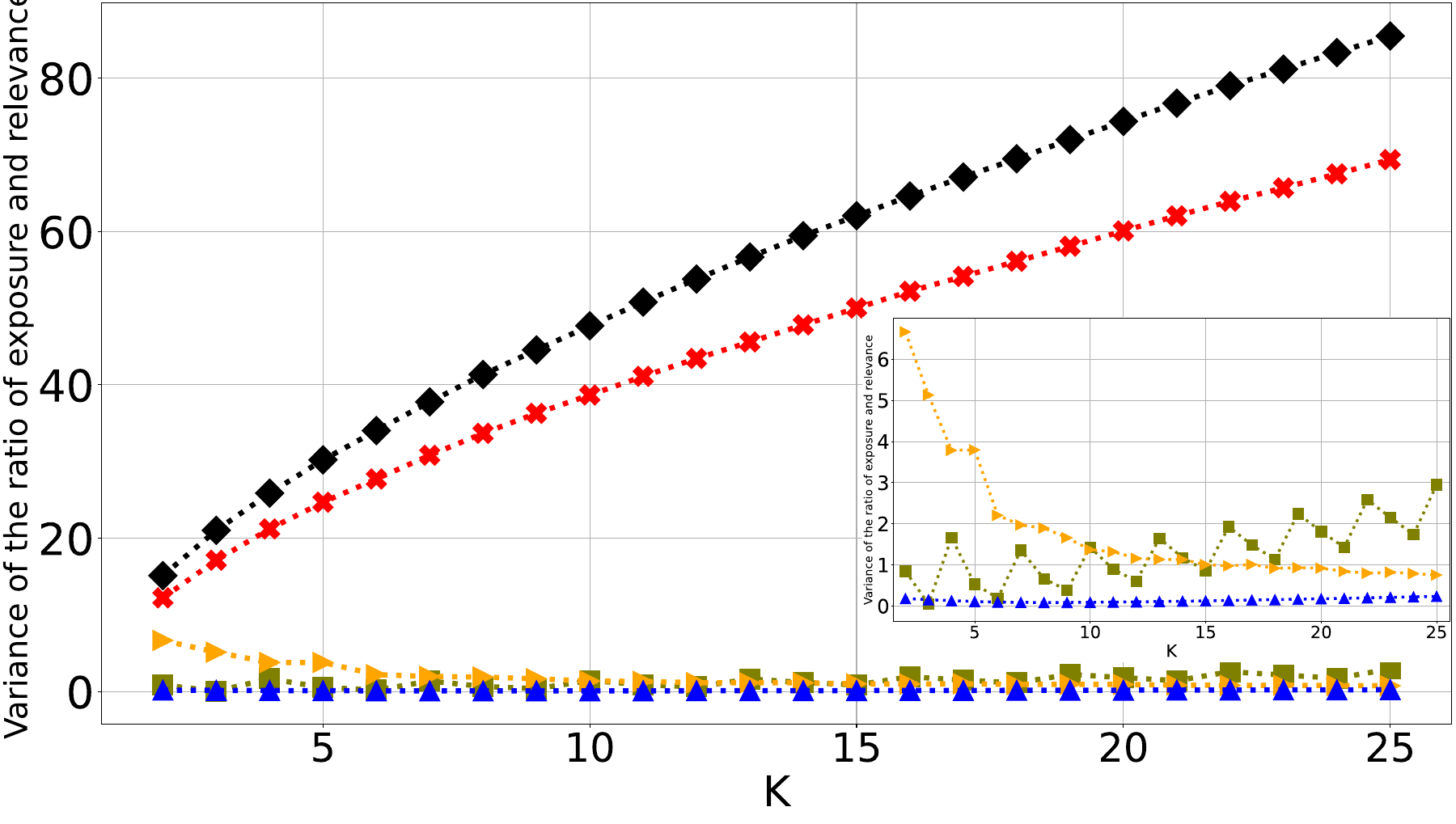}
			\end{minipage}
		}%
		\centering
		\caption{Experiment Results on Google Dataset in the Offline Scenario.}
		\label{offEfig3}
  \vspace{-1\baselineskip} 

	\end{figure*}

\begin{figure*}[!h]
    \centering
    \centerline{ 
    
     \resizebox{\textwidth}{32pt}{
       \includegraphics{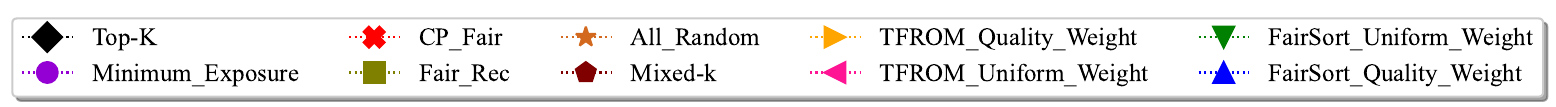}} 
        }

\end{figure*}

\subsection{ \textbf{Compared Approaches}}

            \subsubsection{\textbf{\textit{Top-K}}}This algorithm directly recommends the Top-K items in the original recommendation list $\mathbf{l_{u}^{ori}}$.

             \subsubsection{\textbf{\textit{Mixed-K}}}Choose top ($\left \lceil \frac{K}{2}  \right \rceil$)  relevant products at first and then the remaining (K-$\left \lceil \frac{K}{2}  \right \rceil$) randomly.
            
            \subsubsection{\textbf{\textit{All Random}}}  Randomly select K items from the user’s original recommendation list $\mathbf{l_{u}^{ori}}$ to recommend.

            \subsubsection{\textbf{\textit{Minimum Exposure}}} The model helps UF by recommending the item that currently receives the least amount of exposure each time.

            \subsubsection{\textbf{\textit{FairRec (WWW 2020)}}}This is a state-of-the-art algorithm that guarantees two-sided fairness based on a greedy strategy and heuristic rule, which guarantees EF1 fairness on the user side and guarantees at least Maximin Share (MMS) of exposure for most of the items\cite{8}.

            \subsubsection{\textbf{\textit{CPFair (SIGIR 2022)}}}This is a state-of-the-art algorithm that guarantees two-sided fairness based on a greedy strategy and heuristic rule, which is a mixed-integer linear programming re-ranking algorithm (MILP) that seamlessly integrates fairness constraints from both the consumer and producer-side in a joint objective framework\cite{naghiaei2022cpfair}.

            \subsubsection{\textbf{\textit{TFROM (SIGIR 2021)}}}This is a state-of-the-art algorithm that guarantees two-sided fairness based on a greedy strategy and heuristic rule, which also has two versions for offline and online recommendation\cite{9}.
            \vspace{-0.5\baselineskip} 

\subsection{\textbf{Experiment Results and Analysis}}    

    \subsubsection{\textbf{Some Important Question}}
       
    \begin{itemize}[leftmargin=*]
      
         \item \textbf{RQ1:} Does Top-K recommendation lead to severely unfair exposure distribution on the provider side?
         
          \item \textbf{RQ2:} Can FairSort guarantee that the recommendation quality for each user does not fall below a threshold?
          
         \item \textbf{RQ3:} Can FairSort guarantee fairness for both sides while maintaining a high level of personalized recommendation?
         \end{itemize}
           
    \subsubsection{\textbf{Experiment Seting}}
              \begin{itemize}[leftmargin=*]
                  \item We conducted experiments for the offline situation on three datasets and evaluated the results of the algorithms at different $K$ values.

                  \item By generating a random recommendation sequence to simulate an online scenario. Each user is given 10 chances to be recommended, and the order is randomly disrupted. \textit{FairRec and CPFair} are unsuitable for online recommendation scenarios, so we do not compare them in this experiment.

                  \item To address RQ3 and better illustrate the comprehensive capabilities of \textit{FairSort}, we adopted a metric introduced in the \textit{CPFair}\cite{naghiaei2022cpfair}, referred to as UIR (Un-Fairness Inhibition Rate) in equation(\ref{UIR}). A smaller UIR indicates a model's stronger ability to maintain high personalized recommendation quality while ensuring both-side fairness. Where $w_{1}$ and $w_{2}$ are weighting parameters that the system designer can select depending on the priority for each fairness aspect. $Utility$ is the user's average recommendation quality. In this paper, we select $w_{1}=w_{2}=1$, $\mu_{1}$ and $\mu_{2}$ to represent correction parameters, each adopting the  DCF and DPF of the worst-performing value, respectively. For instance, $\mu_{1}$ selects the DCF of the \textit{Minimum Exposure} model, while $\mu_{2}$ chooses the DPF of the \textit{Top-K} model. In this study, we conducted 150 experimental cases and selected 5 models, i.e., 3 (Datasets) $\times$ 25 (The range of K) $\times$ 2 (QF and UF)  $\times$ 5(Models) to evaluate the overall performance of models w.r.t. two-sided fairness. 
              \end{itemize}

        Next, we will answer the previous three questions, providing evidence for the necessity, reliability, and effectiveness of \textit{FairSort}. Our answer to RQ1 highlights the necessity of fair post-processing reordering for \textit{Top-K} recommendation lists. Answering RQ2 demonstrates the reliability of our \textit{FairSort} in ensuring personalized recommendation quality. Finally, our response to RQ3 validates the effectiveness of \textit{FairSort} in simultaneously ensuring fairness on both sides while pursuing high-quality personalized recommendations.

    \subsubsection{\textbf{\textit{Answer to RQ1}}}To answer this question, we need to examine the experimental results of \textit{Top-K} based on UF and QF in both offline and online scenarios, as shown in fig (\ref{offEfig1}, \ref{offEfig2}, \ref{offEfig3}, \ref{onEfig1}, \ref{onEfig2}, \ref{onEfig3}) at (c, d). 
            We can draw the following conclusions:

            \begin{itemize}[leftmargin=*]
                \item Whether based on the UF or QF concept, \textit{Top-K} in an online scenario, over the course of the recommendations,  \textit{Top-K} exhibits a significant variance, resulting in highly uneven exposure allocation on the provider side.  In the offline scenario, the unfairness of exposure on the provider side becomes more severe with the increasing value of K. Because \textit{Top-K}  is user-centric and focuses solely on personalized recommendations, neglecting provider-side fairness. The experimental results address RQ1 and underscore the importance of fair post-processing reordering for Top-K recommendations. Without it, there's an exponential rise in DPF with increasing recommendations, as demonstrated in figures   (\ref{onEfig1}, \ref{onEfig2}, \ref{onEfig3}) at (c, d).
                \end{itemize}

 \textbf{\textit{(3-1) Provider-Side Fairness:}} we further analyze the performance of each model w.r.t UF and QF respectively.

            \begin{itemize}[leftmargin=*]

                \item Let's examine the UF performance of the \textit{Minimum Exposure} model, which best aligns with the UF concept, as well as the \textit{All Random 
                }model, which can mitigate UF. As shown in fig (\ref{offEfig1}, \ref{offEfig2}, \ref{offEfig3}, \ref{onEfig1}, \ref{onEfig2}, \ref{onEfig3}) at (c), in both online and offline scenarios, in contrast to \textit{Top-K}(user-centric), \textit{Minimum Exposure} (provider-centric) perform the best in terms of UF, converging to 0. \textit{All Random} satisfies UF to some extent. \textit{Minimum Exposure} (provider-centric) is because it follows the UF approach, prioritizing exposure fairness for providers. \textit{All Random} benefits from randomness, ensuring equal exposure opportunities for each item,  indirectly satisfying UF to some extent. However, they do not consider user-side recommendation utility, significantly degrading recommendation quality.

    \begin{figure*}[!h]
		\centering
		\subfigure[Average recommendation quality ($ \uparrow $)]{
			\begin{minipage}[t]{0.25\linewidth}
				\centering
				\includegraphics[width=\textwidth,height=2.5cm]{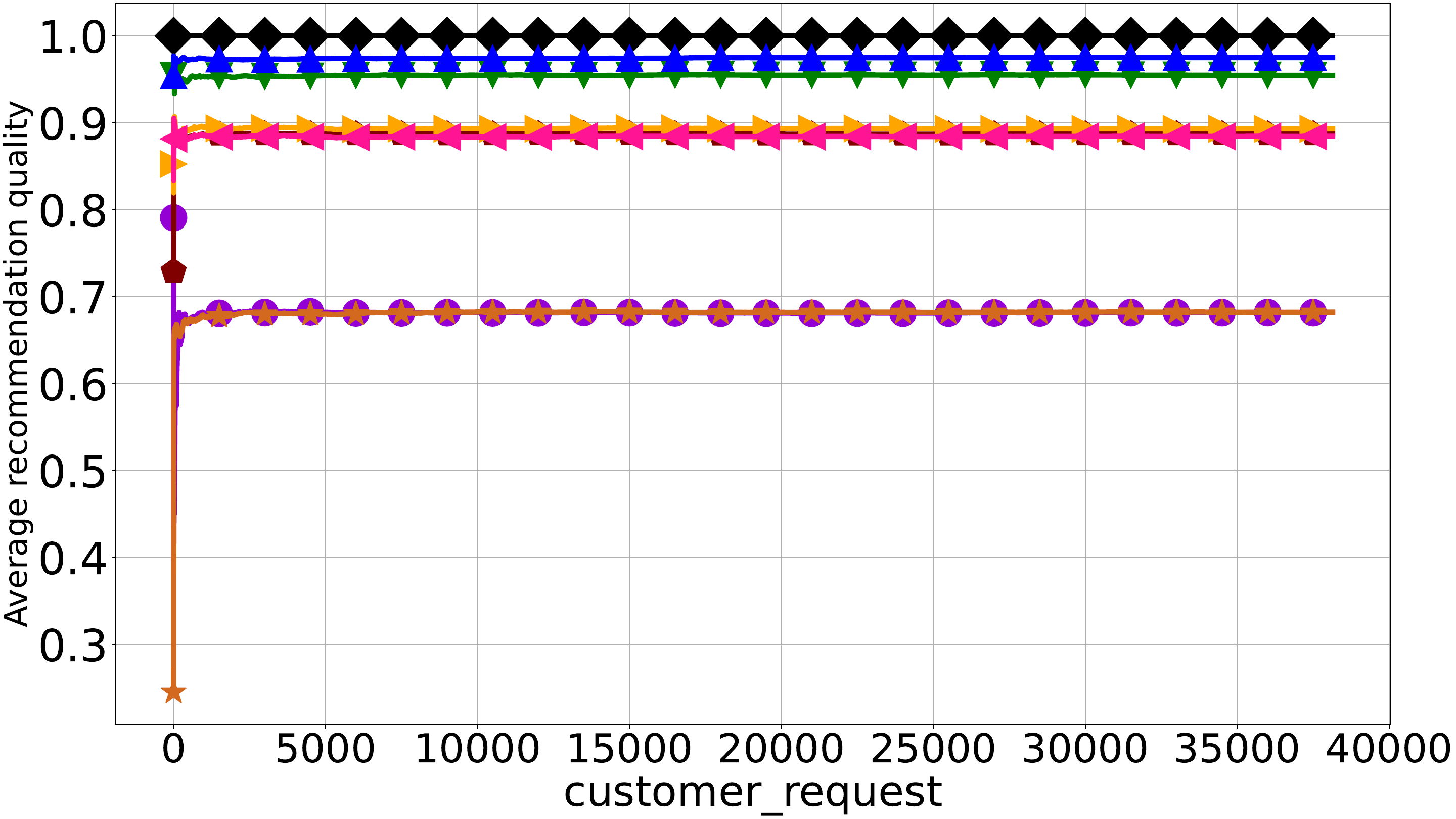}
			\end{minipage}%
		}%
		\subfigure[DCF--Variance of NDCG ($ \downarrow $)]{
			\begin{minipage}[t]{0.25\linewidth}
				\centering
				\includegraphics[width=\textwidth,height=2.5cm]{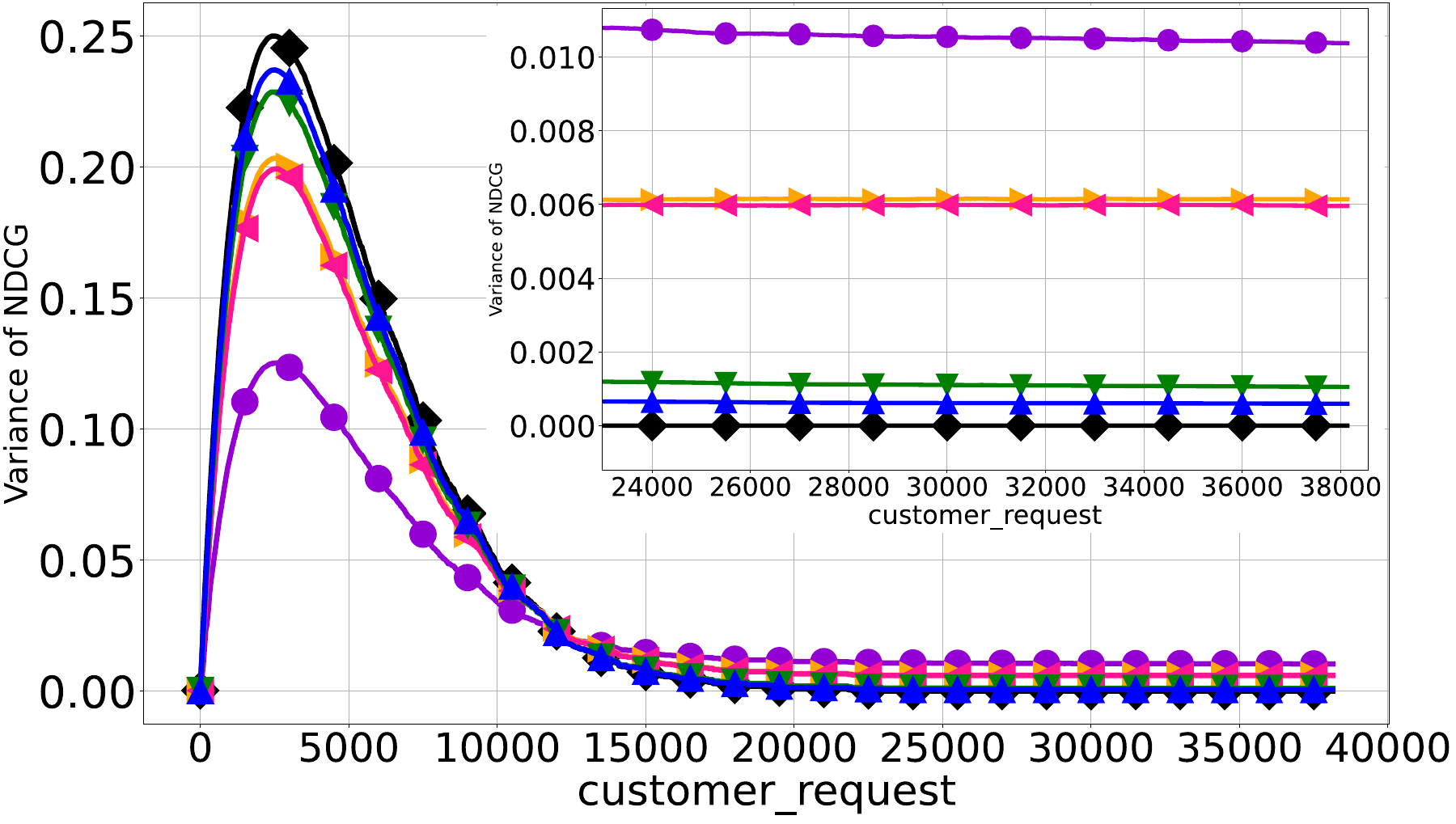}
			\end{minipage}
		}%
		\subfigure[DPF--Uniform Weight Fairness ($ \downarrow $)]{
			\begin{minipage}[t]{0.25\linewidth}
				\centering
				\includegraphics[width=\textwidth,height=2.5cm]{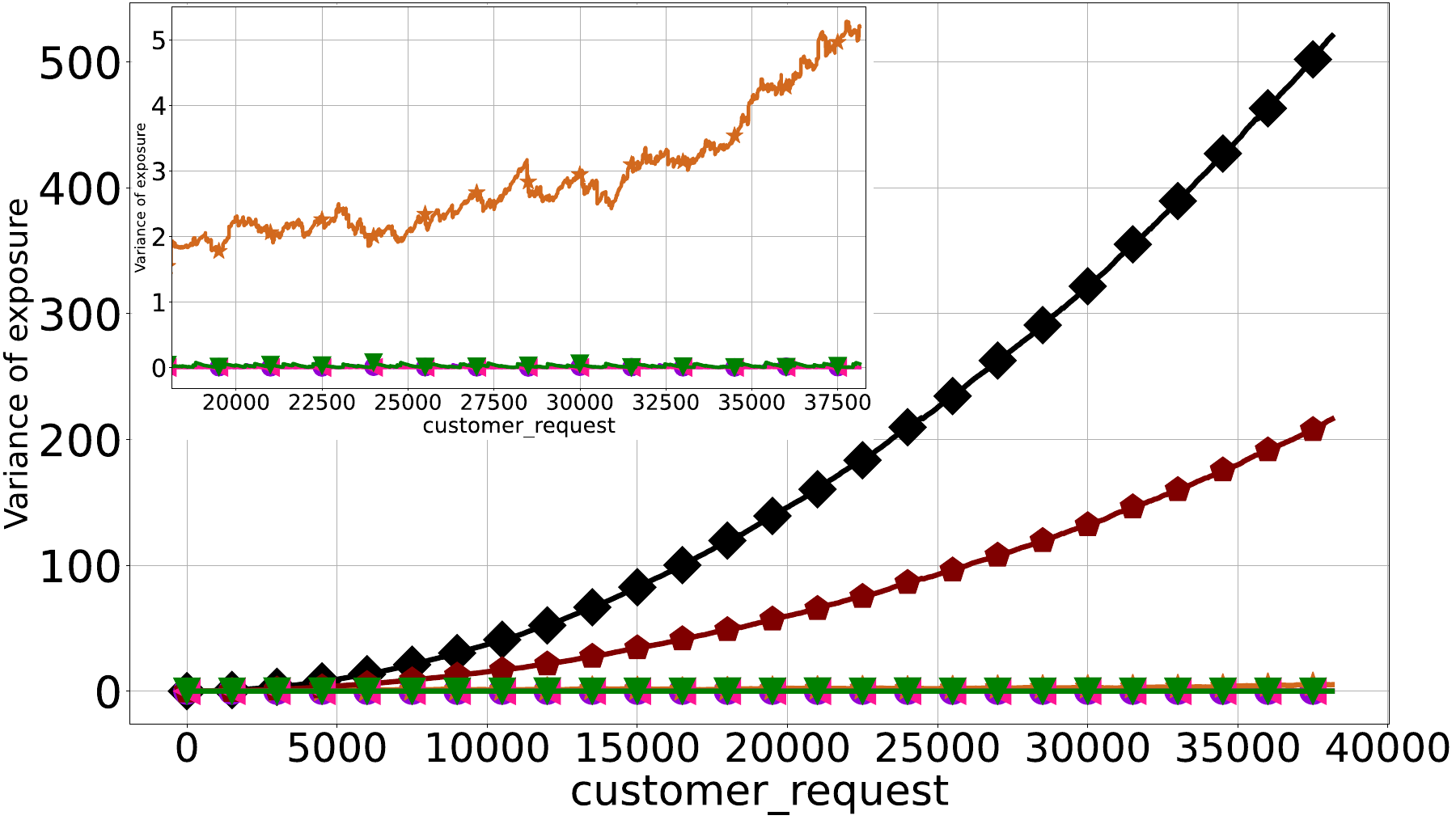}
			\end{minipage}
		}%
		\subfigure[DPF--Quality Weight Fairness ($ \downarrow $)]{
			\begin{minipage}[t]{0.25\linewidth}
				\centering
				\includegraphics[width=\textwidth,height=2.5cm]{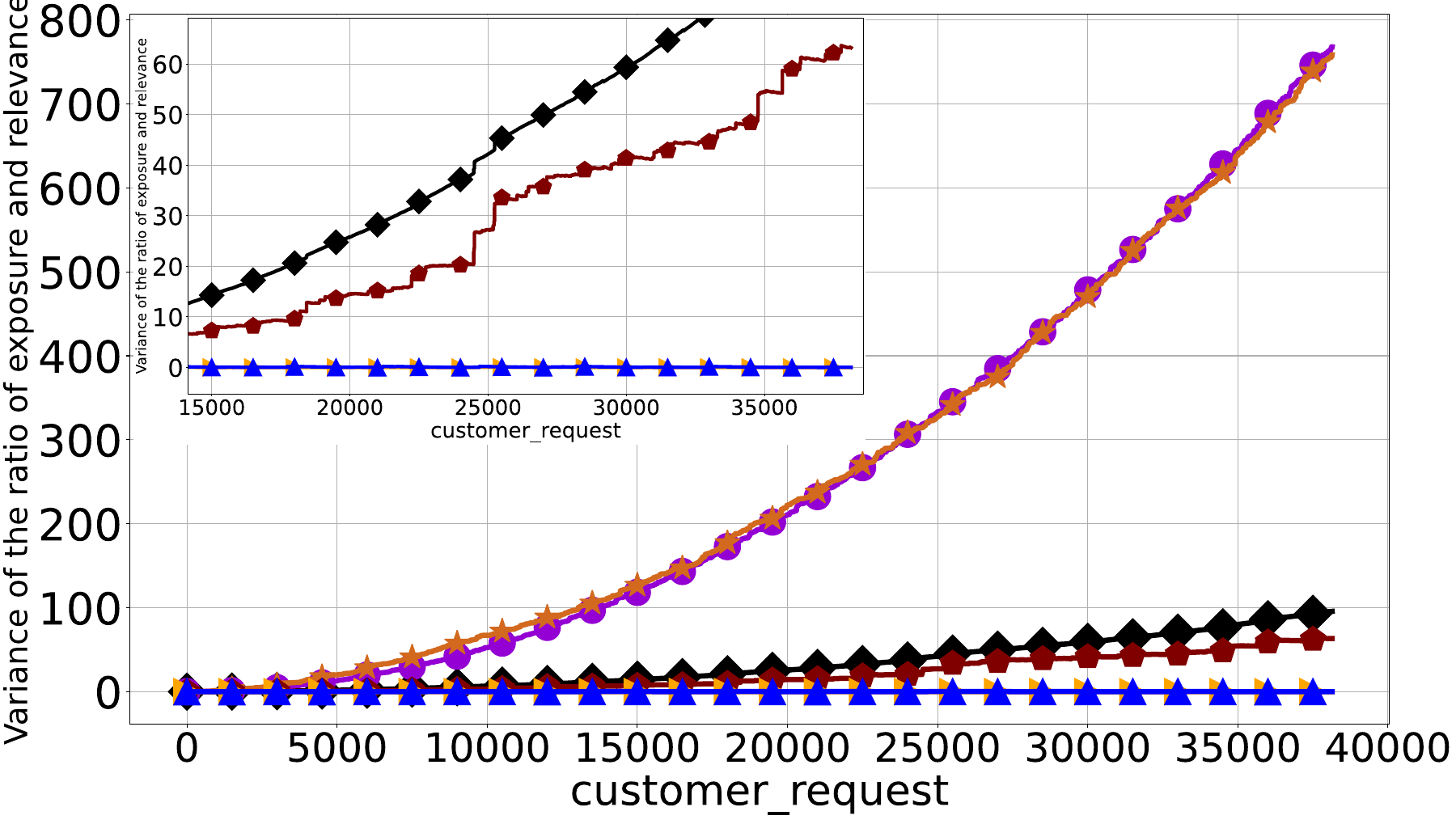}
			\end{minipage}
		}%
		\centering
		\caption{Experiment Results on Ctrip Dataset in the Online Scenario.}
		\label{onEfig1}
	\end{figure*}
	
	\begin{figure*}[!h]
		\centering
		\subfigure[Average recommendation quality ($ \uparrow $)]{
			\begin{minipage}[t]{0.25\linewidth}
				\centering
				\includegraphics[width=\textwidth,height=2.5cm]{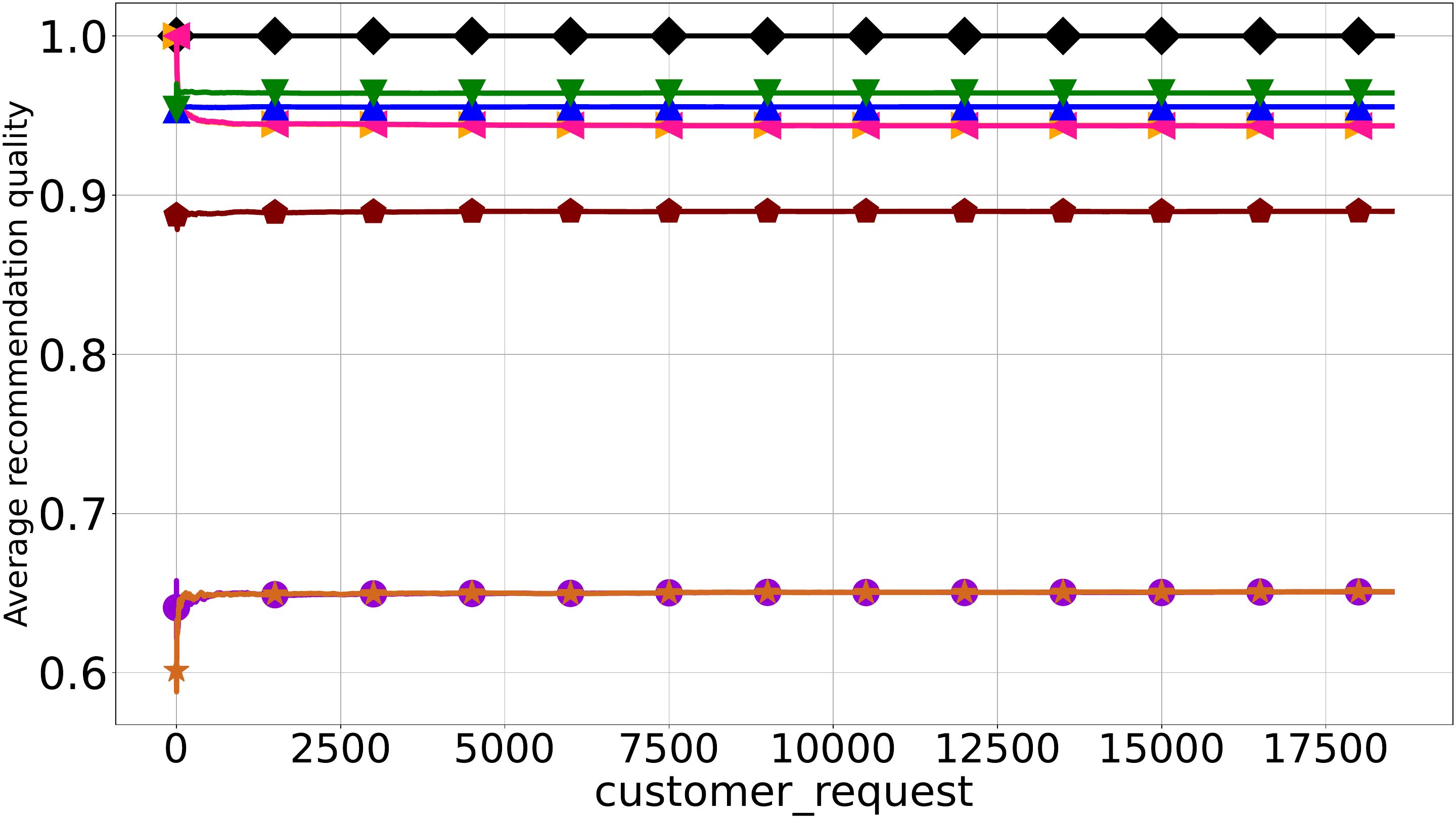}
			\end{minipage}%
		}%
		\subfigure[DCF--Variance of NDCG ($ \downarrow $)]{
			\begin{minipage}[t]{0.25\linewidth}
				\centering
				\includegraphics[width=\textwidth,height=2.5cm]{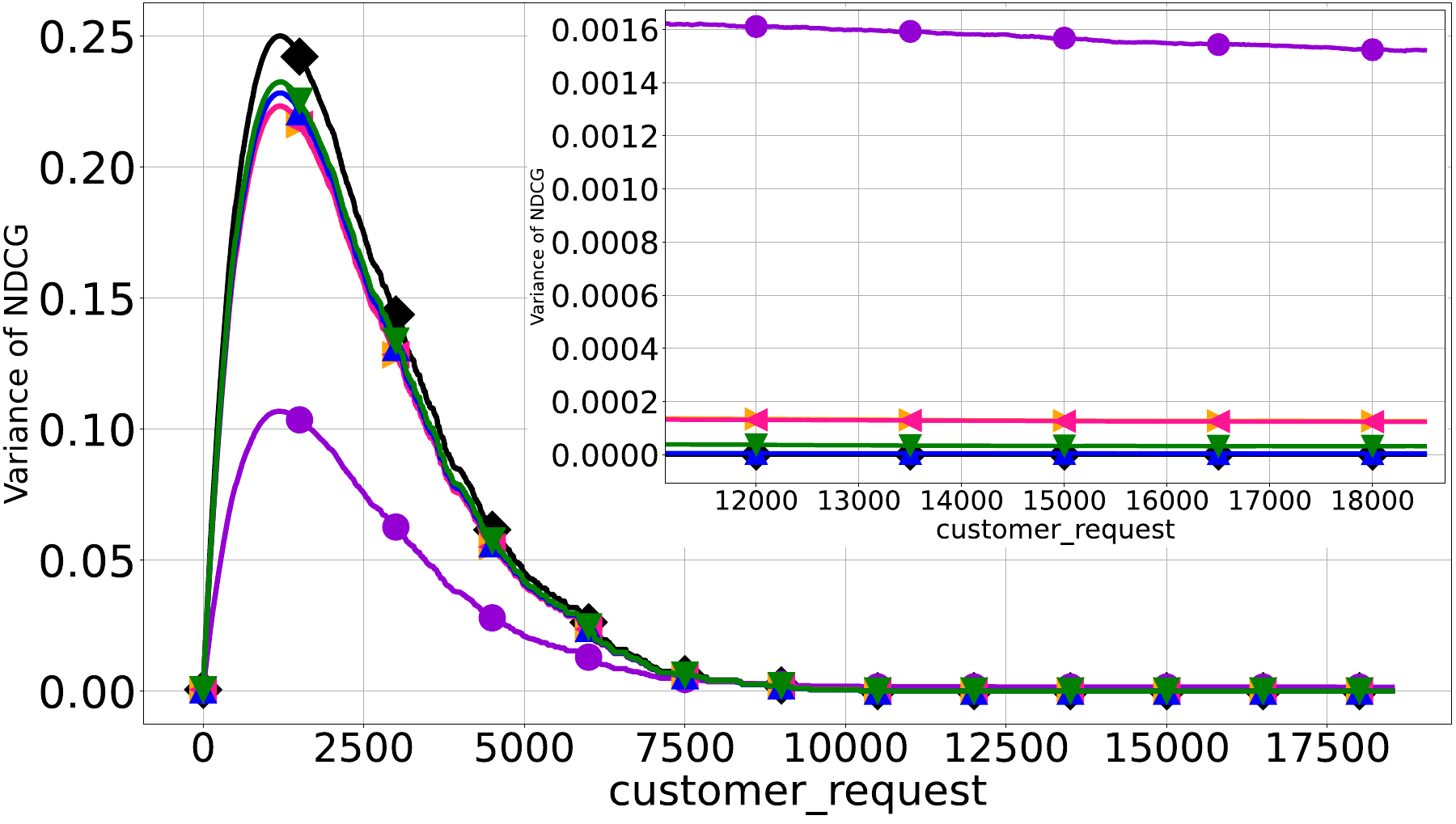}
			\end{minipage}
		}%
		\subfigure[DPF--Uniform Weight Fairness ($ \downarrow $)]{
			\begin{minipage}[t]{0.25\linewidth}
				\centering
				\includegraphics[width=\textwidth,height=2.5cm]{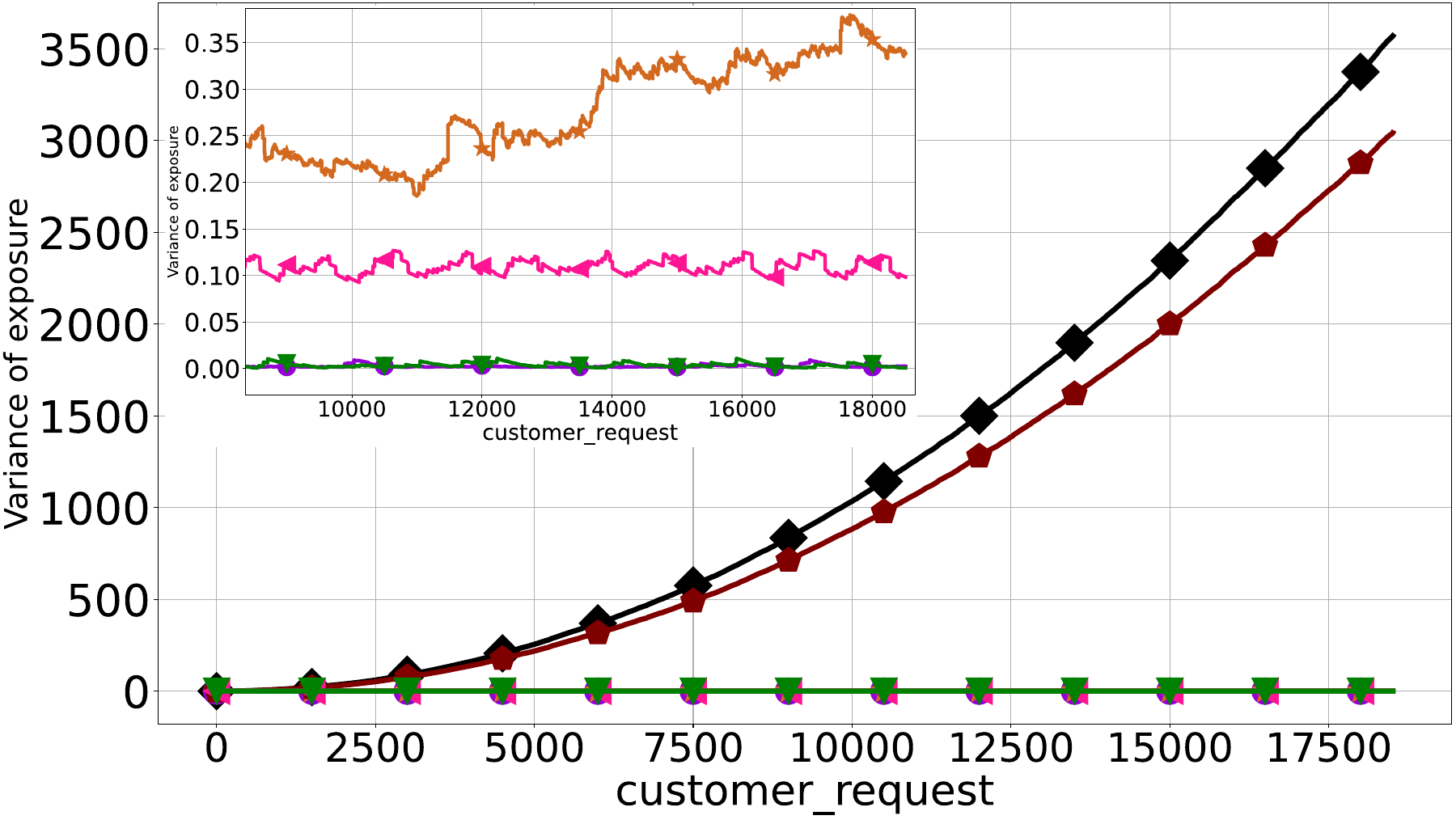}
			\end{minipage}
		}%
		\subfigure[DPF--Quality Weight Fairness ($ \downarrow $)]{
			\begin{minipage}[t]{0.25\linewidth}
				\centering
				\includegraphics[width=\textwidth,height=2.5cm]{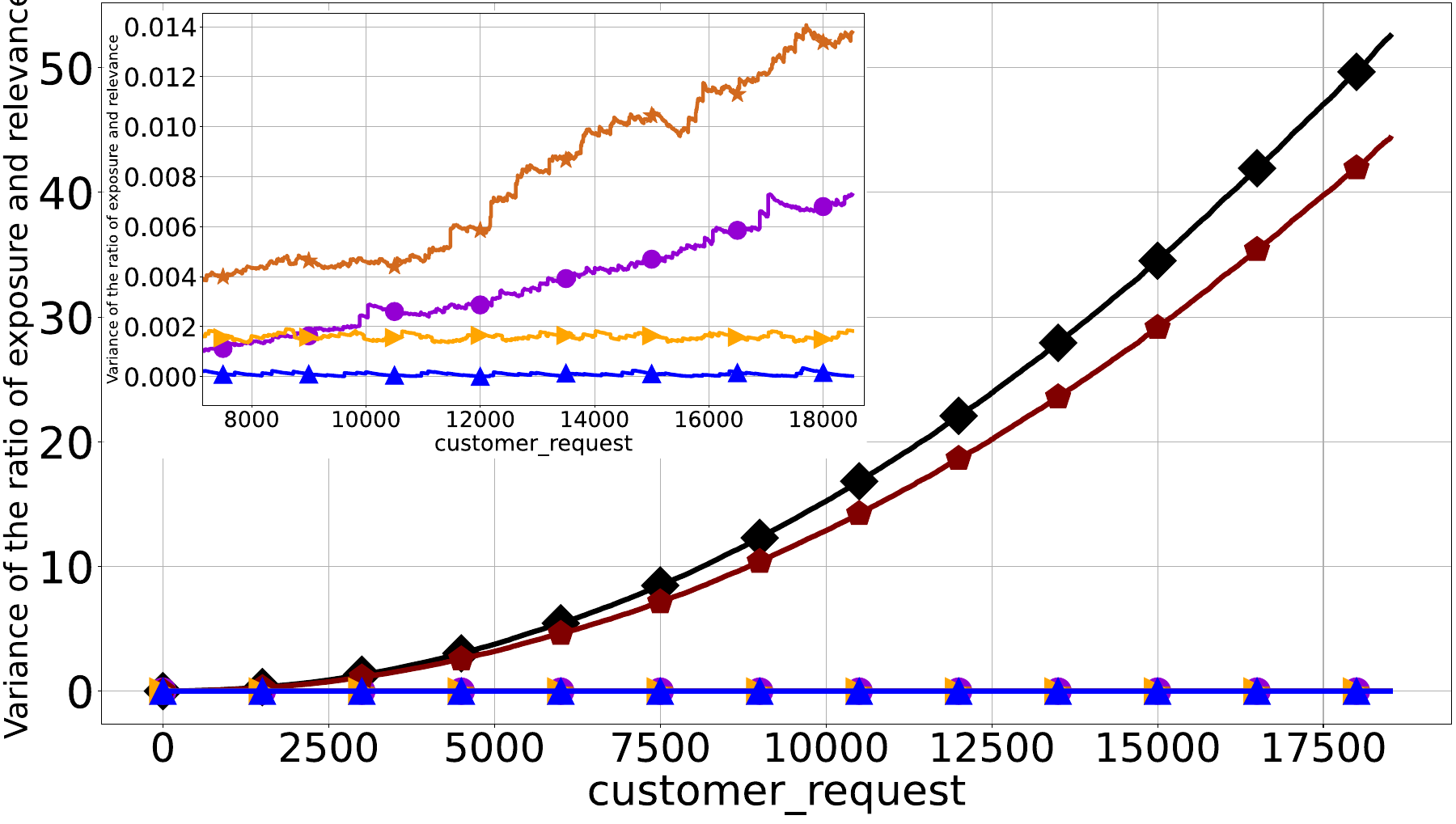}
			\end{minipage}
		}%
		\centering
		\caption{Experiment Results on Amazon Dataset in the Online Scenario.}
		\label{onEfig2}
	\end{figure*}
	
	\begin{figure*}[!h]
		\centering
		\subfigure[Average recommendation quality ($ \uparrow $)]{
			\begin{minipage}[t]{0.25\linewidth}
				\centering
				\includegraphics[width=\textwidth,height=2.5cm]{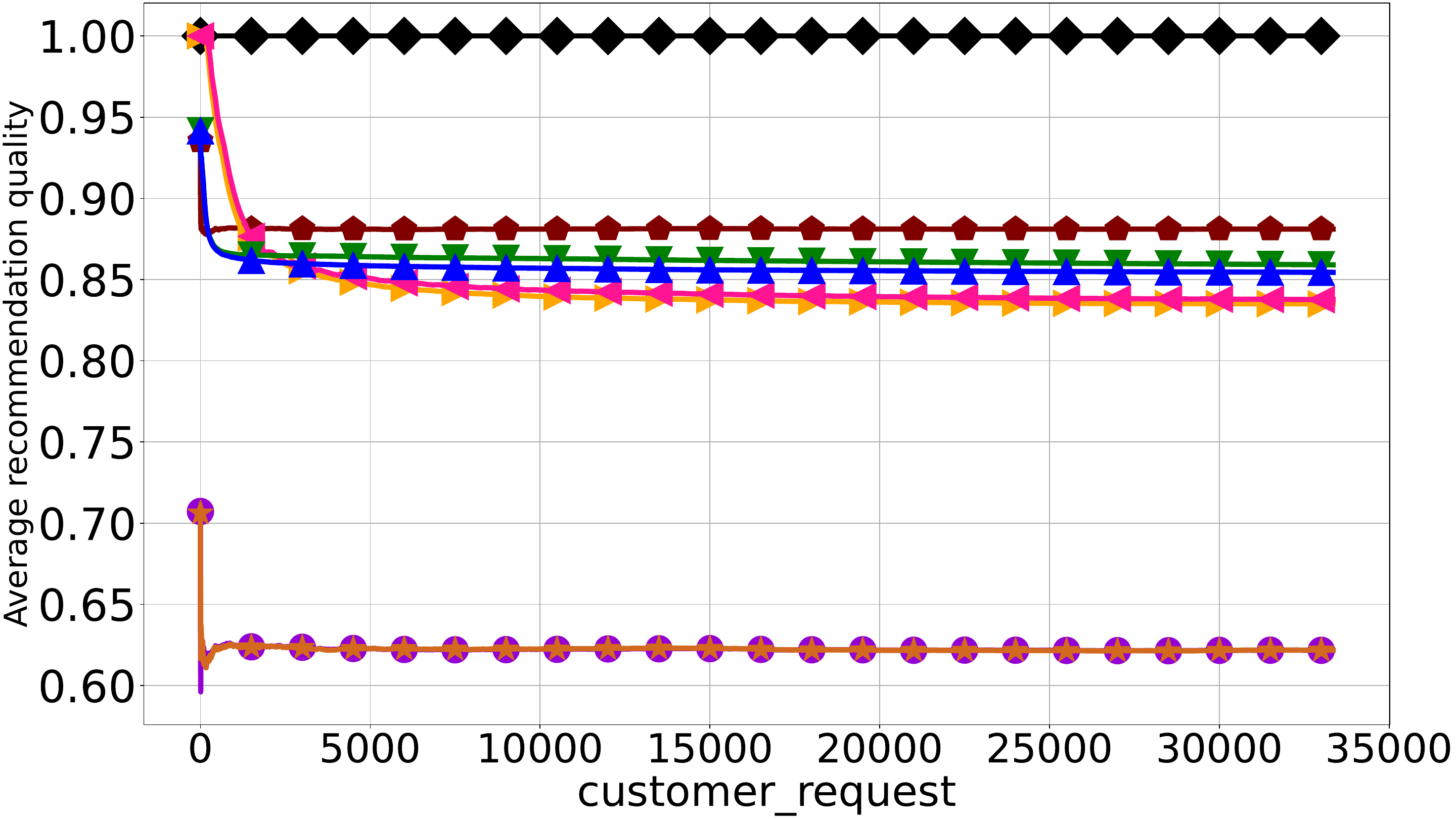}
			\end{minipage}%
		}%
		\subfigure[DCF--Variance of NDCG ($ \downarrow $)]{
			\begin{minipage}[t]{0.25\linewidth}
				\centering
				\includegraphics[width=\textwidth,height=2.5cm]{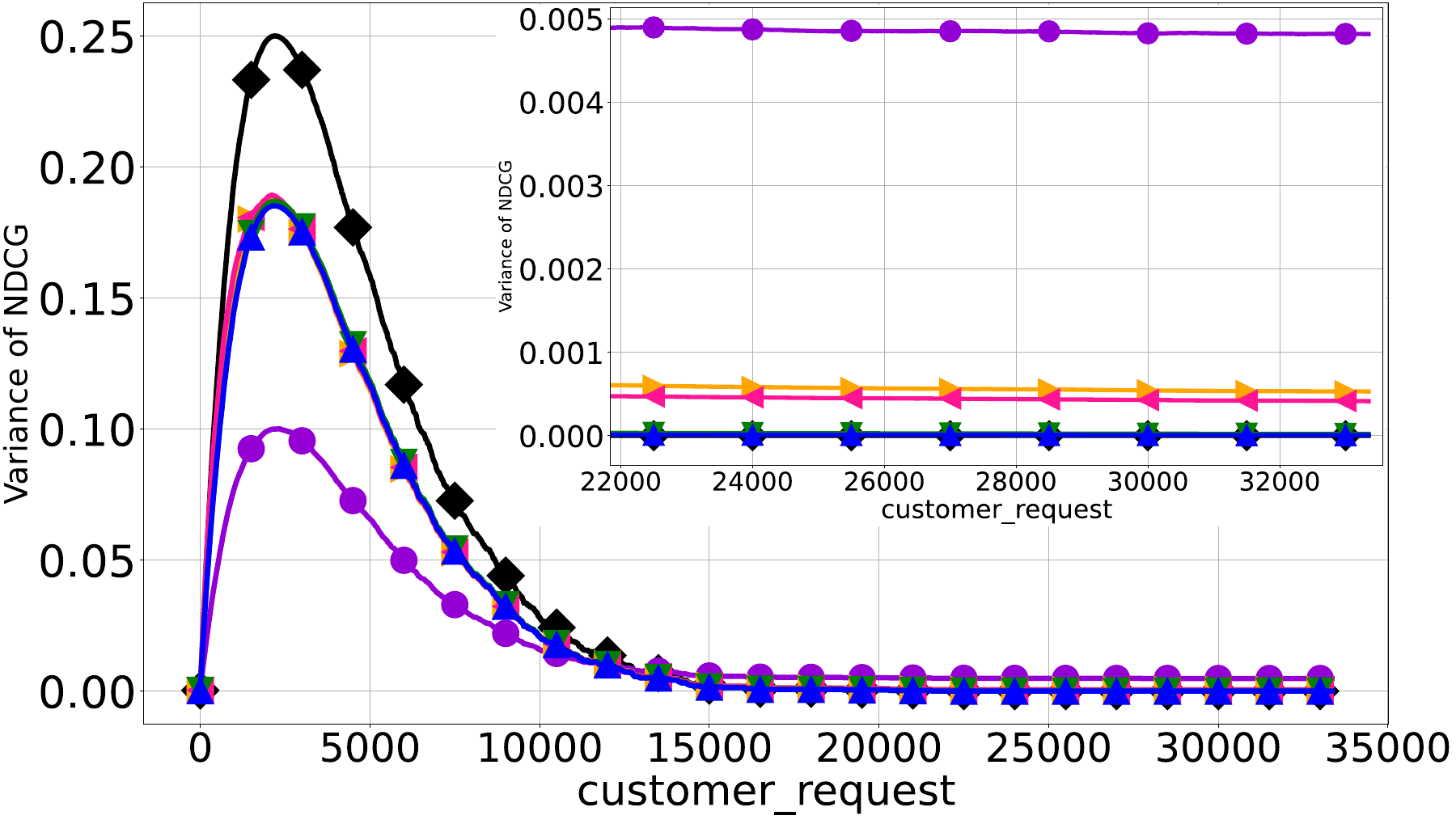}
			\end{minipage}
		}%
		\subfigure[DPF--Uniform Weight Fairness ($ \downarrow $)]{
			\begin{minipage}[t]{0.25\linewidth}
				\centering
				\includegraphics[width=\textwidth,height=2.5cm]{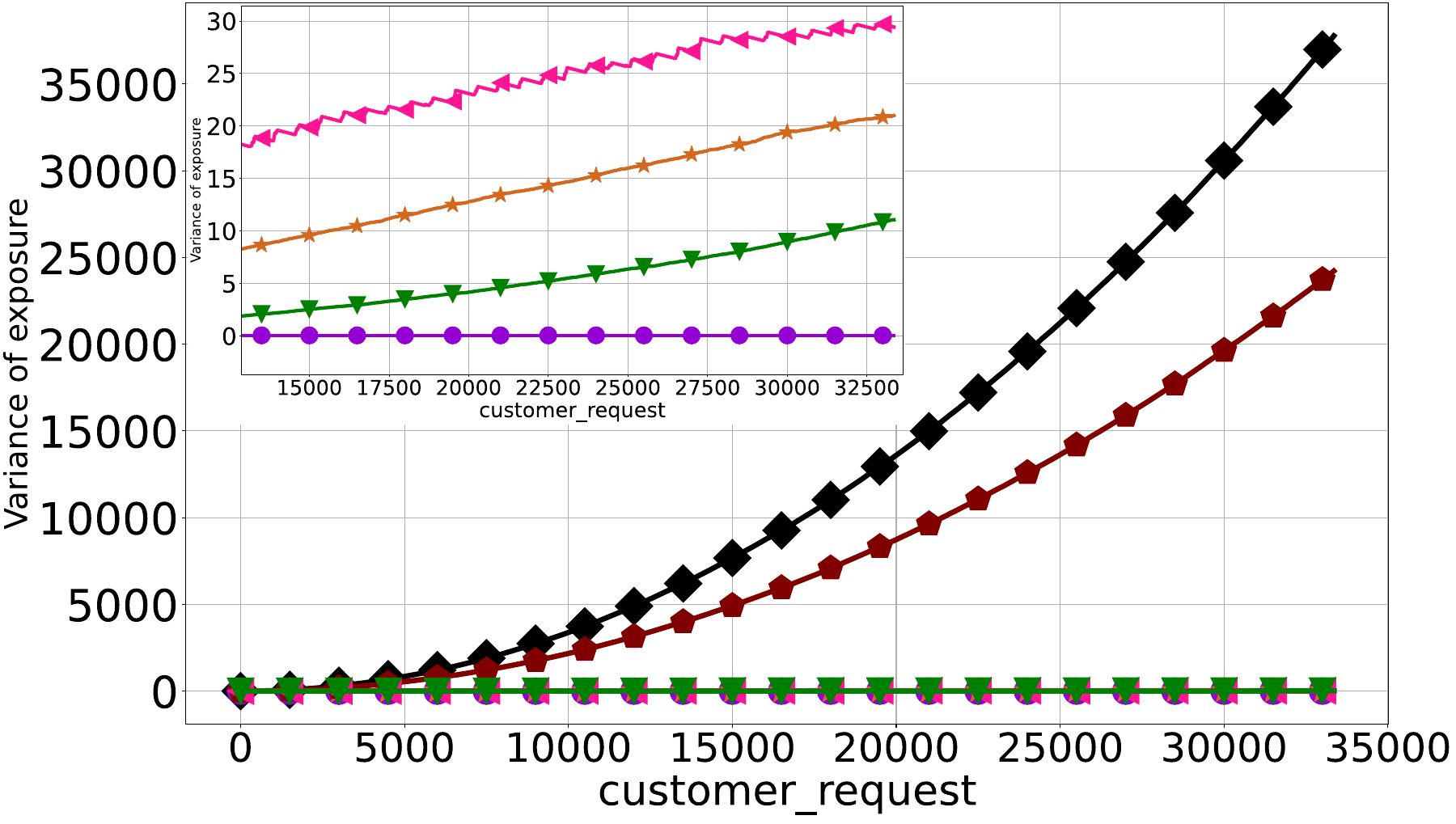}
			\end{minipage}
		}%
		\subfigure[DPF--Quality Weight Fairness ($ \downarrow $)]{
			\begin{minipage}[t]{0.25\linewidth}
				\centering
				\includegraphics[width=\textwidth,height=2.5cm]{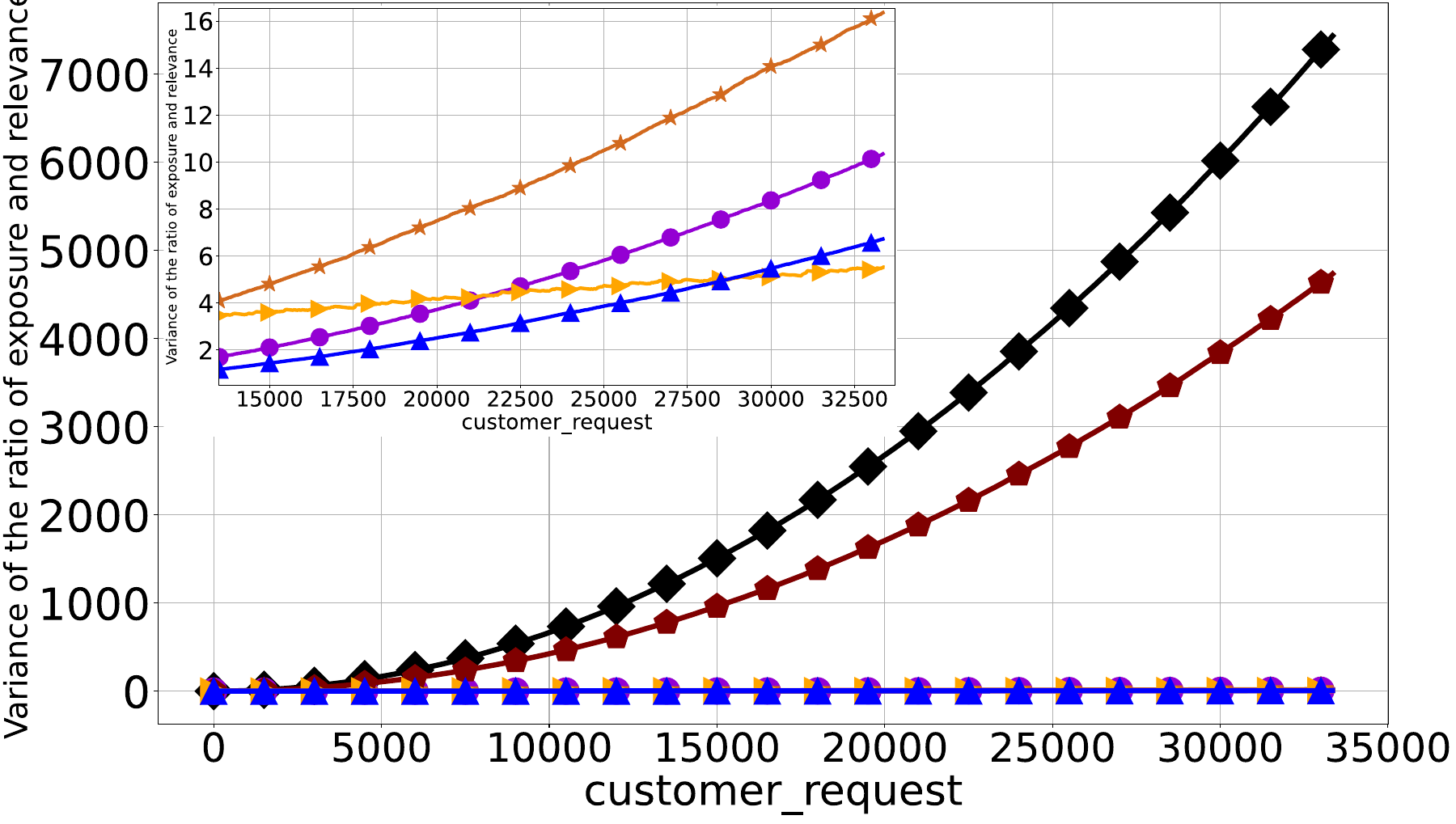}
			\end{minipage}
		}%
		\centering
		\caption{Experiment Results on Google Dataset in the Online Scenario.}
		\label{onEfig3}
  
	\end{figure*} 

\begin{figure*}[!h]
    \centering
    \centerline{ 
    
     \resizebox{0.97\textwidth}{30pt}{
       \includegraphics{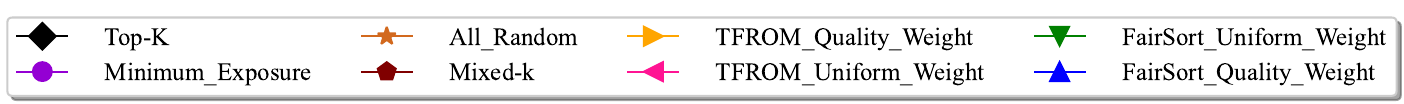}} 
        }

\end{figure*}

                  \item In both UF and QF contexts, in offline scenarios (fig (\ref{offEfig1}, \ref{offEfig2}, \ref{offEfig3}) at (c,d)), \textit{CPFair} attempt to mitigate the provider-side unfairness caused by \textit{Top-K}, but its performance is limited, with even worse results observed on the Ctrip dataset. \textit{CPFair} categorizing items into Active and In-Active groups is not well-suited for fine-grained fairness constraints, particularly when multiple providers are involved. 
                
                \end{itemize}

\begin{itemize}[leftmargin=*]

                \item  As shown in fig (\ref{offEfig1}, \ref{offEfig2}, \ref{offEfig3}, \ref{onEfig1}, \ref{onEfig2}, \ref{onEfig3}) at (c, d). Whether based on QF or UF principles, in both online and offline scenarios,  \textit{TFROM }and\textit{ FairSort} demonstrate clear control over provider-side exposure fairness distribution, with variance converging to 0, in the three datasets. \textit{TFROM} accomplishes this by calculating fair exposure $e_{p_l}^{Fair}$ as benchmarks, employing greedy and heuristic rules to align actual exposure as closely as possible with these benchmarks, ensuring fairness in provider-side exposure. 
                On the other hand, \textit{FairSort} effectively controls fairness because it employs heuristic rules to allocate a lift velocity to each item. Additionally, it uses a Binary Search mechanism to determine an adaptive $\lambda$ for each recommendation list, enabling the Minimum Utility Guarantee. Finally, \textit{FairSort} sacrifices a certain degree of recommendation quality in exchange for both-side fairness.
                \item When in terms of QF, \textit{Minimum Exposure} and \textit{All Random} models are unstable in online scenarios. These models inherently lack the awareness of QF and are only services for the UF ideology. This conclusion can be observed from fig (\ref{onEfig1}, \ref{onEfig2}, \ref{onEfig3}) at (d). Models based on UF awareness can, to some extent, achieve QF. For example, \textit{Minimum Exposure} and \textit{All Random}'s performance is decent on Amazon and Google datasets. However, they might not work as effectively, as observed on the Ctrip dataset, where their performance is even worse than that of the \textit{Top-K} model.

                \item As shown in fig \ref{onEfig3} at (c,d). Another phenomenon worth noting is that, on the Google dataset, except for the \textit{Minimum Exposure}, which demonstrates convergence capability in the UF aspect, the rest of the models, under both UF and QF principles, do not exhibit convergence capability. They only mitigate the unfair distribution of exposure to providers. This is because of the particularity of the Google dataset, where each item represents a provider. Under such fine-grained fairness constraints, there exists an inherent trade-off between recommendation utility and provider fairness. Consequently, as the recommendation process progresses, the unfairness tends to exacerbate, and relevant fairness models only serve a mitigative role.

                \begin{table*}
   
       \centering
        \captionsetup{justification=centering} 
		\caption{The NDCG distribution of  generated recommendation lists in the online  experiment, on  three major dataSets, NDCG: \textbf{[}(0--0.5),(0.5--0.6),(0.6--0.7),(0.7--0.75),(0.75--0.8),(0.8--0.85),(0.85--0.9),(0.9--0.95),( 0.95--1)\textbf{]}}
        \renewcommand\arraystretch{1.7}
        \Large
        \centering
        \resizebox{2\columnwidth}{!}{
        \setlength{\tabcolsep}{3mm}{
        \begin{tabular}{|l|l|l|l|l|}
            \hline  \label{mytable}
         \diagbox[height=2.5em, width=12.2em, font=\Large]{DataSet}{Model} & FairSort\_Quality\_Weighted & TFROM\_Quality\_Weighted (Greedy Policy)& FairSort\_Uniform & TFROM\_Uniform (Greedy Policy) \\ \hline
        Ctrip &
          {[}0, 0, 0, 0, 0, 0, 0, 9595, 28545{]} &
          {[}42, 147, 977, 1457, 2285, 4161, 7146, 11519, 10406{]} &
          {[}0, 0, 0, 0, 0, 0, 7816, 7570, 22754{]} &
          {[}33, 180, 1153, 1535, 2564, 4583, 7963, 12395, 7734{]} \\ \hline
        Amazon &
          {[}0, 0, 0, 0, 0, 0, 0, 0, 18510{]} &
          {[}0, 0, 0, 0, 0, 1, 43, 12260, 6206{]} &
          {[}0, 0, 0, 0, 0, 0, 0, 0, 18510{]} &
          {[}0, 0, 0, 0, 0, 0, 46, 12340, 6124{]} \\ \hline
        Google &
          {[}0, 0, 0, 0, 0, 0, 33267, 80, 3{]} &
          {[}0, 0, 10, 287, 4660, 18423, 8447, 1306, 217{]} &
          {[}0, 0, 0, 0, 0, 0, 33106, 238, 6{]} &
          {[}0, 0, 5, 151, 3320, 19742, 9015, 861, 256{]} \\ \hline
        \end{tabular}}}
        \label{table1}
        
\end{table*}

            \end{itemize}
                
             \subsubsection{\textit{\textbf{Answer to RQ2}}}
             To address this question, We conducted a comparative experiment on recommendation quality between \textit{FairSort} and \textit{TFROM} in the online recommendation scenario.  \textit{TFROM}  employs a greedy algorithm to ensure the recommendation quality, whereas \textit{FairSort} uses a Binary Search mechanism to guarantee recommendation quality. Table \ref{table1} illustrates the distribution of recommendation quality for \textit{FairSort} and \textit{TFROM} across the three major datasets. From these results, it's evident that  \textit{FairSort} consistently ensures that the quality of each generated recommendation list does not fall below a certain threshold. In contrast, \textit{TFROM}'s greedy strategy results in many low-quality recommendation lists. This may potentially harm the user experience and lead to user attrition. Therefore, the reliability of using a greedy algorithm to ensure recommendation utility is subject to question.
             
             In addition to addressing RQ2,  across the three major datasets and in both online and offline recommendation scenarios, we analyse the varied capabilities of all models w.r.t recommendation quality and user-side fairness.\\

                \textbf{\textit{(4-1) Recommendation Quality:}} We examine the experimental results in fig (\ref{offEfig1}, \ref{offEfig2}, \ref{offEfig3}, \ref{onEfig1}, \ref{onEfig2}, \ref{onEfig3}) at (a), we observe that:

            \begin{itemize}[leftmargin=*]
                \item In both online and offline scenarios, the user-centric \textit{Top-K} does not sacrifice any recommendation quality, so it performs optimally. On the contrary, both \textit{Minimum Exposure} (provider-centric) and \textit{All Random} models cause the most significant degradation in recommendation quality. The former is a provider-centric recommendation model that solely focuses on the provider's UF, neglecting personalized recommendation utility. The latter, due to its high degree of randomness, naturally leads to severe damage to the recommendation lists. We calculate the average recommendation quality by  $(total\  quality/c\_num)$ in an online scenario.

                \item In both online and offline scenarios, the rest are fairness-aware models, and their performance in recommendation quality falls between  \textit{Top-K} and \textit{Minimum Exposure} (provider-centric)   these two extremes. The noteworthy point is that \textit{FairSort} and \textit{TFROM} only sacrifice less than 10\% of the total recommendation quality in exchange for fairness on both sides, which can be maintained at a high level of quality recall.  This is attributed to \textit{FairSort}'s implementation of the \textbf{Minimum Utility Guarantee} mechanism, which ensures the user's recommendation quality and enhances the algorithm's reliability. We can confidently entrust \textit{FairSort} to post-processing. In contrast, \textit{TFROM} relies on a greedy rule to ensure recommendation quality. While it maintains a high-quality recall, our experiments have revealed that the greedy rule is sometimes unreliable. When both fairness constraints are considered, some recommendation lists suffer significant declines in quality, as illustrated in Table \ref{table1}.

                \item The reliability of relying on a greedy algorithm to ensure recommendation utility is questionable. Examining the offline scenario in the fig (\ref{offEfig1}, \ref{offEfig2}, \ref{offEfig3}) at (a), it's evident that \textit{FairRec}, \textit{CPFair}, and \textit{TFROM} significantly compromise user recommendation quality when K is small, which gradually alleviates as K increases. In contrast, FairSort exhibits more stability across different K values. This is because the former models employ a greedy strategy, making it hard to balance fairness constraints and maintaining recommendation quality when K is small, resulting in subpar recommendation quality recall. While \textit{FairSort} uses a Binary Search method that guarantees a minimum recommendation quality for each user regardless of the K  and considers both-side fairness constraints. Thus, \textit{FairSort} exhibits stronger robustness and reliability regarding recommendation quality recall. This observation is also reflected in user-side fairness. In fig (\ref{offEfig1}, \ref{offEfig2}, \ref{offEfig3}) at (b), models relying on greedy strategies encounter difficulties in ensuring fair user-side recommendations, especially when K is small, due to significant disruptions in recommendation quality. As a result, their Variance of NDCG shows notable fluctuations in such scenarios, whereas FairSort demonstrates more stable performance.
                
                \item In the offline scenario, illustrated in fig (\ref{offEfig1}, \ref{offEfig2}, \ref{offEfig3}) at (a) \textit{CPFair}'s performance regresses to that of \textit{Top-K}. Despite sacrificing minimal recommendation quality, its metrics for provider-side fairness are comparatively subpar. 
                 In summary, concerning recommendation quality,  \textit{Top-K} outperforms all, whereas \textit{Minimum Exposure} and \textit{All Random} exhibit the worst performance. However, \textit{TFROM} and \textit{FairSort} strike a nuanced trade-off between these two extremes in exchange for both sides fairness.

            \end{itemize} 
    
    \textbf{\textit{(4-2) User-Side Fairness: }}Examining the experimental results in fig (\ref{offEfig1}, \ref{offEfig2}, \ref{offEfig3}, \ref{onEfig1}, \ref{onEfig2}, \ref{onEfig3}) at (b), we conclude that:
    
    
\begin{figure*}[!h]
	\centering

	\subfigure[UIR of Quality Weight on Ctrip ($ \downarrow $)]{
		\begin{minipage}[t]{0.33\textwidth}
			\centering
			\includegraphics[width=0.95\textwidth]{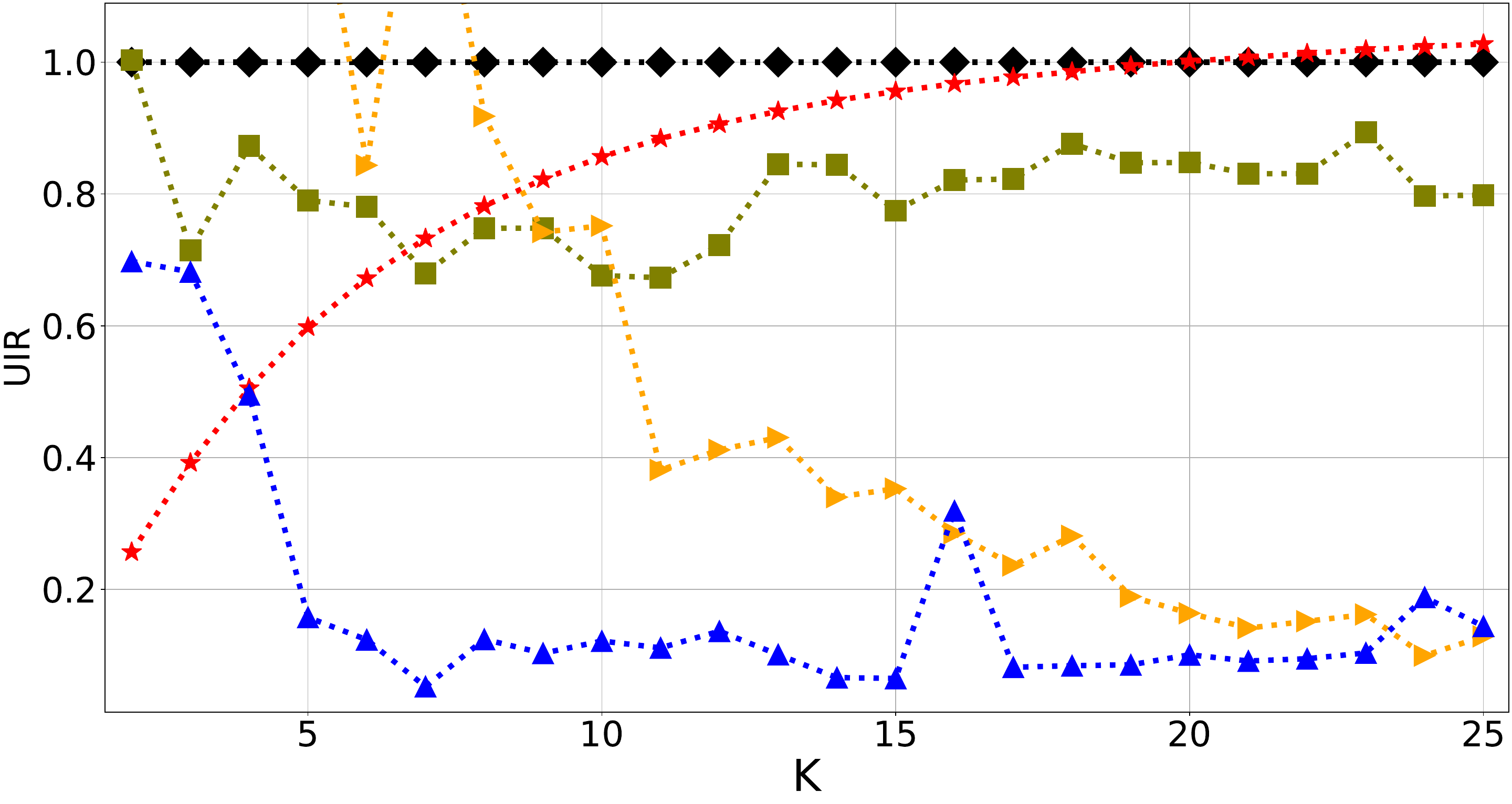}
		\end{minipage}%
	}%
	\subfigure[UIR of Quality Weight on Amazon ($ \downarrow $)]{
		\begin{minipage}[t]{0.33\textwidth}
			\centering
			\includegraphics[width=0.95\textwidth]{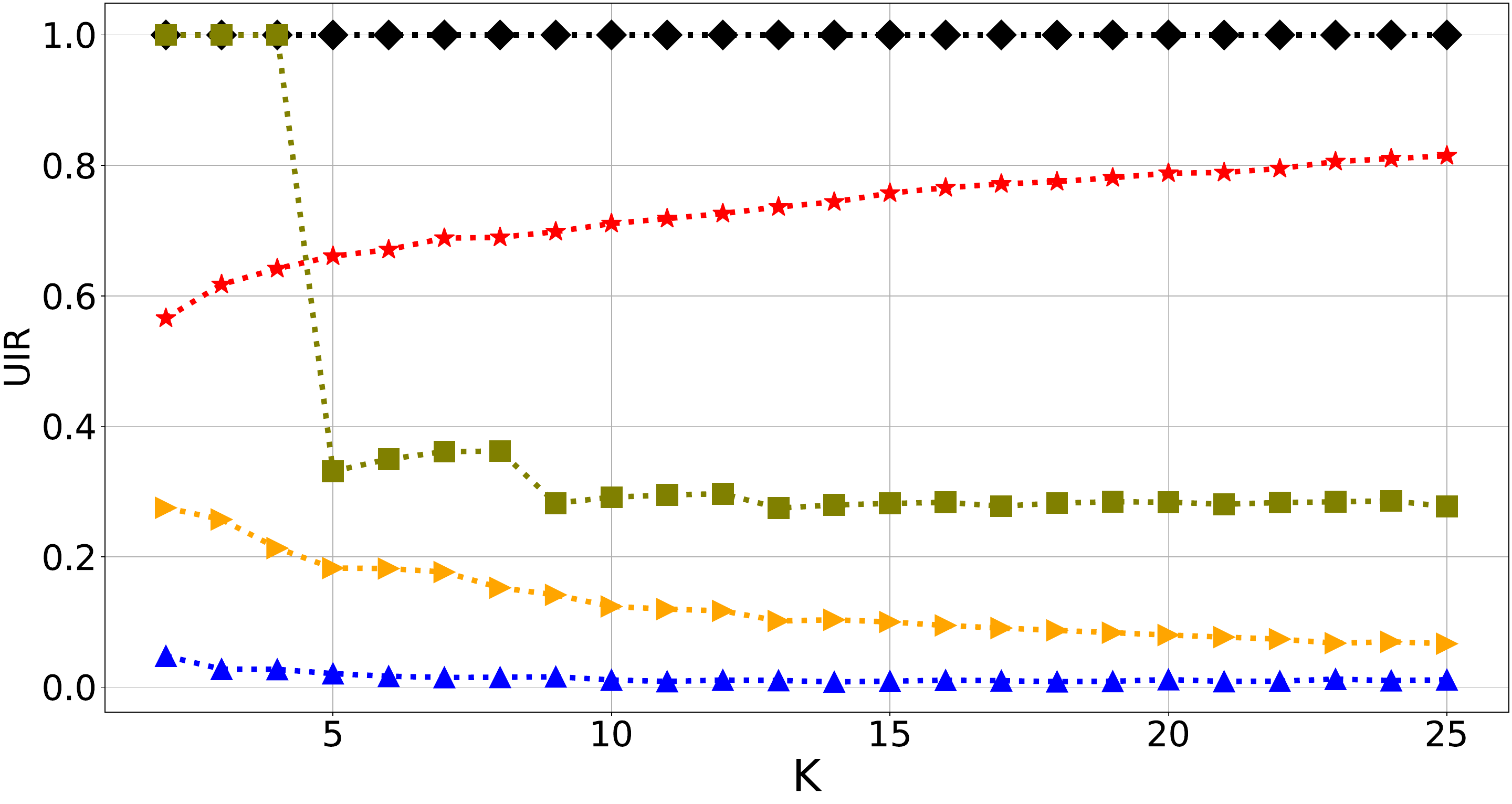}
		\end{minipage}
	}%
	\subfigure[UIR of Quality Weight on Google ($ \downarrow $) ]{
		\begin{minipage}[t]{0.33\textwidth}
			\centering
			\includegraphics[width=0.95\textwidth]{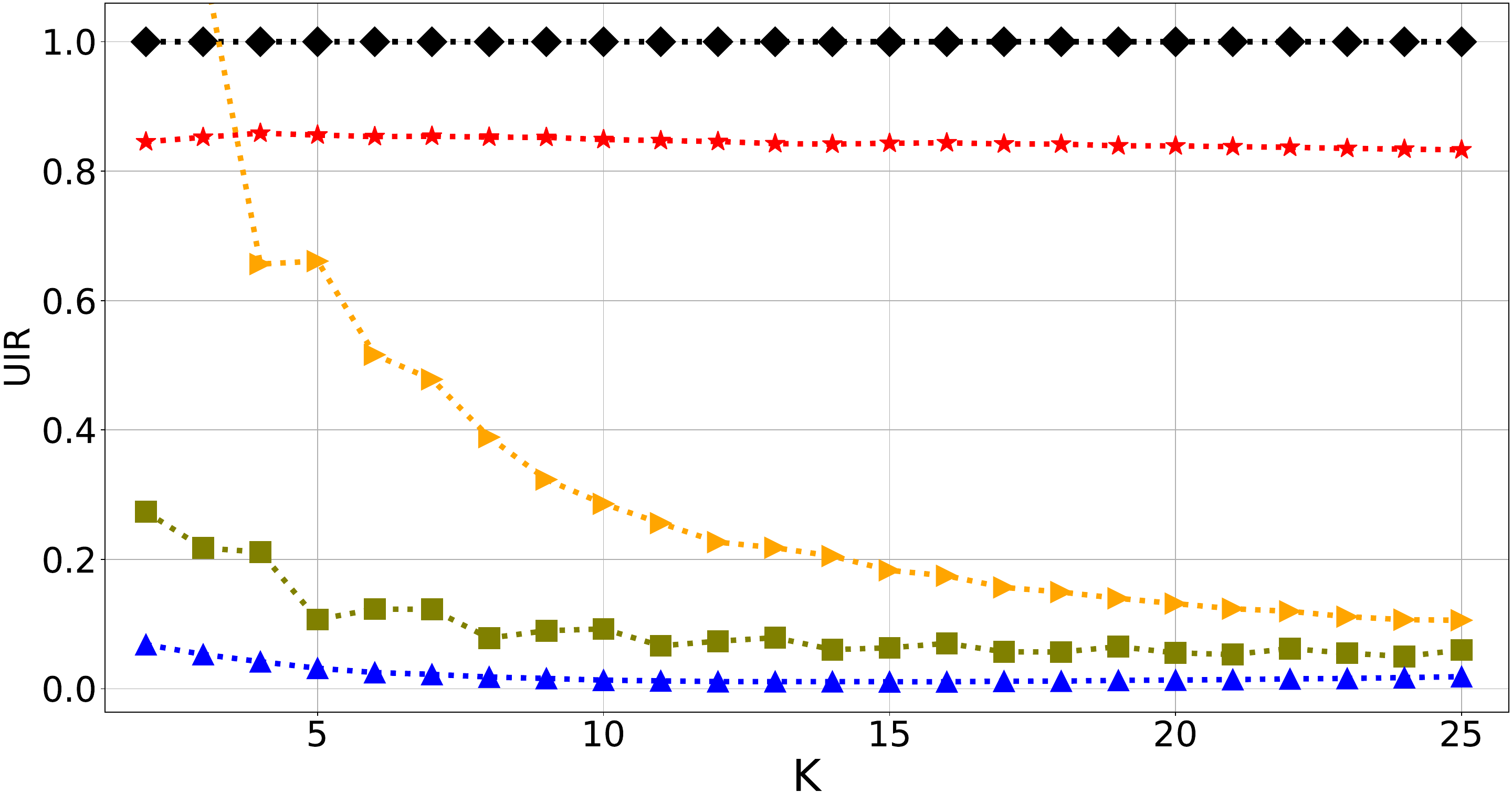}
		\end{minipage}
	}%

	\caption{The UIR Results on Three Datasets in the Offline Scenario.}
	\label{fig5}
\end{figure*}

 \begin{figure*}[!h]
	\centering

	\subfigure[UIR of Uniform Weight on Ctrip ($ \downarrow $) ]{
		\begin{minipage}[t]{0.33\textwidth}
			\centering
			\includegraphics[width=0.95\textwidth]{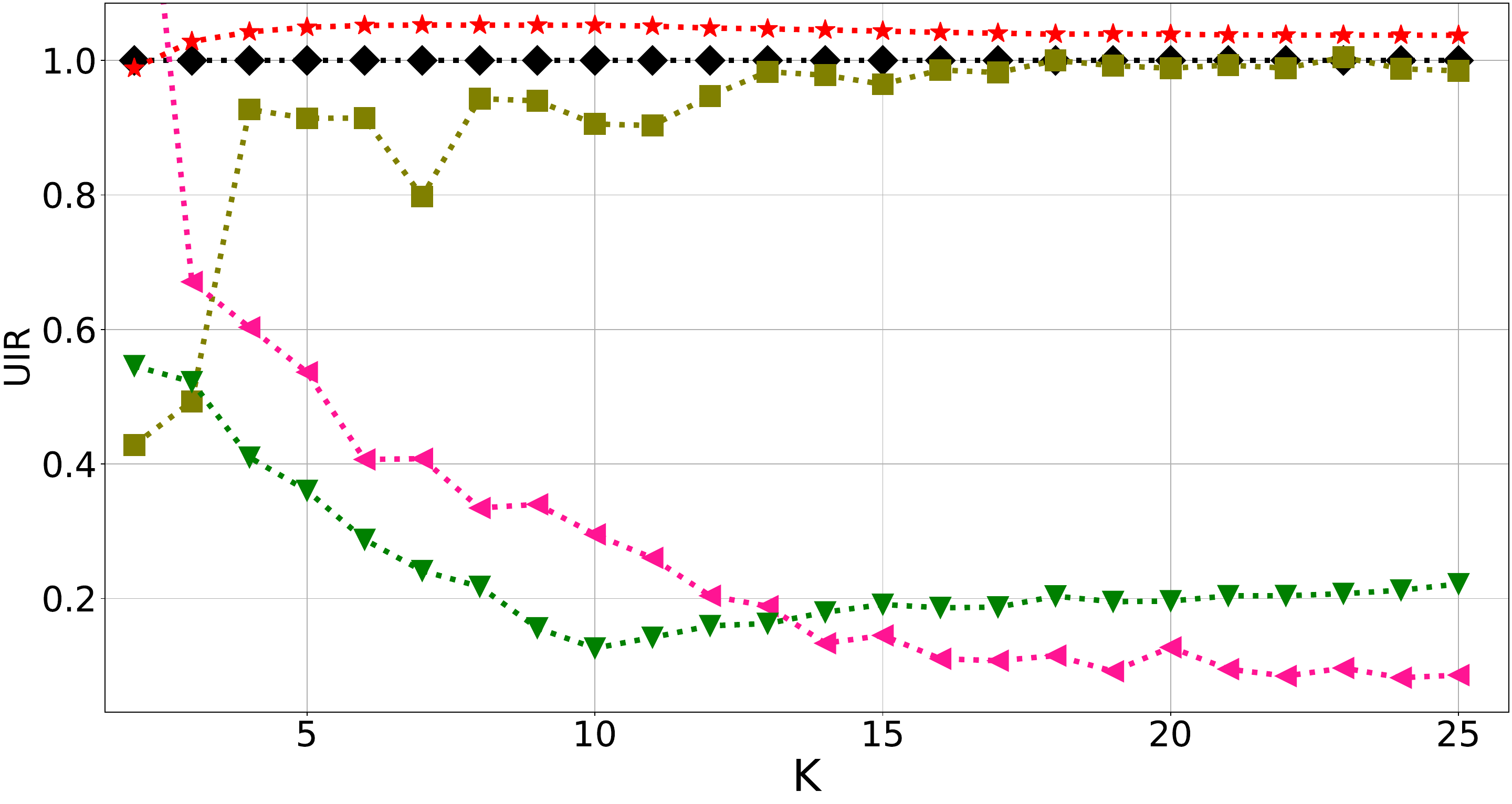}
		\end{minipage}%
	}%
	\subfigure[UIR of Uniform Weight on Amazon ($ \downarrow $) ]{
		\begin{minipage}[t]{0.33\textwidth}
			\centering
			\includegraphics[width=0.95\textwidth]{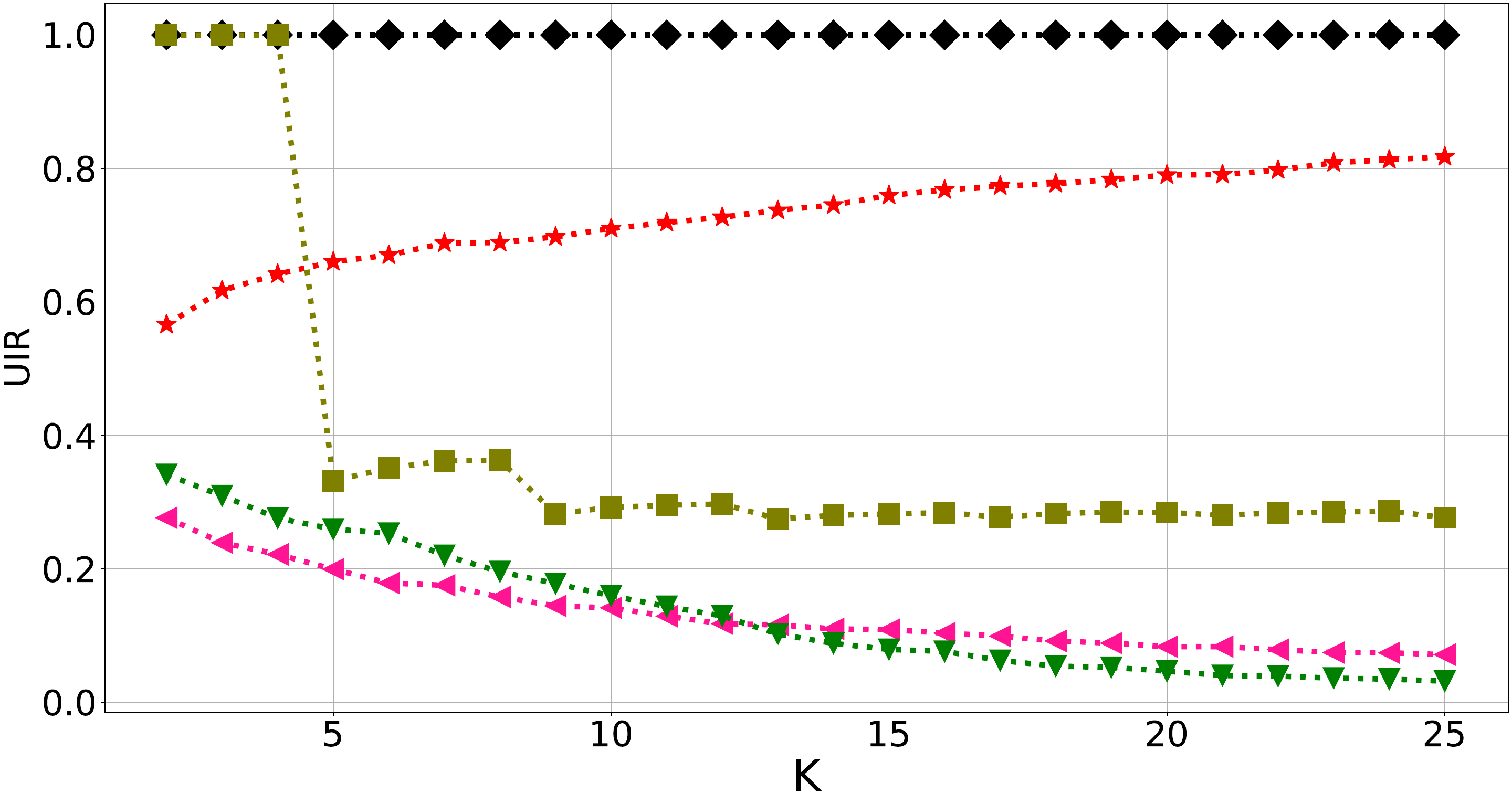}
		\end{minipage}
	}%
	\subfigure[UIR of Uniform Weight on Google ($ \downarrow $) ]{
		\begin{minipage}[t]{0.33\textwidth}
			\centering
			\includegraphics[width=0.95\textwidth]{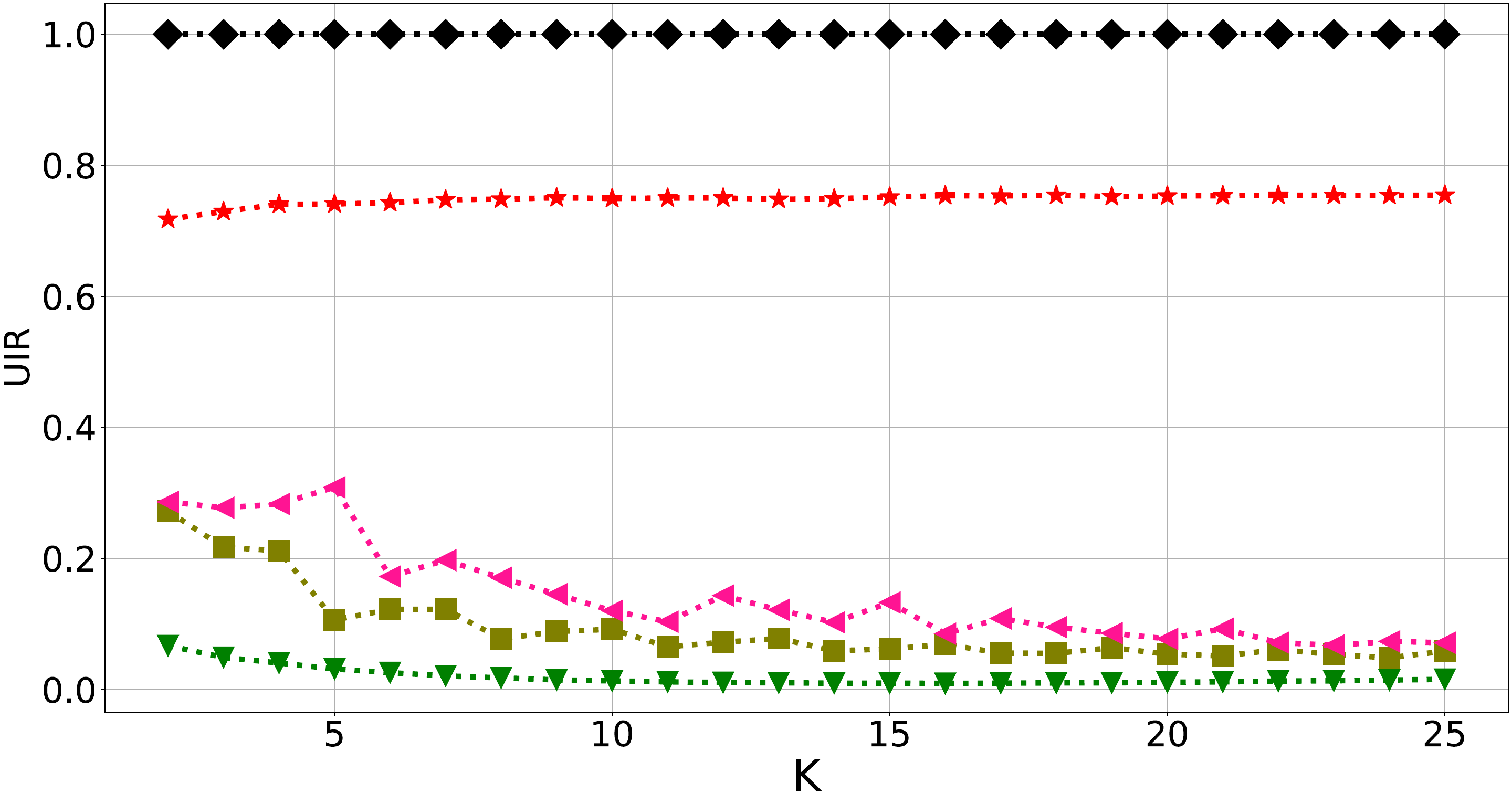}
		\end{minipage}
	}%

	\caption{The UIR Results on Three Datasets in the Offline Scenario.}
	\label{fig6}
 
\end{figure*}

\begin{figure*}[!h]
    \centering
    \centerline{ 
    
     \resizebox{\textwidth}{24pt}{
       \includegraphics{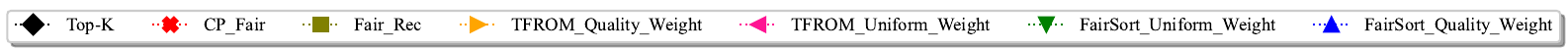}} 
        }

\end{figure*}

             \begin{itemize}[leftmargin=*]
                 \item From the figures (\ref{onEfig1}, \ref{onEfig2}, \ref{onEfig3}) at (b), we also display the late-stage convergence results of each model during the online recommendation process.  In the online scenario, all models exhibit a trend where the variance of NDCG  initially increases and then decreases.  Because, at the beginning of recommendations, the average recommendation quality of all users is 0, leading to a variance of NDCG values of 0. However, as the recommendations proceed, some users receive recommendations first, causing a rapid increase in the NDCG variance. When it reaches its peak,  it begins to decline gradually because, at this time, most of the users have been recommended at least once, and then with the increase in the number of recommendations, the user who did not get the recommendation opportunity to get the recommendation, the variance of the NDCG will decrease. Until all users have obtained at least one recommendation opportunity, as the recommendation process continues, all users' average recommendation quality variance will further converge to 0. However, different algorithms exhibit varying convergence capabilities, reflecting their distinct abilities to ensure user-side fair recommendations. 

                \item From fig (\ref{offEfig1}, \ref{offEfig2}, \ref{offEfig3}, \ref{onEfig1}, \ref{onEfig2}, \ref{onEfig3}) at (b), it can be observed that, in the three major datasets, both online and offline scenario, \textit{Top-K} ultimately achieves the best user-side fairness performance (converging to 0),  because it doesn't compromise on recommendation quality. In contrast, \textit{Minimum Exposure} (provider-centric) focuses solely on provider-side UF, significantly degrading recommendation quality. The experimental results across all three major datasets indicate that it inconsistently impacts the recommendation quality for different users, leading to the highest variance of NDCG. 

                \item In both online and offline scenarios, the recommendation models with a both-sides fairness awareness include \textit{FairSort} (QF, UF), \textit{TFROM} (QF, UF), along with \textit{FairRec} and \textit{CPFair},  All six of them can converge to the \textit{Top-K}'s performance in user-side fairness. Especially in the online scenario, \textit{FairSort-Quality-Weighted} and \textit{FairSort-Uniform-Weighted} stand out. \textit{FairSort}  achieves this performance through a \textit{Binary Search} mechanism, ensuring consistent levels of recommendation quality sacrifice for each user. Thus, \textit{FairSort}  can guarantee user-side fair recommendations while ensuring the quality does not fall below a certain threshold. On the other hand, \textit{TFROM} employs heuristic rules, prioritizing the repair of severely disrupted recommendation lists. This method can ensure user-side fairness. \textit{FairRec}  tailors a recommendation list for each user through a fixed service order. This can achieve EF1 fairness and guarantee user-side fair recommendations.
             \end{itemize}


    \subsubsection{\textbf{\textit{Answer to RQ3}}}To address RQ3 and better illustrate the comprehensive capabilities of our proposed model \textit{FairSort}, this is validated by the UIR metric in fig (\ref{fig5},\ref{fig6}).\\

    \textit{\textbf{(5-1) In the context of QF ideology:}}
        \begin{itemize}[leftmargin=*]
            \item  As illustrated in fig \ref{fig5} at (a, b, c), the \textit{FairSort-Quality-Weighted} model consistently delivers the best performance across all three major datasets. This indicates that our algorithm can significantly ensure fairness for both sides while pursuing higher personalized recommendation utility.
            
            \item The UIR for the \textit{Top-K} model consistently remains at 1. This is attributed to the fact that, in the UIR formula (\ref{UIR}), the DCF is 0, and the $DPF/\mu_{2}$ is 1, where $\mu_{2}$ considers the DPF of the \textit{Top-K} model. The denominator computes the average recommendation quality, and since \textit{Top-K} recommendations do not incur any quality losses, it is also 1. Consequently, the final UIR result remains 1.
            
            \item It's worth noting that when K is relatively small, striking a balance between recommendation quality and both-side fairness is more challenging. \textit{TFROM}, \textit{CPFair}, and \textit{FairRec}, employing greedy strategies, perform poorly under such challenges, as evidenced by their relatively high UIR. In contrast, \textit{FairSort} demonstrates more stable performance. This is attributed to its  Binary Search mechanism, which ensures the minimum recommendation utility while sacrificing quality to achieve fairness. 
            
            \item \textit{CPFair} tends to regress to \textit{Top-K} and exhibits poor overall performance. This is attributed to its consideration of a relatively coarse-grained fairness scenario, focusing on inter-group fairness between users and items, with only two groups being considered. \textit{FairRec} exhibits unstable performance due to its pursuit of EF1 fairness among individual users, a strong constraint that can lead to regression to \textit{Top-K} recommendations and poor provider-side fairness performance. Thus, its overall performance is less stable.\\

        \end{itemize}

     \textit{\textbf{(5-2) In the context of UF ideology:}}

     \begin{itemize}[leftmargin=*]
         \item As depicted in fig \ref{fig6} at (a, b, c), the conclusions align closely with those for QF and will not be reiterated here. However, it's noteworthy that \textit{FairSort} exhibits stronger and more stable capabilities in handling QF compared to UF.
     \end{itemize}

\section{CONCLUSIONS}\label{section VI}
    This paper proposes a re-ranking model FairSort to find a trade-off solution among user-side fairness, provider-side fairness, and personalized recommendations utility. Providing a novel perspective, we analogize the $\mathbf{l_{u} ^{ori}}$ as a runway rather than a conventional knapsack. Aiming at offline and online scenarios, FairSort ensures provider-side fairness by assigning fair lift velocity for each item. Guided by theorem 1, we devise a binary search strategy enabling items on each runway to run at their velocities within a specified time $\lambda$, achieving re-ranking. This strategy ensures user-side fairness and guarantees each user a minimum recommendation utility. In addition to theoretical guarantees, the experimental results from three datasets show that FairSort can guarantee both-side fairness  and provide a more reliable personalized recommendation. 
    
    Several interesting directions need further investigation. Firstly, this paper's item utility estimation method may be relatively simplified. Since fair exposure allocation is built upon item utility estimation, accurately and fairly estimating the utility of items becomes an important issue. Secondly, the implicit function E\ref{E14} requires further research to guarantee bounds on the degree of fairness or unfairness. Finally, based on this novel perspective,  further research can explore multi-granularity fairness across multiple stakeholders and the re-ranking problem under a multidimensional evaluation system, considering factors such as fairness, diversity, and novelty.

\ifCLASSOPTIONcompsoc
  \section*{Acknowledgments}
\else
  \section*{Acknowledgment}
\fi
This work was supported in part by the National Natural
Science Foundation of China (No. 62372323, 62032016, 62102281).

\section{Proof}\label{section VII}
    \label{Proof}
    \subsection{\textbf{Notations\ And\ Definition}}
        \begin{enumerate}
            \item \textbf{K:}The number of items recommended for each user
            
             \item $\mathbf{N D CG _{u}\left(\lambda_{0}\right)  }$:when $\lambda = \lambda_{0} $,  The  NDCG   of user u

            \item $\mathbf{ p_i(\lambda_0) = [V_{ui}+\lambda_{0} getFair(i)]:}$The score of  item i in $\mathbf{l_{u}^{ori}}$ , when$\ \lambda = \lambda_{0} $

            \item $\mathbf{rank(i)_{\lambda _{0}}}:$Get the current position of item i in $\mathbf{l_{u}^{ori}},rank(i)_{\lambda _{0}}\in \left [ 1,len(\mathbf{l_{u}^{ori}})\right],$when $\lambda $ = $\lambda _{0}$

            \item $\mathbf{NDCG_{(i,j)}^u \left[\lambda _0\right]}$ = $\frac{V_{ui}}{log_2(rank(i)_{\lambda _0}+1)}+\frac{V{uj}}{log_2(rank(j)_{\lambda _0}+1)}$: Items i and j,the NDCG components calculated based on their respective positions in list $\mathbf{l_u^{ori}}$
    
            \item  $\mathbf{A_{\lambda_0}=}\{(i,j)|\ \forall i,j$ $ \in \mathbf{I},$
            $i\ne j$
             $,p_i(\lambda_0)>p_j(\lambda _0),p_i(\lambda_0+\triangle  \lambda)<p_{j}(\lambda _{0}+\triangle \lambda ),\ rank(i)_{\lambda_{0}}\le K,\triangle \lambda \rightarrow 0^+\}  \ \forall \lambda _0 \in \left [ 0,+\infty  \right )$:The set A includes the Partial order relationship (i,j) which had been broken when \ $\lambda = \lambda_0,\triangle \lambda \rightarrow 0^+$
        \end{enumerate}
    \subsection{\textbf{Theory  and Proof}} 
    
        \textbf{Lemma 1: } when $\mathbf{A_{\lambda_0}\ne \varnothing ,\forall (i,j)\in A_{\lambda_0}
        \Rightarrow V_{ui} \ge V_{uj}},$ where $ \mathbf{\lambda _0\in \left [ 0,+\infty  \right )}$.

       \begin{proof}
      \ Assume$\ V_{ui}< V_{uj}\Rightarrow $when $\ \lambda = 0,\ p_{i(0)} - p_{j(0)}<0 $,with the increase of $ \mathbf{\lambda\in \left [ 0,+\infty  \right )}$. \\ 
        $p_{i(\lambda )}-p_{j(\lambda )}= \left[V_{ui}+\lambda getFair(i)\right]-
        \left[V_{uj}+\lambda getFair(j)\right]=V_{ui}-V_{uj}+\lambda \left[getFair(i)-getFair(j)\right]$.
        $\because V_{ui}-V_{uj}<0,\lambda \ge 0$
        $\Rightarrow$M($\lambda$) = $\left[p_i(\lambda)-p_j(\lambda)\right]$\  will be constantly less than 0  or have the transition from $<$0 at the begining to constant $\ge$ 0, while according to the defnition of $A_{\lambda _0},$
        M$\left(\lambda _{0}\right)$ = $\left[p_i(\lambda_0)-p_j(\lambda_0)\right] >0,M(\lambda _{0}+\triangle \lambda ) = \left[p_i{(\lambda_0+\triangle \lambda) }-p_j{(\lambda_0+\triangle \lambda) }\right]<0 \Rightarrow(i,j)\notin A_{\lambda _0}$
        Contradict $\Rightarrow$The  assumption is not valid$\Rightarrow  V_{ui}\ge V_{uj}$.
        \end{proof}

    \textbf{Theorem 1: }$\lambda_0 \in \left[0,+\infty\right),\forall u\in \mathbf{U},\mathbf{l_u^{ori}}$,the initial $NDCG_u$ is 1,the NDCG value decreases monotonically as the $\lambda_0$ increases.\label{Theorem1}

        \begin{proof}
         According Lemma 1, $\forall (i,j)\in A_{\lambda _0}\Rightarrow  V_{ui}\ge V_{uj},\lambda _0\in \left[0,+\infty\right)$.
         def:\  F(x) = $\frac{1}{log_2(1+x)},x\in \left[1,+\infty \right)$,
         x$\uparrow\   \Leftrightarrow$ F(x)$\downarrow$
         $\because NDCG_u = \sum_{i = 1}^{K} F(K)V_{ul_{u}^{ori}\left[K\right]}$

           when $A_{\lambda _0}\ne \varnothing$, there must exist an order such that items i,j are swapper by adjacent positions to achieve one-by-one destruction of the partial order relations in the set $A_{\lambda_{0}}$.\\

              $\Rightarrow\mathbf{\forall (i,j)}\in A_{\lambda _0},\Rightarrow $
    $\left\{
    \begin{matrix}
      &p_i(\lambda _0)=p_j(\lambda _0+\triangle \lambda ) & \\
      &p_j(\lambda _0)=p_i(\lambda _0+\triangle \lambda ) &       
    \end{matrix}\right.$
    \\ \\ \\
    $\Leftrightarrow$
    $\left\{
    \begin{matrix}
      &rank(i)_{\lambda _{0}}=rank(j)_{\lambda _{0}+\triangle \lambda }\  \textcircled{1}& \\
      &rank(j)_{\lambda _{0}}=rank(i)_{\lambda _{0}+\triangle \lambda }\  \textcircled{2}&       
    \end{matrix}\right.(exchange\ position)$\\ \\

    $\because   V_{ui}\ge V_{uj}\Rightarrow V_{ui}-V_{uj}\ge 0,
p_i(\lambda _0)-p_j(\lambda _0)\ge 0\Rightarrow F(rank(i)_{\lambda _{0}})-F(rank(j)_{\lambda _{0}})\ge 0\Rightarrow$\\$\left[V_{ui}-V_{uj}\right]\left[ F(rank(i)_{\lambda _{0}})-F(rank(j)_{\lambda _{0}})\right]\ge 0$\\

$\Rightarrow$$V_{ui}$$F\left(rank(i)_{\lambda _{0}}\right)+$$V_{uj}$$F\left(rank(j)_{\lambda _{0}}\right)$$\ge $$V_{ui}F\left(rank(j)_{\lambda _{0}}\right)\\+$$V_{uj}$$F\left(rank(i)_{\lambda _{0}}\right)$\\

$\small\xrightarrow[]{\textcircled{2}\textcircled{1}}V_{ui}F(rank(i)_{\lambda _{0}})+V_{uj}F(rank(j)_{\lambda _{0}})\ge V_{ui}F(rank(i)_{\lambda _{0}+\triangle \lambda })+V_{uj}F(rank(j)_{\lambda _{0}+\triangle \lambda })\\\\$
 $\Rightarrow NDCG_{(i,j)}^u [\lambda _0]\ge NDCG_{(i,j)}^u[\lambda _0+\triangle \lambda ]$\\
 $\therefore \forall \lambda _0\in [0,+\infty),\forall (i,j)\in A_{\lambda _0}\Rightarrow NDCG_{(i,j)}^u[\lambda _0+\triangle \lambda ]- NDCG_{(i,j)}^u[\lambda _0]\le 0$ It means that the loss of NDCG will continuous if exchanging position.\\
 $\Rightarrow \forall  \lambda _0\in [0,+\infty )\ \ \lim_{\triangle \lambda  \to 0^+} \frac{NDCG_u[\lambda _0+\triangle \lambda ]-NDCG_u[\lambda _0]}{\triangle \lambda }\\ \le 0$
 $\Rightarrow NDCG{_{+}^{'}}_u[\lambda ]\le 0. \  \ $ When  $A_{\lambda _0}= \varnothing$, NDCG remains unchanged, thus completing the proof.
 \end{proof}



\bibliographystyle{IEEEtran}
\bibliography{sample-base}

\begin{thebibliography}{10}
\providecommand{\url}[1]{#1}
\csname url@samestyle\endcsname
\providecommand{\newblock}{\relax}
\providecommand{\bibinfo}[2]{#2}
\providecommand{\BIBentrySTDinterwordspacing}{\spaceskip=0pt\relax}
\providecommand{\BIBentryALTinterwordstretchfactor}{4}
\providecommand{\BIBentryALTinterwordspacing}{\spaceskip=\fontdimen2\font plus
\BIBentryALTinterwordstretchfactor\fontdimen3\font minus \fontdimen4\font\relax}
\providecommand{\BIBforeignlanguage}[2]{{%
\expandafter\ifx\csname l@#1\endcsname\relax
\typeout{** WARNING: IEEEtran.bst: No hyphenation pattern has been}%
\typeout{** loaded for the language `#1'. Using the pattern for}%
\typeout{** the default language instead.}%
\else
\language=\csname l@#1\endcsname
\fi
#2}}
\providecommand{\BIBdecl}{\relax}
\BIBdecl

\bibitem{1}
A.~Castelnovo, R.~Crupi, G.~Greco, D.~Regoli, I.~G. Penco, and A.~C. Cosentini, ``A clarification of the nuances in the fairness metrics landscape,'' \emph{Scientific Reports}, vol.~12, no.~1, p. 4209, 2022.

\bibitem{4}
N.~Mehrabi, F.~Morstatter, N.~Saxena, K.~Lerman, and A.~Galstyan, ``A survey on bias and fairness in machine learning,'' \emph{ACM Computing Surveys (CSUR)}, vol.~54, no.~6, pp. 1--35, 2021.

\bibitem{5}
T.~Schnabel, A.~Swaminathan, A.~Singh, N.~Chandak, and T.~Joachims, ``Recommendations as treatments: Debiasing learning and evaluation,'' in \emph{international conference on machine learning}.\hskip 1em plus 0.5em minus 0.4em\relax PMLR, 2016, pp. 1670--1679.

\bibitem{6}
J.~Ding, Y.~Quan, X.~He, Y.~Li, and D.~Jin, ``Reinforced negative sampling for recommendation with exposure data.'' in \emph{IJCAI}.\hskip 1em plus 0.5em minus 0.4em\relax Macao, 2019, pp. 2230--2236.

\bibitem{7}
A.~Vardasbi, M.~de~Rijke, and I.~Markov, ``Cascade model-based propensity estimation for counterfactual learning to rank,'' in \emph{Proceedings of the 43rd International ACM SIGIR Conference on Research and Development in Information Retrieval}, 2020, pp. 2089--2092.

\bibitem{zhao2023fair}
C.~Zhao, L.~Wu, P.~Shao, K.~Zhang, R.~Hong, and M.~Wang, ``Fair representation learning for recommendation: A mutual information perspective,'' in \emph{Proceedings of the AAAI Conference on Artificial Intelligence}, vol.~37, no.~4, 2023, pp. 4911--4919.

\bibitem{zeng2021fair}
Z.~Zeng, R.~Islam, K.~N. Keya, J.~Foulds, Y.~Song, and S.~Pan, ``Fair representation learning for heterogeneous information networks,'' in \emph{Proceedings of the International AAAI Conference on Web and Social Media}, vol.~15, 2021, pp. 877--887.

\bibitem{ge2021towards}
Y.~Ge, S.~Liu, R.~Gao, Y.~Xian, Y.~Li, X.~Zhao, C.~Pei, F.~Sun, J.~Ge, W.~Ou \emph{et~al.}, ``Towards long-term fairness in recommendation,'' in \emph{Proceedings of the 14th ACM international conference on web search and data mining}, 2021, pp. 445--453.

\bibitem{li2021user}
Y.~Li, H.~Chen, Z.~Fu, Y.~Ge, and Y.~Zhang, ``User-oriented fairness in recommendation,'' in \emph{Proceedings of the web conference 2021}, 2021, pp. 624--632.

\bibitem{fu2020fairness}
Z.~Fu, Y.~Xian, R.~Gao, J.~Zhao, Q.~Huang, Y.~Ge, S.~Xu, S.~Geng, C.~Shah, Y.~Zhang \emph{et~al.}, ``Fairness-aware explainable recommendation over knowledge graphs,'' in \emph{Proceedings of the 43rd International ACM SIGIR Conference on Research and Development in Information Retrieval}, 2020, pp. 69--78.

\bibitem{zhu2021fairness}
Z.~Zhu, J.~Kim, T.~Nguyen, A.~Fenton, and J.~Caverlee, ``Fairness among new items in cold start recommender systems,'' in \emph{Proceedings of the 44th International ACM SIGIR Conference on Research and Development in Information Retrieval}, 2021, pp. 767--776.

\bibitem{beutel2019fairness}
A.~Beutel, J.~Chen, T.~Doshi, H.~Qian, L.~Wei, Y.~Wu, L.~Heldt, Z.~Zhao, L.~Hong, E.~H. Chi \emph{et~al.}, ``Fairness in recommendation ranking through pairwise comparisons,'' in \emph{Proceedings of the 25th ACM SIGKDD international conference on knowledge discovery \& data mining}, 2019, pp. 2212--2220.

\bibitem{xu2023p}
C.~Xu, S.~Chen, J.~Xu, W.~Shen, X.~Zhang, G.~Wang, and Z.~Dong, ``P-mmf: Provider max-min fairness re-ranking in recommender system,'' in \emph{Proceedings of the ACM Web Conference 2023}, 2023, pp. 3701--3711.

\bibitem{8}
G.~K. Patro, A.~Biswas, N.~Ganguly, K.~P. Gummadi, and A.~Chakraborty, ``Fairrec: Two-sided fairness for personalized recommendations in two-sided platforms,'' in \emph{Proceedings of the web conference 2020}, 2020, pp. 1194--1204.

\bibitem{9}
Y.~Wu, J.~Cao, G.~Xu, and Y.~Tan, ``Tfrom: A two-sided fairness-aware recommendation model for both customers and providers,'' in \emph{Proceedings of the 44th International ACM SIGIR Conference on Research and Development in Information Retrieval}, 2021, pp. 1013--1022.

\bibitem{naghiaei2022cpfair}
M.~Naghiaei, H.~A. Rahmani, and Y.~Deldjoo, ``Cpfair: Personalized consumer and producer fairness re-ranking for recommender systems,'' in \emph{Proceedings of the 45th International ACM SIGIR Conference on Research and Development in Information Retrieval}, 2022, pp. 770--779.

\bibitem{Karp1972}
R.~M. Karp, \emph{Reducibility among Combinatorial Problems}.\hskip 1em plus 0.5em minus 0.4em\relax Boston, MA: Springer US, 1972, pp. 85--103.

\bibitem{li2022fairgan}
J.~Li, Y.~Ren, and K.~Deng, ``Fairgan: Gans-based fairness-aware learning for recommendations with implicit feedback,'' in \emph{Proceedings of the ACM Web Conference 2022}, 2022, pp. 297--307.

\bibitem{wu2021learning}
L.~Wu, L.~Chen, P.~Shao, R.~Hong, X.~Wang, and M.~Wang, ``Learning fair representations for recommendation: A graph-based perspective,'' in \emph{Proceedings of the Web Conference 2021}, 2021, pp. 2198--2208.

\bibitem{wu2022selective}
Y.~Wu, R.~Xie, Y.~Zhu, F.~Zhuang, A.~Xiang, X.~Zhang, L.~Lin, and Q.~He, ``Selective fairness in recommendation via prompts,'' in \emph{Proceedings of the 45th International ACM SIGIR Conference on Research and Development in Information Retrieval}, 2022, pp. 2657--2662.

\bibitem{wang2022make}
J.~Wang, W.~Ma, J.~Li, H.~Lu, M.~Zhang, B.~Li, Y.~Liu, P.~Jiang, and S.~Ma, ``Make fairness more fair: Fair item utility estimation and exposure re-distribution,'' in \emph{Proceedings of the 28th ACM SIGKDD Conference on Knowledge Discovery and Data Mining}, 2022, pp. 1868--1877.

\bibitem{heuss2022fairness}
M.~Heuss, F.~Sarvi, and M.~de~Rijke, ``Fairness of exposure in light of incomplete exposure estimation,'' in \emph{Proceedings of the 45th International ACM SIGIR Conference on Research and Development in Information Retrieval}, 2022, pp. 759--769.

\bibitem{liu2023toward}
J.~Liu, ``Toward a two-sided fairness framework in search and recommendation,'' in \emph{Proceedings of the 2023 Conference on Human Information Interaction and Retrieval}, 2023, pp. 236--246.

\bibitem{wang2024intersectional}
Y.~Wang, P.~Sun, W.~Ma, M.~Zhang, Y.~Zhang, P.~Jiang, and S.~Ma, ``Intersectional two-sided fairness in recommendation,'' in \emph{Proceedings of the ACM on Web Conference 2024}, 2024, pp. 3609--3620.

\bibitem{wang2023two}
C.~Wang, Y.~Liu, Y.~Yu, W.~Ma, M.~Zhang, Y.~Liu, H.~Zeng, J.~Feng, and C.~Deng, ``Two-sided calibration for quality-aware responsible recommendation,'' in \emph{Proceedings of the 17th ACM Conference on Recommender Systems}, 2023, pp. 223--233.

\bibitem{14}
E.~Budish, ``The combinatorial assignment problem: Approximate competitive equilibrium from equal incomes,'' \emph{Journal of Political Economy}, vol. 119, no.~6, pp. 1061--1103, 2011.

\bibitem{15}
A.~J. Biega, K.~P. Gummadi, and G.~Weikum, ``Equity of attention: Amortizing individual fairness in rankings,'' in \emph{The 41st international acm sigir conference on research \& development in information retrieval}, 2018, pp. 405--414.

\bibitem{wang2021user}
L.~Wang and T.~Joachims, ``User fairness, item fairness, and diversity for rankings in two-sided markets,'' in \emph{Proceedings of the 2021 ACM SIGIR International Conference on Theory of Information Retrieval}, 2021, pp. 23--41.

\bibitem{zafar2019fairness}
M.~B. Zafar, I.~Valera, M.~Gomez-Rodriguez, and K.~P. Gummadi, ``Fairness constraints: A flexible approach for fair classification,'' \emph{The Journal of Machine Learning Research}, vol.~20, no.~1, pp. 2737--2778, 2019.

\bibitem{goh2016satisfying}
G.~Goh, A.~Cotter, M.~Gupta, and M.~P. Friedlander, ``Satisfying real-world goals with dataset constraints,'' \emph{Advances in neural information processing systems}, vol.~29, 2016.

\bibitem{zehlike2017fa}
M.~Zehlike, F.~Bonchi, C.~Castillo, S.~Hajian, M.~Megahed, and R.~Baeza-Yates, ``Fa* ir: A fair top-k ranking algorithm,'' in \emph{Proceedings of the 2017 ACM on Conference on Information and Knowledge Management}, 2017, pp. 1569--1578.

\bibitem{biega2018equity}
A.~J. Biega, K.~P. Gummadi, and G.~Weikum, ``Equity of attention: Amortizing individual fairness in rankings,'' in \emph{The 41st international acm sigir conference on research \& development in information retrieval}, 2018, pp. 405--414.

\bibitem{wan2021modeling}
M.~Wan, D.~Zha, N.~Liu, and N.~Zou, ``Modeling techniques for machine learning fairness: A survey,'' \emph{arXiv preprint arXiv:2111.03015}, 2021.

\bibitem{caton2020fairness}
S.~Caton and C.~Haas, ``Fairness in machine learning: A survey,'' \emph{arXiv preprint arXiv:2010.04053}, 2020.

\bibitem{pollin2008measure}
R.~Pollin, M.~Brenner, S.~Luce, and J.~Wicks-Lim, \emph{A measure of fairness: The economics of living wages and minimum wages in the United States}.\hskip 1em plus 0.5em minus 0.4em\relax Cornell University Press, 2008.

\bibitem{green2010minimum}
D.~A. Green and K.~Harrison, ``Minimum wage setting and standards of fairness,'' IFS working papers, Tech. Rep., 2010.

\bibitem{falk2006fairness}
A.~Falk, E.~Fehr, and C.~Zehnder, ``Fairness perceptions and reservation wages—the behavioral effects of minimum wage laws,'' \emph{The Quarterly Journal of Economics}, vol. 121, no.~4, pp. 1347--1381, 2006.

\bibitem{engbom2022earnings}
N.~Engbom and C.~Moser, ``Earnings inequality and the minimum wage: Evidence from brazil,'' \emph{American Economic Review}, vol. 112, no.~12, pp. 3803--47, 2022.

\bibitem{lin2016effects}
C.~Lin and M.-S. Yun, ``The effects of the minimum wage on earnings inequality: Evidence from china,'' in \emph{Income Inequality Around the World}.\hskip 1em plus 0.5em minus 0.4em\relax Emerald Group Publishing Limited, 2016, vol.~44, pp. 179--212.

\bibitem{10}
T.~Joachims and F.~Radlinski, ``Search engines that learn from implicit feedback,'' \emph{Computer}, vol.~40, no.~8, pp. 34--40, 2007.

\bibitem{jarvelin2002cumulated}
K.~J{\"a}rvelin and J.~Kek{\"a}l{\"a}inen, ``Cumulated gain-based evaluation of ir techniques,'' \emph{ACM Transactions on Information Systems (TOIS)}, vol.~20, no.~4, pp. 422--446, 2002.

\bibitem{hoare1962quicksort}
C.~A. Hoare, ``Quicksort,'' \emph{The computer journal}, vol.~5, no.~1, pp. 10--16, 1962.

\bibitem{gu2019addressing}
Q.~Gu, J.~Cao, Y.~Zhao, and Y.~Tan, ``Addressing the cold-start problem in personalized flight ticket recommendation,'' \emph{IEEE Access}, vol.~7, pp. 67\,178--67\,189, 2019.

\bibitem{he2016ups}
R.~He and J.~McAuley, ``Ups and downs: Modeling the visual evolution of fashion trends with one-class collaborative filtering,'' in \emph{proceedings of the 25th international conference on world wide web}, 2016, pp. 507--517.

\bibitem{13}
D.~Liang, J.~Altosaar, L.~Charlin, and D.~M. Blei, ``Factorization meets the item embedding: Regularizing matrix factorization with item co-occurrence,'' in \emph{Proceedings of the 10th ACM conference on recommender systems}, 2016, pp. 59--66.

\bibitem{he2014location}
P.~He, J.~Zhu, Z.~Zheng, J.~Xu, and M.~R. Lyu, ``Location-based hierarchical matrix factorization for web service recommendation,'' in \emph{2014 IEEE international conference on web services}.\hskip 1em plus 0.5em minus 0.4em\relax IEEE, 2014, pp. 297--304.

\end{thebibliography}


\begin{IEEEbiography}[{\includegraphics[width=1in,height=1.25in,clip,keepaspectratio]{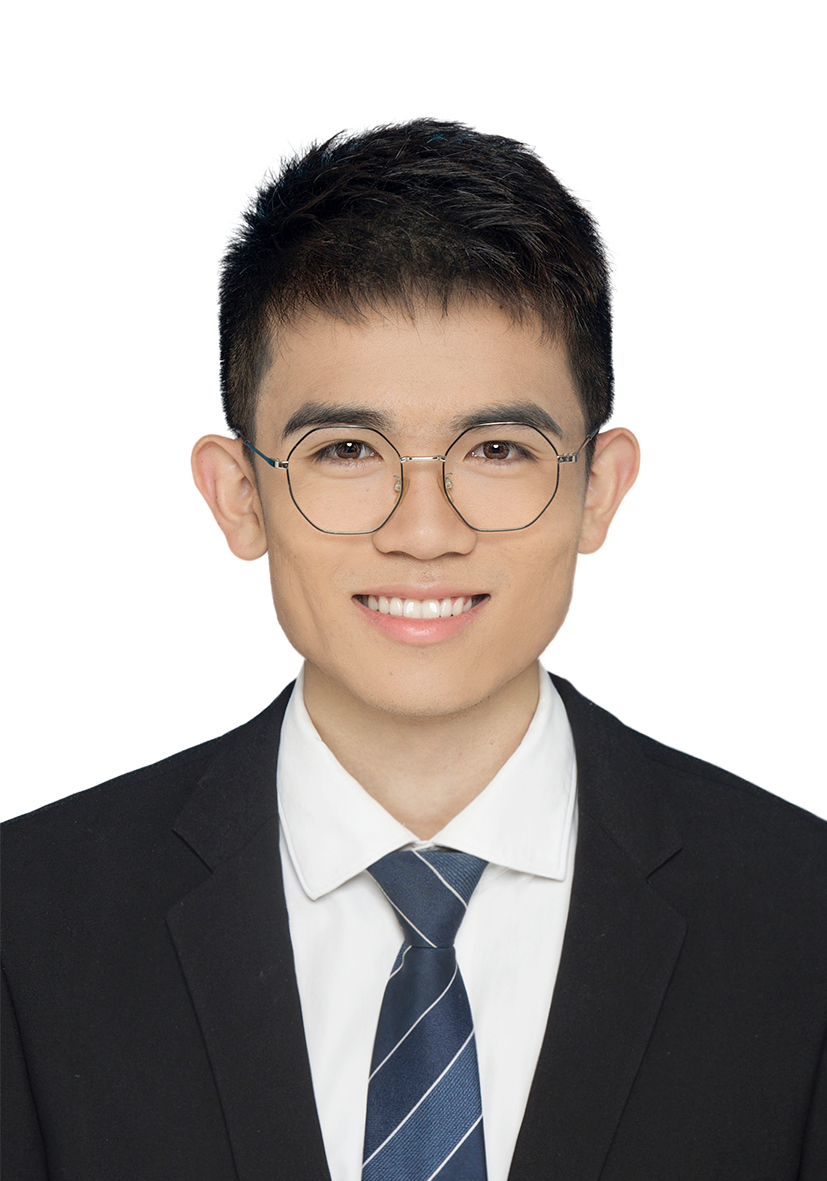}}]{Guoli Wu}
 was born in Guangdong, China, in 1999.
He received his master's degree from Tianjin University, Tianjin,
China, in 2024.

With the goal to
build a socially efficient and fair recommender system. His current research interests include service computing, recommender system, and fair machine learning algorithms.

\end{IEEEbiography}

\begin{IEEEbiography}[{\includegraphics[width=1in,height=1.25in,clip,keepaspectratio]{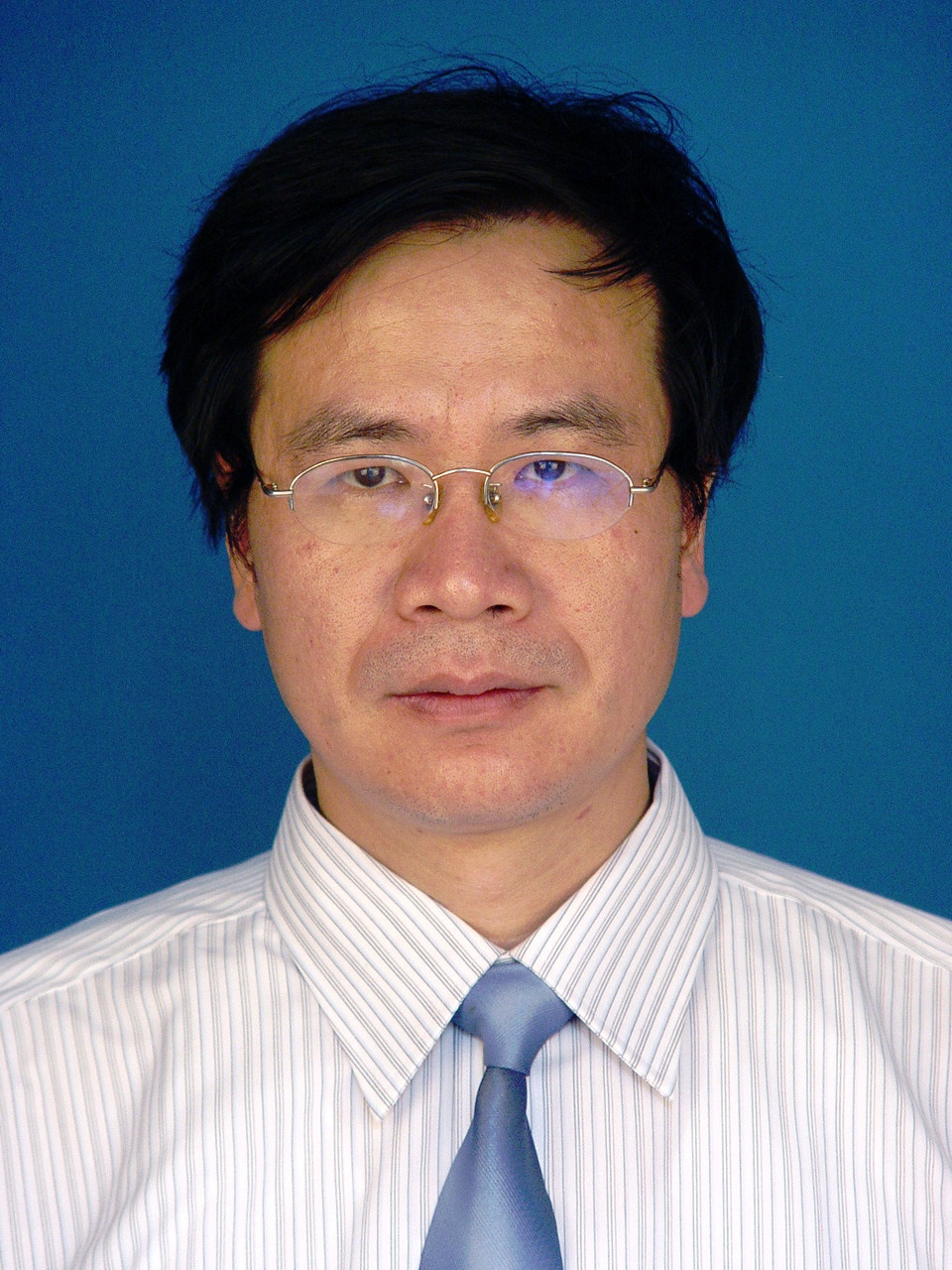}}]{Zhiyong Feng}
(Member, IEEE)  was born in 1965.
He received the Ph.D. degree from Tianjin University, Tianjin, China, in 1996.

He is currently a Professor with the College of Intelligence and Computing, Tianjin University, Tianjin, China.
He has authored more than 200 articles, one book and 40 patents.
His research interests include service computing, knowledge engineering, and software engineering.

Dr. Feng is a distinguished member of China Computer Federation (CCF),
a member of the Association for Computing Machinery (ACM),
and the Chairman of ACM China Tianjin Branch.
He has served as General Chair and Program Committee Chair of numerous international conferences, such as ICWS, ICSS, APWeb-WAIM, etc.

\\
\\
\end{IEEEbiography}
 \vspace{-4\baselineskip}

\begin{IEEEbiography}[{\includegraphics[width=1in,height=1.25in,clip,keepaspectratio]{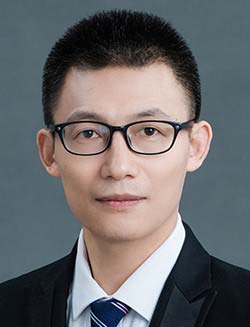}}]{Shizhan Chen}
(Member, IEEE) was born in 1975.
He received the bachelor's degree in engineering and the Ph.D. degree in computer applications from Tianjin University, Tianjin, China, in 1998 and 2010, respectively.

He is currently a Professor with the College of Intelligence and Computing, Tianjin University.
He has authored or co-authored 50 academic articles in Chinese and international academic journals, magazines, and conferences,
and applied for 40 national invention patents and 11 software copyrights.
His research interests include service computing and service-oriented architecture.

\end{IEEEbiography}
 \vspace{-4\baselineskip}
  
\begin{IEEEbiography}[{\includegraphics[width=1in,height=1.25in,clip,keepaspectratio]{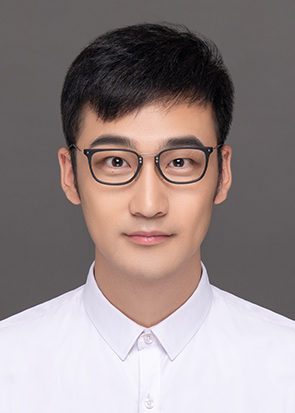}}]{Hongyue Wu}

 is currently an associate professor at the College of Intelligence and Computing, Tianjin University, China. He received his Ph.D. degree from Zhejiang University, China, in 2018. From 2017 to 2018, he was a joint Ph.D. student at the University of Sydney, Australia. His research interests include service computing, mobile edge computing, and cloud computing. He has published more than 50 papers and won the Best Paper Award at the ICSOC 2017. He is a member of the IEEE.
\end{IEEEbiography}

\begin{IEEEbiography}[{\includegraphics[width=1in,height=1.25in,clip,keepaspectratio]{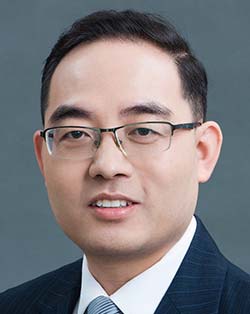}}]{Xiao Xue}
(Member, IEEE) was born in 1979.
He received the bachelor's degree in engineering from North China Electric Power University, Beijing, China, in 2001,
and the Ph.D. degree in engineering from the Institute of Automation, Chinese Academy of Sciences (CAS), Beijing, in 2007.

He is currently a Professor with the College of Intelligence and Computing, Tianjin University, Tianjin, China.
He has authored or co-authored more than 50 IEEE TRANSACTIONS and other high-quality articles and authored a book, Computational Experiment Method of Complex System.
His research interests include service computing, computational experiment, and swarm intelligence.

Dr. Xue is an Associate Editor of the IEEE TRANSACTIONS ON INTELLIGENT VEHICLES,
and an Editorial Board Member of the International Journal of Crowd Science.
He is one of the Chairs of Young Experts in Services Computing (YESC) Committee.

\end{IEEEbiography}
\vspace{-1\baselineskip}

\begin{IEEEbiography}[{\includegraphics[width=1in,height=1.25in,clip,keepaspectratio]{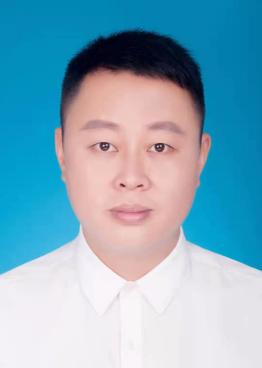}}]{Jianmao Xiao }
received the Ph.D. from the College of Intelligence and Computing, Tianjin University. He is an assistant professor at Jiangxi Normal University, is the deputy director of the Jiangxi Provincial Engineering Research Center of Blockchain Data Security and Governance. He is a member of the CCF TCSC. His main research interests include blockchain, service computing, intelligent software engineering. 
Dr. Xiao has authored more than 30 high-level academic papers, served as MONAMI 2022 Web Chair and ICSS 2022 PC Member. He also served as a reviewer for many domestic and international high-level journals and conferences in related fields.

\end{IEEEbiography}

\begin{IEEEbiography}[{\includegraphics[width=1in,height=1.25in,clip,keepaspectratio]{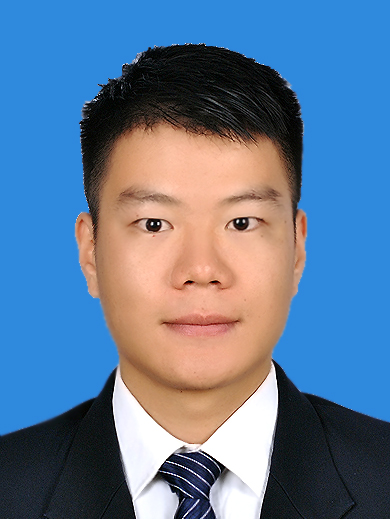}}]{Guodong Fan}
is currently pursuing a Ph.D. degree in Computer Science and Technology with the College of Intelligence and Computing, Tianjin University. 
He is working on cognitive services. His main research interests are representation learning, service computing, and software repository mining. 

\end{IEEEbiography}
\vspace{-1\baselineskip}

\begin{IEEEbiography}[{\includegraphics[width=1in,height=1.25in,clip,keepaspectratio]{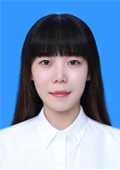}}]{Hongqi Chen}
was born in Rizhao, China, in 1994.
She received the master's degree in software engineering from China University of Petroleum (East China), Qingdao, China, in 2019.

She is currently pursuing a Ph.D. degree in software engineering with Tianjin University, Tianjin, China.
Her current research interests include service computing, graph neural networks and recommender system.
\end{IEEEbiography}

\begin{IEEEbiography}[{\includegraphics[width=1in,height=1.25in,clip,keepaspectratio]{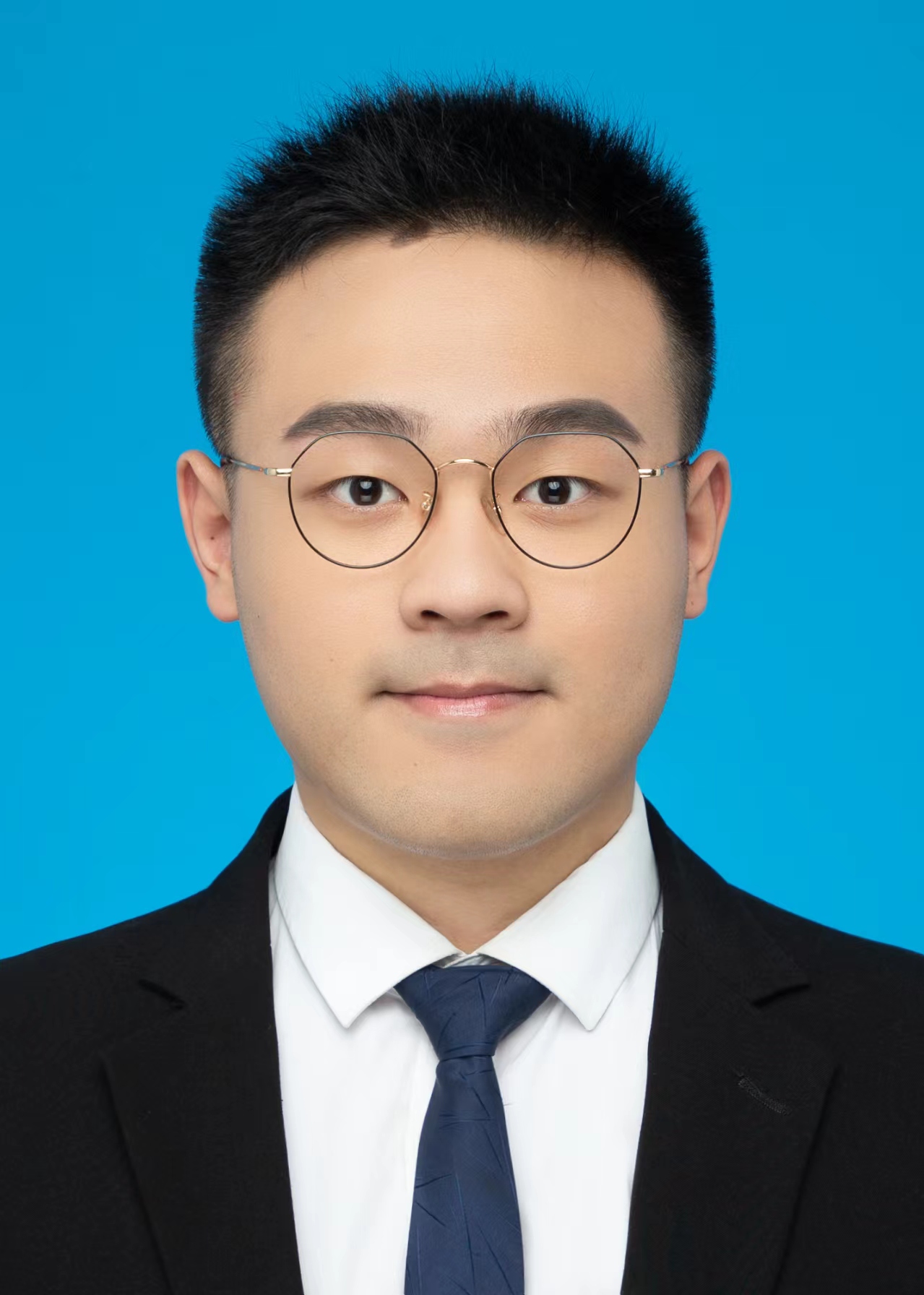}}]{Jingyu Li}
Ph.D student at College of Intelligence of Computing, Tianjin University. In 2022, He recieved bachelor's degree in Computer Science and Technology at Hohai University, Nanjing.  His current research interests include service computing, reinforcement learning, interactive recommendation.
\end{IEEEbiography}

\vfill
\end{document}